\documentclass[twocolumn,rmp,english,reprint, longbibliography, breaklinks=true, showkeys, showpacs=false, showkeywords=false]{revtex4-2}
\usepackage[T1]{fontenc}
\usepackage[utf8]{inputenc}
\usepackage{color,footnote}
\usepackage{babel}
\usepackage{amsmath,amssymb,mathrsfs}
\usepackage{subfigure}
\usepackage{graphicx}
\usepackage{physics,braket}
\usepackage{comment}
\usepackage{mathtools,bbm}
\usepackage[varg]{txfonts} 
\usepackage{natbib}
\usepackage[unicode=true,pdfusetitle,bookmarks=true,bookmarksnumbered=false,bookmarksopen=false,breaklinks=true,pdfborder={0 0 0},backref=false,colorlinks=true]
 {hyperref}
 
\makeatletter
\@ifundefined{textcolor}{}
{%
 \definecolor{BLACK}{gray}{0}
 \definecolor{WHITE}{gray}{1}
 \definecolor{RED}{rgb}{1,0,0}
 \definecolor{GREEN}{rgb}{0,1,0}
 \definecolor{BLUE}{rgb}{0,0,1}
 \definecolor{CYAN}{cmyk}{1,0,0,0}
 \definecolor{MAGENTA}{cmyk}{0,1,0,0}
 \definecolor{YELLOW}{cmyk}{0,0,1,0}
}

\pdfoutput=1
\hypersetup{colorlinks=true,citecolor=blue,linkcolor=cyan,urlcolor=blue,filecolor= green, breaklinks=true}
\usepackage{url}
\usepackage{breakurl}
\usepackage{enumitem}
\makeatother

\begin{document}

\author{Diego S. Starke\href{https://orcid.org/0000-0002-6074-4488}{\includegraphics[scale=0.05]{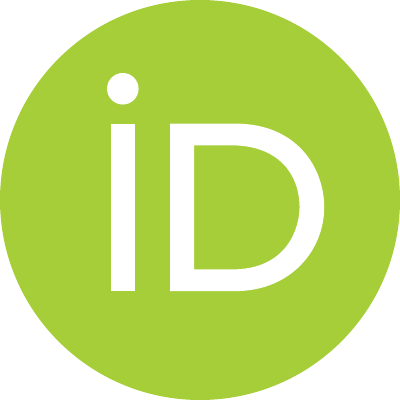}}}
%\email{starkediego@gmail.com}
\address{Physics Department, 
Federal University of Santa Maria, 97105-900,
Santa Maria, RS, Brazil}

\author{Marcos L. W. Basso\href{https://orcid.org/0000-0001-5456-7772}{\includegraphics[scale=0.05]{orcidid.pdf}}}
%\email{marcoslwbasso@hotmail.com}
\address{Department of Applied Mathematics, State University of Campinas, 13083-859, Campinas, S\~ao Paulo, Brazil}

\author{Tabish Qureshi\href{https://orcid.org/0000-0002-8452-1078}{\includegraphics[scale=0.05]{orcidid.pdf}}}
%\email{tqureshi@jmi.ac.in}
\address{Centre for Theoretical Physics, Jamia Millia Islamia, New Delhi-110025, India}

\author{Jonas Maziero\href{https://orcid.org/0000-0002-2872-986X}{\includegraphics[scale=0.05]{orcidid.pdf}}}
%\email{jonas.maziero@ufsm.br}
\address{Physics Department, 
Federal University of Santa Maria, 97105-900,
Santa Maria, RS, Brazil}

\selectlanguage{english}

\title{Bohr's complementarity}

\begin{abstract}
Quantum complementarity is a fundamental feature of quantum systems and has captivated the physics research community for nearly a century, with significant advancements emerging in recent decades.
This review traces the historical evolution of the concept of complementarity, beginning with Bohr's original formulation. It then explores its modern quantification through complementarity relations and its profound connection to the foundational postulates of quantum theory. Furthermore, it delves into key related developments, such as the operational definition of complementarity in the context of incompatible observables, its potential links with quantum uncertainty relations and contextuality, its various applications, and other pertinent topics.
This review aims to serve physicists interested in quantum resources, quantum correlations, and the foundational principles of quantum mechanics.
\end{abstract}

%\keywords{Bohr's complementarity principle; Quantum complementarity principle; Uncertainty relations}

%\date{\today}

\maketitle

\tableofcontents

%
%---------------------------
%
\section{Introduction}
\label{sec:intro}

Earliest studies of light led to the emergence of two distinct theories. While \citet{Huygens1690} elucidated optical phenomena through a wave theory,  \citet{Newton1704} proposed a corpuscular model, suggesting that light is composed of a stream of high-speed particles. While Newton's corpuscular model seemed attractive at first, at the dawn of the 19th century, \citet{Young1804} revealed the wave characteristics of light by demonstrating interference with his famous two-slit experiment. Further experiments corroborated the wave nature of light~\cite{Grangier1986}, particularly with the identification of electromagnetic waves, of which light is a specific instance \cite{Maxwell1985}.

The narrative shifted once more when \citet{Einstein1905} explained the photoelectric effect through his proposition that light comprises energy quanta that move without segmentation and can only be absorbed or produced as a whole. These massless light particles, termed photons, possess a fixed amount of energy $E = h\nu$, where $h$ is Planck's constant and $\nu$ is the frequency of light. \citet{Taylor1909} demonstrated a Young-type interference, with extremely low intensity light, an experiment lasting three months. This long span of time pointed to mostly one photon at a time, but still successfully yielded interference.

According to \citet{deBroglie1924}, all massive particles also possess wave-like properties.
The wavelength associated with a particle of momentum $p$ is expressed as $\lambda = h/p$. The demonstration of diffraction of an electron beam at a nickel crystal by \citet{Davisson1927, Davisson1928} and \citet{Estermann1930} confirmed the wave nature of massive particles, and validated de Broglie's hypothesis.
More recent experiments confirmed these effects in increasingly large composite systems~\cite{Arndt1999, Gerlich2011, Fein2019, Pedalino2026}.
All these experiments called for a revision of the mental picture of microscopic particles, whether of light or matter. It was a curious observation that different experiments revealed the microscopic entities acting either as waves or as particles, never both at the same time.

Niels Bohr was deeply intrigued by this observation and interpreted it as a manifestation of the fundamental nature of quantum systems. He emphasized that wave-like behavior---revealed through interference---and particle-like behavior---associated with approximately well-defined trajectories---are mutually exclusive descriptions. According to Bohr, the type of behavior observed depends on the experimental arrangement, and in any given experiment only one of these complementary aspects can be manifested.~\citet{Bohr1928} elevated this idea to the status of a general principle, now known as Bohr's complementarity principle (BCP). In modern, more colloquial terms, this principle is often referred to simply as wave-particle duality.

However, in Bohr's original formulation, complementarity is not merely a statement about the mutual exclusivity of wave-like and particle-like descriptions, but rather a deeper assertion about the limits of classical concepts when applied to quantum phenomena. The impossibility of simultaneously attributing definite values to complementary properties reflects the contextual nature of quantum measurements, where the experimental arrangement plays an essential role in defining what can be meaningfully observed. On this matter,~\citet{Bohr1935} asserted that ``\textit{it is only the mutual exclusion of any two experimental procedures, permitting the unambiguous definition of complementary physical quantities, which provides room for new physical laws}''. In this sense, Bohr’s complementarity goes beyond a simple dichotomy between waves and particles, pointing instead to a structural feature of quantum theory: the necessity of employing mutually exclusive, yet jointly exhaustive, descriptions to account for physical reality.

Over the years, this qualitative insight has been refined and reinterpreted within different frameworks, ranging from operational and information-theoretic approaches to more formal developments based on the mathematical structure of quantum mechanics. In particular, one of the directions emphasizes the role of incompatible observables and the limitations imposed on their joint measurability, providing a natural bridge between complementarity, uncertainty relations, and quantum correlations~\cite{Bush2006}. In parallel, significant progress has been achieved through the development of quantitative complementarity relations, which will constitute the central focus of this review. Early steps in this direction can be traced back to the seminal work of~\citet{Wootters1979}, who first formulated complementarity in quantitative terms using information-theoretic concepts, establishing a trade-off between the acquisition of which-path information and the visibility of interference fringes. Building on these ideas, quantitative duality relations were later formalized in influential works by~\citet{Greenberger1988} and~\citet{Englert1996}, providing precise trade-offs between wave-like and particle-like properties. More recent developments have extended these relations in multiple directions and, in addition, have shown how such complementarity relations can be systematically obtained from the mathematical structure of quantum mechanics. These include extensions to multi-path interference, entropic formulations, and connections with quantum coherence and correlations, offering a unified and quantitative framework for understanding complementarity. These developments suggest that complementarity should not be regarded as an isolated postulate or as something derived from the uncertainty principle, but rather as a manifestation of constraints encoded in quantum theory.

In addition to the scientific literature, several articles and books provided extensive historical and conceptual discussions of Bohr's ideas and their development. Notable examples include Bohr's own writings~\cite{Bohr1935, Bohr1937, Bohr1949, Bohr1950}, as well as historical and philosophical studies~\cite{Jammer1966, Jammer1974, Faye1991, Honner1987, Murdoch1987, Folse1988, Holton1988, Howard1994, Beller1999, Katsumori2011}. Collectively, these works explore the historical, epistemological, and philosophical aspects of complementarity. While these contributions are essential for a deeper understanding of the conceptual foundations of the subject, the present review will not address these aspects in detail, focusing instead on the physical, operational, and quantitative developments of complementarity within quantum theory.

Finally, we stress that this review does not aim to provide an exhaustive account of all developments related to complementarity and wave-particle duality. Given the vast and continuously growing literature on the subject, the discussion presented here necessarily reflects a particular perspective shaped by the themes and approaches emphasized by the authors. Our goal is therefore to offer a coherent and structured overview of general results and concepts, rather than a complete survey of the subject.

The remainder of this review is organized as follows. In Sec.~\ref{sec:twoslit}, we revisit the two-slit experiment as the paradigmatic setting for understanding complementarity. In Sec.~\ref{sec:1st}, we present an overview of quantitative complementarity relations, focusing on the development of duality relations and their generalizations. In particular, we discuss different approaches to quantifying particle-like behavior, including predictability and distinguishability, as well as the evolution of wave-like quantifiers, from interferometric visibility to more general measures based on quantum coherence. Building on these ideas, we review triality relations in multi-path interference, which incorporate additional physical features such as entanglement, and discuss other relevant extensions that contribute to a more unified understanding of complementarity. In Sec.~\ref{sec:QCP}, we adopt a more formal perspective and show how complementarity relations can be systematically derived from the structure of quantum mechanics. In particular, we demonstrate how duality and triality relations emerge directly from the properties of the density operator, highlighting the role of purity, coherence, and correlations in constraining the behavior of quantum systems. This approach provides a unified framework in which different complementarity relations can be obtained and compared, clarifying the connections between predictability, distinguishability, coherence, and entanglement, and establishing a more systematic route to their generalization. In Sec.~\ref{sec:retro}, we turn to retro-inference scenarios, including delayed-choice and quantum eraser experiments, which provide particularly striking illustrations of complementarity. These setups challenge the standard definitions of wave-like and particle-like behavior, showing that their characterization depends sensitively on the choice of observables and measurement schemes. In Sec.~\ref{sec:furthertopics}, we explore broader aspects and applications of complementarity, extending the discussion beyond standard interferometric settings. In particular, we examine connections with contextuality and realism, resource-theoretic approaches to predictability and coherence, and applications in quantum information protocols, including entanglement swapping and quantum thermodynamics. We also discuss the role of complementarity in relation to incompatible observables and consider extensions beyond non-relativistic quantum mechanics. The review concludes with a discussion of perspectives and open questions in Sec.~\ref{sec:persp}, highlighting possible directions for future research and emphasizing the continuing relevance of complementarity in modern quantum theory. Finally, in Sec.~\ref{sec:conc}, we give our conclusions and summarize what has been achieved.

%
%---------------------------
%
\section{Two-slit interference}
\label{sec:twoslit}

In this section, we revisit the historical and conceptual foundations of quantum complementarity, focusing on its most emblematic manifestation: wave-particle duality in interferometric scenarios, particularly in the double-slit experiment. We begin by recalling the early discussions that shaped the principle, including the well-known debates between Einstein and Bohr, which played a central role in clarifying the limits imposed by quantum mechanics on the simultaneous description of complementary properties. We then review subsequent developments in the literature that sought to identify the physical mechanisms underlying the loss of interference, including approaches based on momentum transfer (``momentum kicks''), entanglement with which-path detectors, and uncertainty relations. These perspectives provide complementary insights into the origin of wave-particle trade-offs and set the stage for the quantitative formulations discussed in the following sections.

%---------------------------
\subsection{Two-slit experiment and Bohr's complementarity}

The double-slit experiment conducted with
massive objects serves as a testing ground for numerous foundational concepts in quantum theory. It has become a cornerstone in illustrating BCP as the wave-particle duality. The double-slit experiment encapsulates the core of quantum theory in such a profound manner that \citet{Feynman1965} asserted it is a phenomenon ``...\textit{which has in it the heart of quantum mechanics; in reality it contains the \emph{only} mystery}'' of the theory. Even today, the two-slit interference experiment conducted with electrons \cite{Merli1976} may be considered the most beautiful experiment in physics, according to~\citet{Crease2002}. BCP emerges in a two-slit experiment when one tries to find out which of the two slits each 
object passes through. The knowledge of which slit a
quantum system passed through illustrates its particle nature. On the other hand, the interference pattern that unfolds from the cumulative effect of individual objects passing through the double-slit and hitting the screen illustrates its wave nature. BCP asserts that in a two-slit experiment, acquiring knowledge of which slit each object passed through would destroy the interference pattern.

%---------------------------
\subsection{Bohr--Einstein debate}

\begin{figure}
\centerline{\resizebox{8.5cm}{!}{\includegraphics{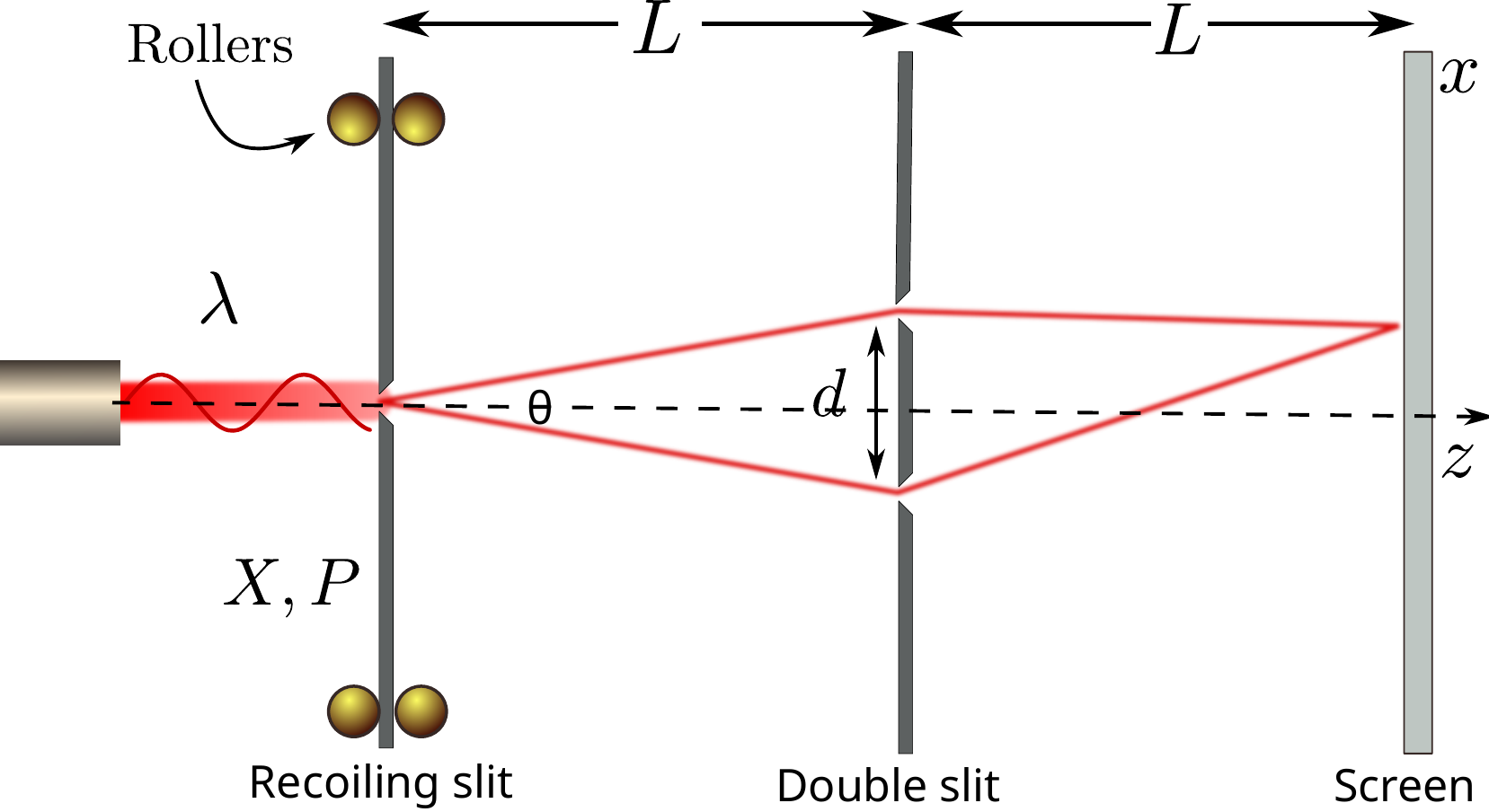}}}
\caption{Schematic diagram of Einstein's recoiling slit experiment. 
Objects traveling along the $z$-axis, with average zero momentum in the $x$-direction, pass through a single slit which is free to slide along the $x$-axis, before reaching the double-slit. Particles reaching the upper (lower) slit would have acquired a net momentum in the positive (negative) $x$-direction, causing a recoil of the recoiling slit in the negative (positive) $x$-direction. This way the momentum of the recoiling slit carries the information about which slit the particle went through.} 
\label{rslit}
\end{figure}

Historically, the debate on BCP and wave-particle duality was set in motion with Einstein challenging the principle~\cite{Maleki2023} at the 5th Solvay conference on quantum mechanics in 1927 in Brussels. Einstein proposed his famous ``recoiling slit'' experiment to argue that one could get an interference pattern in a two-slit experiment and still extract information about which slit each 
object passed through~\cite{Bohr1949}. The basic idea of his proposal was that before passing through the double-slit, each object passed through a single slit that was free to move perpendicular to the direction of motion of the
object, and was light enough to experience a recoil if it slightly changed the course of an object during its passage, as depicted in Fig.~\ref{rslit}. Einstein argued that after an object passed through the double-slit and was registered on the screen, one could measure the recoil of the single slit and infer the path information of the object. Einstein's proposal initially caught Bohr off guard, but the next day he came up with a brilliant rebuttal, arguing that in such an experiment, the recoiling slit itself must be treated quantum mechanically. He showed that each different \emph{position} of the recoiling slit would lead to a slightly shifted interference pattern. An accurate value of the recoil \emph{momentum} of the single-slit would imply an uncertain value of its position, due to the uncertainty principle. The uncertain position of the single slit would in turn lead to an uncertain position of the interference pattern. This uncertainty in the position of the interference pattern is enough to wash it out \cite{Bohr1949,Qureshi2013,Maleki2023}.

Bohr's rebuttal can be demonstrated by a simple back-of-the-envelope calculation as follows. In the simplified recoiling slit setup in Fig.~\ref{rslit}, the central position on the screen with zero path difference between the two paths is determined by the position of the single recoiling slit. If the objects move essentially along the $z$-axis with a momentum $p$, with a small momentum spread along the $x$-axis, the difference in the momenta of the object in the two slits is 
$\Delta p_x = 2p\sin(\theta/2) \approx p\theta = \frac{h}{\lambda}\theta=\frac{h}{\lambda}\frac{d}{L}$. The recoil momentum of the
slit should be measured at least as accurately as $\Delta p_x = \frac{h}{\lambda}\frac{d}{L}$. The position of the single-slit, and consequently the position of the interference, is uncertain at least by an amount 
$\Delta x = \frac{\hbar}{2\Delta p_x} = \frac{\lambda L}{4\pi d}$. Since this uncertainty is of the same order as the fringe separation $\frac{\lambda L}{d}$, the interference will be washed out. Although Einstein's challenge was refuted, it served to put BCP on a stronger foundation \cite{Bohr1949}.

Einstein's recoiling slit experiment continues to be a subject of interest. Initially, it was merely a thought experiment put forward to illustrate a point. However, with modern technological progress, it has been experimentally realized in several different ways.
The recoiling slit is implemented at the quantum-classical boundary by~\citet{Bertet2001}, where which-way information is encoded in the state of microwave cavities interacting with the atoms. The excitation of a cavity mode plays the role of the slit's recoil, storing path information and thereby suppressing interference.~\citet{Utter2007} were the first to realize the experiment using photons scattered from a harmonically trapped ion. \citet{Schmidt2013} went a step further by measuring the momentum transferred to a free-floating molecular double slit and the momentum change of the atom scattering from it.~\citet{Liu2015} implemented this \textit{gedanken} experiment using resonant X-ray photoemission from molecular oxygen. Near equilibrium, the two atomic centers act as coupled slits, and interference is observed. In contrast, in the dissociative regime, where the centers behave as independent (decoupled) slits, the electron momentum transfer encodes which-path information through Doppler shifts, leading to the suppression of interference.
\citet{Fedoseev2025} realized this experiment by scattering photons off a free, minimum-uncertainty atomic wave packet. \citet{Zhang2025} also performed the experiment using a single trapped atom as a movable slit whose momentum uncertainty can be controlled. This allows a direct observation of how photon-slit entanglement governs the trade-off between interference visibility and which-path information, as well as the transition from quantum to classical behavior.

%---------------------------
\subsection{Uncertainty, entanglement and momentum kicks}
\label{sec:IIC_ucertainty}

Bohr's analysis of Einstein's recoiling slit experiment by invoking Heisenberg's uncertainty principle led many to believe that complementarity stemmed from the uncertainty principle, or at least was fundamentally connected to it \cite{Tan1993}. Some even believed that BCP may just be a restatement of the uncertainty principle in a different language.
\citet{Scully1991} proposed a two-slit which-path experiment and argued that the complementarity principle is usually enforced by the position-momentum uncertainty relation, as originally pointed out by Bohr. However, their setup provided a way around it, and pointed to other mechanisms that enforce complementarity. The other mechanism was quantum correlations between the particle and the which-way detector. \citet{Storey1994} countered this by proving that a minimum amount of momentum should be transferred to the interfering particle for the interference to be destroyed. This point of view appears to be the result of interpreting Bohr's reply to Einstein's recoiling slit experiment as meaning that since the recoiling slit experienced a momentum recoil, the particle passing through would also experience an equal but opposite back action, or a ``momentum kick''. These momentum kicks were believed to lead to the loss of interference. The controversy raged for decades without any resolution \cite{Englert1995,Wiseman1995,Wiseman1998,Liu2012,Durr2000}. However, \citet{Storey1994} were unable to explain how these momentum kicks could arise in the setup proposed by \citet{Scully1991}, as there was no recoiling slit involved. \citet{Wiseman1997} introduced the idea of ``nonlocal'' momentum kicks in order to reconcile the momentum kick picture with \citet{Scully1991}'s experiment. \citet{Tanimura2015} showed that although the discussion on Einstein's recoiling slit experiment involved momentum kicks and back action, the uncertainty relations that are involved are not of the \citet{Ozawa2003} type, but of the \citet{Kennard1927} type. On the other side, \citet{Scully1991} were unable to point out quantum correlations in Einstein's recoiling slit experiment. Interestingly, it was shown later that Einstein's recoiling slit experiment can be explained without invoking the uncertainty principle, but by recognizing that the momentum states of the recoiling slit necessarily get entangled with the two paths of the particle \cite{Qureshi2013}. Further related experimental studies were carried out by \citet{Wiseman2003,Mir2007,Wiseman2019}.

The controversy finally came to an end after it was demonstrated that the two apparently conflicting views were actually complementary and could be reconciled within the same quantum formalism \cite{Qureshi2018,Pathania2021}. The basic idea behind the resolution is as follows. If $\psi_1(x)$ and $\psi_2(x)$ represent the wavefunctions of the particle emerging from the two slits, respectively, and $|d_1\rangle, |d_2\rangle$ are the two states of a which-way detector, then according to the basic formulation of quantum measurement~\cite{Neumann1955}, the two should necessarily get entangled. The entangled state may be represented as
\begin{eqnarray}
    \psi(x) &=& \tfrac{1}{\sqrt{2}}(\psi_1(x)|d_1\rangle + \psi_2(x)|d_2\rangle).
    \label{entstate}
\end{eqnarray}
The particle then evolves in time, traveling to the screen, and is registered at any random position $x$ on the screen, with a probability density given by
\begin{align}
    & |\psi(x,t)|^2 = \tfrac{1}{2}(|\psi_1(x,t)|^2\langle d_1|d_1\rangle + |\psi_2(x,t)|^2\langle d_2|d_2\rangle  \nonumber\\
    & + \psi_1^*(x,t)\psi_2(x,t)\langle d_2|d_1\rangle + \psi_1(x,t)\psi_2^*(x,t)\langle d_1|d_2\rangle).
\end{align}
It is straightforward to see that if $|d_1\rangle, |d_2\rangle$ are orthogonal, the two cross terms, which represent interference, would vanish. This argument represents the view of \citet{Scully1991}, namely that in a which-way experiment the interference is destroyed by the correlations between the particle and the which-way detector. However, by choosing another basis for the which-way detector, $|d_{\pm}\rangle =\tfrac{1}{\sqrt{2}}(|d_1\rangle \pm |d_2\rangle)$, the entangled state in Eq.~(\ref{entstate}) can be written as
\begin{eqnarray}
\Psi(x) &=& \tfrac{1}{2} [ \psi_1(x)+ \psi_2(x)]|d_+\rangle + \tfrac{1}{2} [ \psi_1(x) - \psi_2(x)]|d_-\rangle.
\label{ent2}
\end{eqnarray}
The two terms in the above expression differ only in that in the second term there is a ``phase flip'' between the two paths. Interestingly, this phase flip can be interpreted as a momentum kick \cite{Qureshi2018}. If we assume that one slit is located at $x=0$, and the other at $x=d$, and the wavefunction of the particle, as it emerges from the double-slit, is sharply localized at the two slit positions, the second term can also be written as
\begin{eqnarray}
\tfrac{1}{2} e^{ip_{0}x/\hbar} [\psi_1(x) + \psi_2(x)]|d_-\rangle =
\tfrac{1}{2} [ \psi_1(x) + e^{ip_{0}d/\hbar} \psi_2(x)]|d_-\rangle,~~~
\end{eqnarray}
where $p_0=\hbar\pi/d$ is a momentum kick the particle receives when the which-way detector state is $|d_{-}\rangle$. This is due to the fact that $e^{ip_{0}d/\hbar} = -1$, making this term identical to the second one in Eq.~(\ref{ent2}). This is not a fortuitous case, and the idea can be generalized to multi-slit interference too \cite{Qureshi2018}. Thus, the loss of interference in \emph{any} which-way two-path interference experiment can be interpreted either as arising from entanglement or due to the tiny \emph{virtual} momentum kicks that the particle appears to receive, depending on which basis of which-way detector states one chooses to consider. This is \emph{always} true in any such experiment, contrary to the opposite belief that none of these views is universal \cite{Englert2000b}. However, it should be emphasized that these momentum kicks are not real, but only virtual, again arising from the entanglement of the particle with the which-way detector. So, there is no momentum transfer to the particle from anywhere.

%
%---------------------------
%
\section{Quantum complementarity relations}
\label{sec:1st}

In this section, we present a historically motivated overview of the development of quantitative complementarity relations. We begin by tracing the early formulations, highlighting the different approaches adopted to quantify particle-like behavior, in particular through predictability and distinguishability. We also discuss the evolution of wave-like quantifiers, starting from interferometric visibility and naturally leading to the more general and robust framework based on quantum coherence. Building on these developments, we review the emergence of triality relations, which incorporate additional physical features beyond the standard wave-particle trade-offs, as well as other relevant approaches that have contributed to a deeper and more unified understanding of complementarity.

%---------------------------
\subsection{Duality relations in two-slit interference}
\label{sec:duality}

While formulating the principle of complementarity, \citet{Bohr1928} \emph{qualitatively} stated that one can observe the wave or particle character of a quantum system in a given experimental setup, but not both simultaneously.
However, Bohr's reply to Einstein's recoiling slit experiment implies that if one tries to obtain path information about a particle passing through a double-slit, it would degrade the interference. 

\citet{Wootters1979} decided to probe this aspect in detail, aiming to provide a quantitative statement of BCP. They assumed the recoiling slit to be constrained by a harmonic oscillator potential, and carried out a quantitative analysis. They quantified the wave-nature by the sharpness of interference, which in modern parlance is called  interferometric visibility. The particle nature was quantified in terms of the information one has obtained about which of the two paths the particle followed. They showed that these could be observed in the same experiment, but increasing one would decrease the other. More importantly, they used Shannon information theory to quantify the information we lack regarding the path of the particle, denoting it by $H$. Denoting the sharpness of interference by $S$, they evaluated the minimum path information one has to give up in order to obtain that sharpness of interference, and denoted it by $H(S)$. They derived an inequality:
\begin{equation}
    H \ge H(S) .
\end{equation}
This result expresses that the residual uncertainty about the particle’s path must be at least as large as the minimal uncertainty compatible with an interference pattern of sharpness $S.$

Following the formulation introduced by \citet{Wootters1979}, \citet{Bartell1980} proposed two additional experimental configurations to enable a more detailed examination of the phenomena.~\citet{Mittelstaedt1987} showed that, in a modified Mach--Zehnder interferometer, path information and interference visibility can be jointly accessed as unsharp particle-like and wave-like features.~\citet{Sanders1989}, using a Kerr-based quantum non-demolition which-path measurement, further demonstrated that increasing the accuracy of path inference---quantified via the probe signal-to-noise ratio---leads to a corresponding reduction in interference visibility. Taken together, these results establish an explicit quantitative trade-off between wave and particle aspects, thereby laying the groundwork for the modern formulation of complementarity relations in terms of information-theoretic and operational measures.

%---------------------------
\subsubsection{Predictability}

By 1988, two-slit interference experiments with cold neutrons ($\lambda \approx 2\mathrm{nm}$) had already been successfully realized by~\citet{Zeilinger1988}. Building on this experimental context,~\citet{Greenberger1988} introduced a framework to quantify wave-particle duality in neutron interference setups. They argued that in the absence of any path measuring device, if the probabilities of a neutron passing through the two slits were unequal, one could correctly guess which slit the neutron passed through with more than $50$ percent success. They introduced the \textit{a priori} path predictability to quantify the corpuscular character and used fringe-contrast to characterize the 
wave behavior.
They assumed that in two-slit neutron interference experiments, one may represent the wavefunction at the screen by
\begin{eqnarray}
    \psi = (a e^{ik_xx} + b e^{-ik_xx}e^{i\phi}) e^{ik_zz},
\end{eqnarray}
where $a,b$ are real parameters and $k_x, k_z$ are determined by the Bragg scattering condition, $\phi$ being a phase factor. Before we continue, let's define the  Michelson fringe visibility \cite{Michelson1890} as
\begin{equation}
\mathcal{V} \equiv \frac{I_{\max}-I_{\min}}{I_{\max}+I_{\min}} \equiv \frac{\Pr_{\max} - \Pr_{\min}}{\Pr_{\max} + \Pr_{\min}},
\label{eq:vis}
\end{equation}
where $I_{\max}$ is the intensity at a bright fringe (maximum), and $I_{\min}$ is the minimum intensity at the dark fringe (minimum), or the equivalence using the maximum $(\Pr_{\max})$ and minimum $(\Pr_{\min})$ probabilities, respectively.
The probability density of a neutron hitting the screen at a position $x$ is then given by
\begin{equation}
    |\psi|^2 = a^2 + b^2 + 2ab\cos(2k_xx+\phi).
\end{equation}
With this, it follows that the visibility in the experiment is obtained as $\mathcal{V} = 2ab/(a^2+b^2)$. The path predictability is defined as $\mathcal{P} = (a^2-b^2)/(a^2+b^2)$ (assuming $a>b$). One can then write a wave-particle duality relation as $\mathcal{P}^2 + \mathcal{V}^2 = 1$. For partially coherent beams, one obtains an inequality
\begin{equation}
\mathcal{P}^2 + \mathcal{V}^2 \le 1 .   
\label{dualityG}
\end{equation}

\citet{Jaeger1995} further refined this duality relation.
Though this was the first duality relation, it was not of much use in the kind of experiments discussed while formulating the BCP. For example, in Einstein's recoiling slit experiment, since a particle is equally probable to pass through either slit, the path predictability turns out to be zero, even though the interference is degraded.

%---------------------------
\subsubsection{Distinguishability: minimum-error discrimination}

A major breakthrough came in 1996 when \citet{Englert1996} addressed the problem of quantifying BCP and introduced the concept of \emph{path distinguishability}. We briefly outline how this path distinguishability can be derived, although Englert did it somewhat differently and for the symmetric case.

\begin{figure}
\centerline{\resizebox{8.5cm}{!}{\includegraphics{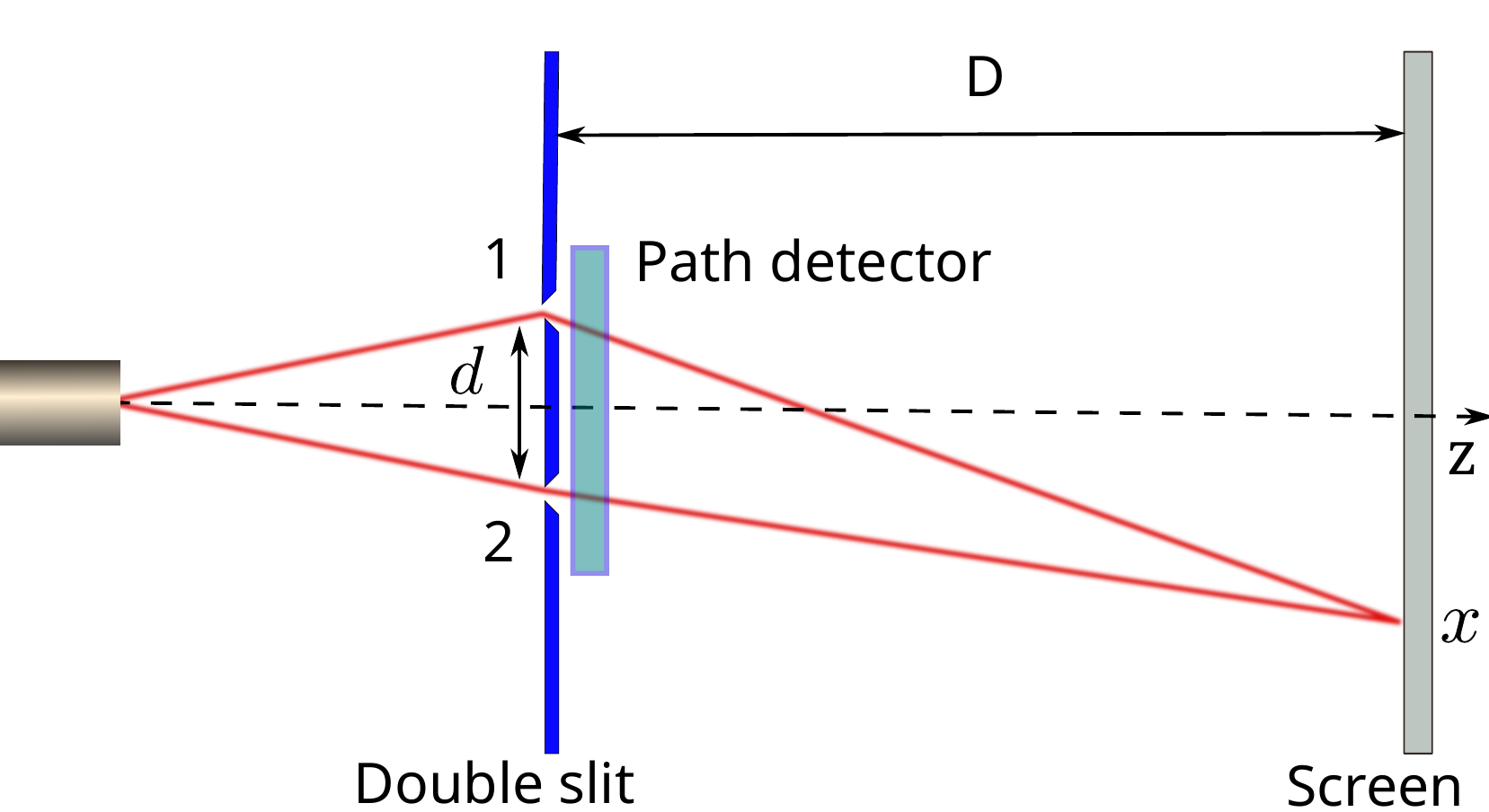}}}
\caption{Schematic diagram of a two-slit interference experiment in the presence of a which-way (or path) detector.}
\label{twoslit}
\end{figure}

Let us consider a quantum particle (let us call it a \emph{quanton} for brevity) passing through a double-slit, as shown in Fig. \ref{twoslit}. The state of the quanton, assuming it has a different probability of passing through each slit, can be written as
\begin{equation}
    |\psi\rangle = \sqrt{p_1}|\psi_1\rangle + \sqrt{p_2}|\psi_2\rangle ,
    \end{equation}
where $p_1,p_2$ are the probabilities of the quanton passing through slit 1 and 2, respectively. This form of the state is quite general, as any phase factors can be absorbed in $|\psi_1\rangle,|\psi_2\rangle$. Let us now introduce a which-way detector just after the double-slit. Without specifying the nature of this path-detector, we just assume that if the quanton passes through slit 1 (slit 2), the path-detector acquires the state $|d_1\rangle$ ($|d_2\rangle$). The states $|d_1\rangle,|d_2\rangle$, though normalized, are in general not orthogonal. The combined state can then be written as
\begin{equation}
    |\psi\rangle = \sqrt{p_1}|\psi_1\rangle|d_1\rangle + \sqrt{p_2}|\psi_2\rangle|d_2\rangle ,
    \label{ent3}
\end{equation}
which is an entangled state, a necessary requirement for the path-detector to be able to tell which way the quanton went. If one is interested in the state of the path-detector irrespective of what happens to the quanton, one can trace over the quanton states and get a reduced density operator:
\begin{equation}
    \rho_d = \text{Tr}_q(|\psi\rangle\langle\psi|)
     = p_1 |d_1\rangle\langle d_1| + p_2 |d_2\rangle\langle d_2| .
\end{equation}
Now we have a mixture of two path-detector states, and the problem of distinguishing between the two paths of the quanton reduces to successfully discriminating between these two non-orthogonal states. If one uses the minimum-error discrimination of quantum states, the Helstrom bound \cite{Helstrom1976} on the success probability of doing so is given by
\begin{equation}
    P_s \le \frac{1}{2} + \frac{1}{2}\sqrt{1 - 4p_1p_2|\langle d_1|d_2\rangle|^2} .
\end{equation}
In order to extract a meaningful distinguishability quantifier based on measurement, one should subtract the random guess probability $1/2$ from it. Then a normalized path distinguishability is simply given by
\begin{equation}
    \mathcal{D} = \sqrt{1 - 4p_1p_2|\langle d_1|d_2\rangle|^2} .
    \label{DE}
\end{equation}
Interference visibility (in a realistic scenario) for the state in Eq.~(\ref{ent3}) is \cite{Menon2018}
\begin{equation}
    \mathcal{V} \le 2\sqrt{p_1p_2}|\langle d_1|d_2\rangle| .
    \label{V2}
\end{equation}
Combining Eqs. (\ref{DE}) and (\ref{V2}), we arrive at Englert's well known duality relation:
\begin{equation}
    \mathcal{D}^2 + \mathcal{V}^2 \le 1 .
    \label{dualityE}
\end{equation}
This inequality is saturated for all pure states. This elegant relation can be considered a quantitative statement of BCP. It tells us precisely how much of the particle and wave natures it is possible to observe simultaneously, in principle. More specifically, it tells us, if a specific amount of path information about the quanton is available, what is the maximum visibility of interference one can get. It was beautifully verified in an atom interferometry experiment by \citet{Durr1998}, demonstrating the validity of BCP for massive particles. It was also experimentally verified using single photons \cite{Jacques2008a,Schwindt1999}. It has even been verified in nuclear magnetic resonance experiments \cite{Peng2003}. \citet{Abe2026} extended the formalism of minimum error discrimination in two-path interference to situations where the path detector may be in a mixed state. They introduced a modified distinguishability which yields a duality relation which is an equality even for mixed path-detector states, and experimentally demonstrated the same with unpolarized single photons. Wave-particle duality has been subsequently explored in various generalized scenarios \cite{Vaccaro2011,Vaccaro2006}. Interestingly, while state discrimination can lead to quantitative duality relations,~\citet{Lu2020a} demonstrated the converse: quantitative wave-particle duality itself can be used to infer the fundamental limits of state discrimination, namely the Helstrom bound.

%---------------------------
\subsubsection{Distinguishability: unambiguous state discrimination}

Since Englert derived his duality relation mostly for symmetric beam interference, wave-particle duality continued to be explored in \emph{asymmetric} beam interference \cite{LiLi2012,LiuL2017}. In a rather recent development, a different approach to dealing with wave-particle duality in asymmetric two-path interference was taken~\cite{Menon2018}. The view followed in this approach was that BCP deals with the knowledge of which of the two paths the quanton followed, and minimum-error state discrimination often provides a wrong answer. One would then like to explore an approach in which one can unambiguously state that the quanton followed a particular path. Unambiguous quantum state discrimination (UQSD) \cite{Ivanovic1987, Dieks1988, Peres1988} does just that, namely, it unambiguously distinguishes between two non-orthogonal states. However, the caveat is that discrimination sometimes fails, but the experimenter finds out that the process has failed. The end result is that the discrimination process does not succeed every time, but when it does, it provides the answer without error. The only task that remains is to minimize the probability of failure. In UQSD, if two non-orthogonal states $|d_1\rangle$ and $|d_2\rangle$ occur with probabilities $p_1$ and $p_2$, respectively, the maximum probability of unambiguously distinguishing between them is given by $1 - 2\sqrt{p_1p_2}|\langle d_1|d_2\rangle|$ \cite{Ivanovic1987}. 

In the scenario described by Eq.~(\ref{ent3}), unambiguously distinguishing between $|d_1\rangle$ and $|d_2\rangle$ would mean unambiguously distinguishing between the two paths the quanton can take. A path distinguishability can then be defined as the maximum probability of unambiguously discriminating between the two path-detector states \cite{Menon2018}
\begin{equation}
    \mathcal{D}_Q = 1 - 2\sqrt{p_1p_2}|\langle d_1|d_2\rangle| .
    \label{DQ}
\end{equation}
Combining Eqs. \eqref{V2} and  (\ref{DQ}), we obtain a \emph{linear} duality relation:
\begin{equation}
    \mathcal{D}_Q + \mathcal{V} \le 1 .
    \label{dualityQ}
\end{equation}
This linear duality relation and Englert's quadratic duality relation may look deceptively similar, but the process of obtaining distinguishability is very different in the two. When the asymmetry is very high, a different duality relation holds
\cite{Menon2018}
\begin{equation}
    \frac{\mathcal{D}_Q}{\tfrac{1}{2}(1+\mathcal{P}_0)} + \frac{\mathcal{V}^2}{\mathcal{V}_0^2} \le 1 ,
    \label{dualityQ2}
\end{equation}
where $\mathcal{P}_0,\mathcal{V}_0$ are the \textit{a priori} predictability and \textit{a priori} visibility used in the duality relation of \citet{Greenberger1988}, given by Eq.~(\ref{dualityG}).

Using a UQSD based distinguishability allows an interesting interpretation of the duality relation. All the quantons passing through a two-path interferometer and registering on the screen can be split into two sub-ensembles depending on whether the UQSD process succeeded or failed. The fraction of quantons, for which one unambiguously finds out which of the two paths they took, is just the distinguishability. These quantons do not contribute to interference, rather they degrade it. The fraction of the quantons, for which the UQSD fails, is just the interference visibility. Such quantons contribute to sharp interference. These two fractions together constitute the duality relation in Eq.~(\ref{dualityQ}) \cite{Qureshi2016}.

\citet{Chen2022} investigated both these duality relations in a designed asymmetric beam interference experiment, and by utilizing the polarization degree of freedom of the photon as a which-way detector, they experimentally confirmed both forms of the duality relations. They also showed that the
difference between the UQSD strategy and the minimum error discrimination strategy can be understood by calculating the mutual information gained
through the measurements. Their conclusion was that the linear form in Eq.~(\ref{dualityQ}) is tighter than the quadratic form in Eq.~(\ref{dualityE}).

%---------------------------
\subsection{Duality relations in multi-slit interference}
\label{sec:duality_multi-slits}

After the BCP was firmly established for two-slit interference, it was natural to expect that it would also apply to multi-slit interference. Substantial effort in that direction has been put forth over the years trying to find a duality relation that would apply to multi-slit experiments well.~\citet{Durr2001} and \citet{Englert2008} initiated a detailed analysis of multi-slit interference, also prescribing what conditions any new predictability and interference visibility measures should satisfy on physical grounds.

For \textit{predictability} quantifiers, the criteria are as follows:
\begin{enumerate}[label=(\arabic*), leftmargin=*, labelsep=0.4em]
    \item[P.1] It must be a continuous function of the diagonal elements of the density matrix of the quanton, i.e. the initial probabilities $\rho_{kk}$.
    \item[P.2] It must be be invariant under permutations of the states indexes. 
    \item[P.3] It should reach its global maximum when one has perfect knowledge of the path of the quanton, i.e., $\rho_{kk} = 1$ for some $k$.
    \item[P.4] It should reach its global minimum when there is equal probability for the quanton to go through any path, i.e., $\rho_{kk}=1/n$ for all $k$.
    \item[P.5] Any attempt towards the equalization of the probabilities, the value of the predictability cannot increase. More concretely, if $\rho_{jj}>\rho_{kk}$ for some pair $(j,k)$, the predictability measure cannot be increased by setting $\rho_{jj}\rightarrow\rho_{jj}-\epsilon$ and $\rho_{kk}\rightarrow\rho_{kk}+\epsilon$, for $\epsilon\in\mathbb{R}_{+}$ and $\epsilon\ll1$.
    \item[P.6] It must be a convex function of the density matrix of the quanton, i.e., $\mathcal{P}( \lambda \rho + (1-\lambda) \sigma)\le \lambda \mathcal{P}(\rho) + (1-\lambda)\mathcal{P}(\sigma)$, with $\lambda \in [0,1]$ and $\rho, \sigma$ being valid density matrices.  
\end{enumerate}

For \textit{wave nature} (or visibility) quantifiers, a similar set of conditions has been specified for any new quantifier:
\begin{enumerate}[label=(\arabic*), leftmargin=*, labelsep=0.4em]
    \item[V.1] It should be possible to give a definition of $\mathcal{V}$ that is based only on the interference pattern, i.e., it must be invariant under permutations of the states' indexes.
    \item[V.2] It should vary continuously as a function of the matrix elements of the reduced density operator of the quanton.
    \item[V.3] If the system shows no interference, visibility should reach its global minimum.
    \item[V.4] If the state of the quanton is pure and all paths are equally probable, the visibility should reach its global maximum.
    \item[V.5] Visibility should be independent of our choice of the coordinate system, i.e., insensitive to resetting the zero points of the phases and insensitive to changing the numbering of the beams.
    \item[V.6] It must be a convex function of the density matrix of the quanton, i.e., $\mathcal{V}( \lambda \rho + (1-\lambda) \sigma)\le \lambda \mathcal{V}(\rho) + (1-\lambda)\mathcal{V}(\sigma)$, with $\lambda \in [0,1]$ and $\rho, \sigma$ being valid density matrices. 
\end{enumerate}

\textit{Distinguishability} measure. In the presence of a path-detector device, an entangled state should also be formed in a multi-slit experiment, i.e.,
\begin{equation}
|\Psi\rangle = c_1|\psi_1\rangle|d_1\rangle+c_2|\psi_2\rangle|d_2\rangle
+ \dots + c_n|\psi_n\rangle|d_n\rangle,
\label{entn}
\end{equation}
where $n$ paths are available to the quanton, and $|\psi_k\rangle$ represents the state of the particle in the $k$'th path. The \emph{a priori} probability for the quanton to be in the $k$'th path is $|c_k|^2$. Correspondingly, the path-detector will be in the state $|d_k\rangle$. Based on this, we observe that any new measure of distinguishability must meet the following basic requirements:

\begin{enumerate}[label=(\arabic*), leftmargin=*, labelsep=0.4em]
    \item[D.1] It should be a continuous function of the initial probabilities $|c_k|^2$.
    \item[D.2] It should reach its global maximum when one has perfect knowledge of the path of the quanton, i.e., say $|c_j|^2=1$ and the rest are zero.
    \item[D.3] It should reach its global minimum when there is equal probability for the quanton to go through any path, and all path-detector states are parallel. 
    \item[D.4] Any attempt towards the equalization of the probabilities ($|c_{11}|^2, |c_{11}|^2, \dots,|c_{nn}|^2$) or parallelization of detector states ($|d_1\rangle, |d_2\rangle, \dots, |d_n\rangle$), should decrease the distinguishability.
    \item[D.5] For any state, the distinguishability should be greater than or equal to the corresponding \emph{a priori} predictability.
    \item[D.6] It must be a convex function of the density matrix of the quanton, i.e., $\mathcal{D}( \lambda \rho + (1-\lambda) \sigma)\le \lambda \mathcal{D}(\rho) + (1-\lambda)\mathcal{D}(\sigma)$, with $\lambda \in [0,1]$ and $\rho, \sigma$ being valid density matrices.
\end{enumerate}

Over the years, several attempts were made  to formulate a duality relation for multi-slit interference~\cite{Bimonte2003a,Bimonte2003b,Englert2008,Zawisky2002,DeRaedt2012,Siddiqui2015}. The breakthrough came when \citet{Bera2015} demonstrated a tight duality relation between path distinguishability and a newly formulated measure of coherence, often called $\ell_1$-coherence \cite{Baumgratz2014}. The new measure of coherence, motivated from an information-theoretic point of view, is simply the sum of the absolute values of the off-diagonal elements of the density operator corresponding to the state under consideration. If the density operator is given by $\rho$ and the reference basis is $\{|j\rangle\}$, the coherence is given by $\mathcal{C}(\rho)=\sum_{j\neq k} |\langle j|\rho|k\rangle|$. It is obvious that this measure depends on the choice of the basis. For the state in Eq.~(\ref{entn}), if one is only interested in the quanton, irrespective of what happens to the path-detector, one can trace over the path-detector states to get a reduced density operator for the quanton:
\begin{eqnarray}
    \rho_q = \text{Tr}_d(|\Psi\rangle\langle\Psi|)
    = \sum_{j,k=1}^n c_jc_k^*|\psi_j\rangle\langle\psi_k|~ \langle d_k|d_j\rangle.
\end{eqnarray}
Using the different path states of the quanton as the basis states, the \emph{normalized} coherence for this state comes out to be \cite{Bera2015}:
\begin{eqnarray}
    \mathcal{C}(\rho_q) =  \tfrac{1}{n-1}\sum_{j\neq k} |c_jc_k| |\langle d_k|d_j\rangle| ,
    \label{C}
\end{eqnarray}
where the factor $\tfrac{1}{n-1}$ was introduced to normalize the measure. This normalized coherence satisfies all the criteria for a good measure of wave-nature, as elaborated in the preceding discussion, and can be used as $n$-path visibility \cite{Qureshi2019a}. Not just that, for $n=2$ it reduces to the visibility given by Eq.~(\ref{V2}). In order to distinguish between the $n$ non-orthogonal path-detector states, and hence between the $n$ paths of the quanton, a $n$-path distinguishability was suggested earlier by \citet{Siddiqui2015} as
\begin{eqnarray}
    \mathcal{D}_Q =  1 - \tfrac{1}{n-1}\sum_{j\neq k} |c_jc_k| |\langle d_k|d_j\rangle| ,
    \label{DQn}
\end{eqnarray}
where the right hand side is just an upper bound on the probability of unambiguously distinguishing between the $n$ non-orthogonal states $\{|d_j\rangle\}$ \cite{Zhang2001,Qiu2002}. Using Eqs.~(\ref{C}) and (\ref{DQn}) one gets an elegant duality relation for $n$-path interference \cite{Bera2015} as
\begin{equation}
    \mathcal{D}_Q + \mathcal{C} =  1 .
    \label{dualityQn}
\end{equation}
This duality relation is the most general quantitative statement of BCP. For $n=2$ it reduces to Eq.~(\ref{dualityQ}). \citet{Machado2020} experimentally explored this duality relation using a 3-slit interference experiment. 

\citet{Bagan2016} endeavored to establish a multi-path duality relation of the quadratic form using coherence, based on minimum-error discrimination of states. Nevertheless, their motivation seems to arise from a flawed perspective, specifically the belief that since Eq.~(\ref{dualityQn}) is linear, it overlooks a certain region above the line $\mathcal{D}_Q + \mathcal{C} =  1$ where both wave and particle characteristics coexist \cite{Bagan2016}. It was later shown that the quadratic duality relation of \citet{Bagan2016} could not be saturated for a  lot of pure states, and in many situations, their distinguishability and coherence increased and decreased together, violating the very spirit of BCP.
It was also shown that Eq.~(\ref{dualityQn}) can indeed be modified into a $n$-path duality relation of a quadratic form \cite{Qureshi2017} as
\begin{equation}
    \mathcal{D}^2 + \mathcal{C}^2 =  1 ,
    \label{dualityQnq}
\end{equation}
where $\mathcal{D}^2=\mathcal{D}_Q(2-\mathcal{D}_Q)$. So, as things stand now, minimum-error discrimination is not suitable for formulating duality relations for $n > 2$.

Coherence can also be used to generalize the two-slit predictability based duality relation in Eq.~(\ref{dualityG}) of \citet{Greenberger1988} to $n$-path interference. Let us define a $n$-path predictability as
\begin{equation}
\mathcal{P} \equiv \sqrt{1 - \left(\tfrac{1}{n-1}\sum_{j\neq k} \sqrt{\rho_{jj}}\sqrt{\rho_{kk}} \right)^2},
\label{Pn}
\end{equation}
where $\rho$ is the density operator for a quanton passing through $n$ paths. Then it can be shown that the following duality relation holds \cite{Roy2019}
\begin{equation}
    \mathcal{P}^2 + \mathcal{C}^2 \le  1 ,
    \label{dualityRQ}
\end{equation}
where the inequality is saturated by pure quanton states. It reduces to Eq.~(\ref{dualityG}) for $n=2$. The duality relation in Eq.~(\ref{dualityRQ}) was experimentally verified in multi-path interference implemented by large-scale silicon-integrated multi-path interferometers \cite{Chen2021}.

%---------------------------
\subsection{Coherence and interference visibility}

While it is welcoming that the path-distinguishability and coherence satisfy a tight duality relation in multi-path interference, one would like to know how coherence can be measured in an interference experiment. For two-slit interference, the Michelson fringe contrast from Eq.~\eqref{eq:vis} is sufficient to capture the sharpness of interference. One problem is that while in two-slit interference all maxima are of equal height, in multi-slit interference there are primary maxima as well as secondary maxima. Could there be another way to measure the sharpness of interference? \citet{Paul2017} carried out a wave-packet analysis of multi-slit interference, and showed that if one measures the intensity at a primary maximum, and then the intensity at the same position by making the beam incoherent (and thus destroying interference), then the coherence is given by
\begin{equation}
    \mathcal{C}_{exp} = \frac{1}{n-1}\frac{I_{\max}-I_{\text{inc}}}{I_{\text{inc}}} ,
    \label{Cexp}
\end{equation}
where $n$ is the number of slits,  $I_{\max}$ the intensity at a primary maximum, and $I_{\text{inc}}$ is the intensity at the position of the same primary maximum, but with \emph{incoherent} light. $I_{\text{inc}}$ may also be obtained, particularly by measuring intensities at that point by opening only one slit at a time and then averaging over them, if the setup so permits.

\begin{figure}
\centerline{\resizebox{8.5cm}{!}{\includegraphics{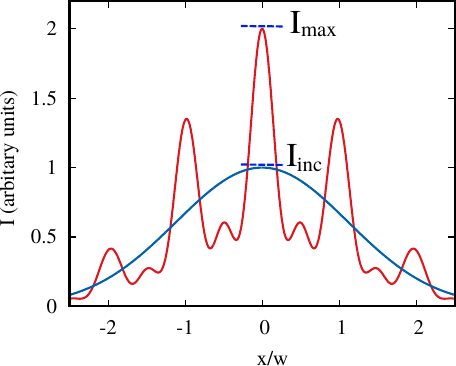}}}
\caption{A typical multi-slit interference pattern with the intensity plotted against position in the units of fringe-width, indicating $I_{\max}$ and $I_{\text{inc}}$.}
\label{Cint}
\end{figure}

Fig. \ref{Cint} shows $I_{\max}$ and $I_{\text{inc}}$ in a typical multi-slit interference pattern. Interestingly, in a two-slit interference, $\mathcal{C}_{exp}$ turns out to be identical to the fringe visibility from Eq.~\eqref{eq:vis}.
So the quantity $\mathcal{C}_{exp}$, given by Eq.~(\ref{Cexp}), can replace visibility in all interference experiments. Coherence has been experimentally measured using this method by 
some research groups \cite{Chen2021,Machado2020}.

More than two decades ago, \citet{Mei2001} carried out an interesting multiple beam Ramsey interference experiment with cesium atoms, which produced some surprising results. They carried out 4-path interference experiments with an atomic beam. By applying controlled decoherence~\cite{Zurek2003, Schlosshauer2007} to a selected path, they observed a loss of contrast in the interference. Furthermore, the phase of that beam was flipped by $\pi$. By applying decoherence in that situation, they surprisingly observed an \emph{increase} in the contrast of the interference. Decoherence in such an experiment can also be interpreted as photons carrying away some path information about the atoms. According to BCP, this should lead to a decrease in the wave-nature of the atoms. The results of these experiments appear to contradict BCP. Some authors have contended that the visibility of interference is not an effective indicator of either interference or the wave nature of light, based on this outcome \cite{Luis2001,Bimonte2003a}. Furthermore, it has been suggested that there are path measurements that do not diminish interference \cite{Luis2001}. These experiments have been theoretically examined in detail by \citet{Mishra2019}. They have shown that indeed interference visibility does increase with increasing decoherence. They have shown that the same problem persists even in 3-path interference, and a duality relation formulated for three slit interference \cite{Siddiqui2015} does not hold in such situations. Interestingly, \citet{Mishra2019}  demonstrated that while the traditional interferometric visibility increases with increasing decoherence in such experiments, quantum coherence decreases. This is further evidence that quantum coherence is a good quantifier of the wave-nature of quantons. One might think that the problem is solved because one can just measure coherence in such experiments using the technique described in the preceding discussion.
However, it turns out that this technique doesn't work in scenarios where the interference pattern itself is distorted by selectively phase-flipping paths. This problem remained unsolved for about two decades.

A rather elegant solution to this problem was provided in 2019, where it was pointed out that, in a multi-path interference experiment, coherence can be measured by opening only pairs of paths at a time and measuring visibility (fringe contrast) for all pairs of paths \cite{Qureshi2019b}. Coherence is then simply the average of those visibilities. Measuring multi-path distinguishability in Eq.~\eqref{DQn} is also a problem, while two-path distinguishability can be measured easily. It was shown that multi-path distinguishability is just the average of two-path distinguishabilities of all path pairs. The results of \citet{Qureshi2019b} can be summarized as  follows
\begin{eqnarray}
\mathcal{D}_Q = \tfrac{2}{n(n-1)}\sum_{\text{pairs}} {\mathcal{D}_Q}_{ij},
~~~~~~~~
\mathcal{C} = \tfrac{2}{n(n-1)}\sum_{\text{pairs}} \mathcal{V}_{ij}.
\label{connect}
\end{eqnarray}
Several outstanding problems in multi-path wave-particle duality are resolved with these equations.
This approach illustrates that the wave nature can consistently be defined through interference, and that it is complementary to path information. The concept of path distinguishability in the context of multi-path interference is redefined in terms of the distinguishability of pairs of paths.
With this, the concept of path distinguishability in multi-path interference remains valid even when the states of the path detectors constitute a linearly dependent set. It provides a way to measure multi-path distinguishability. \citet{Machado2020} experimentally demonstrated measuring coherence in 3-slit interference by measuring visibilities of pairs of paths at a time.

%---------------------------
\subsection{Triality relations}
\label{sec:triality}

As discussed in section \ref{sec:duality}, for two-path interference, two kinds of duality relations have been common in the literature. One is based on predictability, represented by Eq. (\ref{dualityG}), and the other is based on distinguishability, represented by Eqs.~(\ref{dualityE}) and (\ref{dualityQ2}). Despite being different in experimental contexts, the differentiation between the two is frequently overlooked in scholarly works. Furthermore, the link between these two relations remained unclear for an extended period, which also applies to the relationship between predictability and distinguishability.
\citet{Jakob2007,Jakob2010} paved the way for resolving this issue by studying two entangled qubits, and looking at single-partite and bipartite properties. They defined predictability and visibility as single-partite properties $\mathscr{P}$ and $\mathscr{V}$, respectively. The entanglement between the two qubits was quantified by the concurrence $\mathscr{C}$, which is a bipartite property. The distinguishability $\mathscr{D}$ is also a bipartite property. They showed that the duality relation $\mathscr{D}^2 + \mathscr{V}^2 \le 1$ can be written as a triality relation
\begin{equation}
    \mathscr{P}^2 + \mathscr{C}^2 + \mathscr{V}^2 \le 1. \label{JBrelation}
\end{equation}
The distinguishability $\mathscr{D}$ is related to the predictability $\mathscr{P}$ as $\mathscr{D}^2 = \mathscr{P}^2 + \mathscr{C}^2$, thus having a single-partite component and a bipartite component. This interesting result, unfortunately, remained confined to studies on entangled qubits, and its usefulness to interference of quantons was not explored for a long time. Some effort was made to extend this idea to multidimensional systems by \citet{Hiesmayr2008}.

A decade later, the results of \citet{Jakob2007,Jakob2010} were applied to the study of interference in classical optics \cite{Qian2018}. The view taken in those studies was that the traditional duality relations were incomplete without accounting for entanglement. Consequently, the term \emph{complete complementarity relations} has emerged in the literature as an alternative designation for triality relations. In what follows, we will use both nomenclatures interchangeably. That study of interference in classical optics spurred an interest in exploring triality relations in \emph{multi-path} interference of quantons, as BCP was already quantified in multi-path interference by Eq.~(\ref{dualityQn}). In an investigation of multi-path interference of quantons, \citet{Qureshi2021b} showed that using the multi-path predictability of Eq.~(\ref{Pn}), the duality relation  in Eq.~(\ref{dualityQnq}) can be written as
\begin{equation}
    \mathcal{P}^2 + \mathcal{E}^2 + \mathcal{C}^2 = 1 ,
\label{pce}
\end{equation}
where 
\begin{eqnarray}
\mathcal{E}^2 = \tfrac{1}{(n-1)^2}\Big[ \Big( \sum_{i\neq j}
 \sqrt{\rho_{ii}\rho_{jj}}\Big)^2 - \Big(\sum_{i\neq j}
\sqrt{\rho_{ii}\rho_{jj}} |\langle d_i|d_j\rangle|\Big)^2\Big] 
\end{eqnarray}
is a quantity that can be considered a quantifier of entanglement. Although not proven to be an entanglement monotone in general, for $n=2$ the quantity $\mathcal{E}$ does reduce to concurrence, which is an entanglement monotone. BCP can then be quantified by a triality relation between predictability, normalized coherence, and a quantifier of entanglement, $\mathcal{E}$. The multipath distinguishability can be written in terms of predictability and entanglement as $\mathcal{D}^2 = \mathcal{P}^2 + \mathcal{E}^2$.

Alternatively, one can define a simpler form of predictability,  as shown in classical optics \cite{Paul2020}:
\begin{equation}
\mathcal{P}_Q \equiv 1 -\tfrac{1}{n-1}\sum_{i\neq j}
\sqrt{\rho_{ii}\rho_{jj}}. \label{PQ}
\end{equation}
Using this quantity, the duality relation in Eq.~(\ref{dualityQn}) can be reformulated as
\begin{eqnarray}
\mathcal{P}_Q + \mathcal{E}_Q + \mathcal{C} = 1 ,
\label{pceq}
\end{eqnarray}
where 
\begin{eqnarray}
\mathcal{E}_Q &=& \tfrac{1}{n-1}\sum_{i\neq j}
\big(\sqrt{\rho_{ii}\rho_{jj}} - \sqrt{\rho_{ii}\rho_{jj}}
|\langle d_i|d_j\rangle|\big) . \label{eq:E_Q}
%&=& \tfrac{1}{n-1}\sum_{i\neq j}
%\big(\sqrt{{\rho_r}_{ii}{\rho_r}_{jj}} - |{\rho_r}_{ij}| \big),
\end{eqnarray}
Again, $\mathcal{E}_Q$ is a quantity that has all the characteristics of a quantifier of entanglement, although it has not been proven to be an entanglement monotone. The relation in Eq.~(\ref{pceq}) is a triality relation quantifying BCP. The distinguishability in Eq.~(\ref{DQn}) is simply $\mathcal{D}_Q = \mathcal{P}_Q + \mathcal{E}_Q$, indicating that distinguishability is connected to predictability in a precise way through entanglement. The triality relation given by Eq.~(\ref{pce}) was experimentally investigated by \citet{Yoon2021}.

In the preceding analysis, although a good measure was used to describe the coherence of a quanton, the entanglement could not be described using an entanglement monotone. In another work on similar lines \cite{Roy2022}, the wave aspect of a quanton, passing through a multipath interferometer, was described using the normalized Hilbert--Schmidt coherence
\begin{equation}
\mathcal{V}^2 \equiv \frac{n}{n-1} \sum_{j\neq k} |\rho_{jk}|^2,
\end{equation}
where the subscripts are the path labels. Path predictability was defined \cite{Durr2001} as 
\begin{equation}
\mathcal{P}^2 \equiv 1 -\tfrac{n}{n-1}\sum_{j\neq k}
\rho_{jj}\rho_{kk}.
\label{PD}
\end{equation}
If the entanglement between the quanton and the path detector is quantified using the well known monotone I-concurrence, denoted by $\mathcal{E}$, it was shown \cite{Roy2022} that the following triality relation holds
\begin{eqnarray}
\mathcal{P}^2 + \mathcal{E}^2 + \mathcal{V}^2 = 1 . \label{Eq:trialityl_2}
\label{pve}
\end{eqnarray}
This can be considered a tight wave-particle triality relation quantifying BCP. Some remarks are in order here. While the normalized Hilbert--Schmidt coherence $\mathcal{V}$ does not serve as a strict measure of quantum coherence, it adequately characterizes interference visibility in multipath interference scenarios and proves to be resilient even in contexts such as the experiment conducted by \citet{Mei2001}. Interestingly, if in the multipath interference experiment, one opens only a pair of paths at a time, and measures the \emph{fringe contrast} of the resultant two-path interference, the normalized Hilbert--Schmidt coherence $\mathcal{V}$ is equal to the root mean square of the fringe contrasts
of interference from all such path pairs \cite{Roy2022}. The path distinguishability in this situation is related to the predictability and entanglement by 
$\mathcal{D}^2 = \mathcal{P}^2 + \mathcal{E}^2$. The triality relation in Eq.~(\ref{pve}) beautifully connects the two kinds of duality relations in Eqs.~(\ref{dualityE}) and (\ref{dualityG}), even for multi-path interference. In the absence of entanglement of the quanton with a path detector, Eq.~(\ref{pve}) reduces to a duality relation involving predictability. The relation in Eq.~(\ref{pve}) was tested in different experimental setups, including hBN-quantum-dot-emitted single photon~\cite{Qian2020a}, superconducting qubits in the IBM quantum computer~\cite{Schwaller2021}, and silicon-integrated nanophotonic quantum chips~\cite{Ding2025}.

%---------------------------
\subsection{Duality games}

In an interesting development, \citet{Bagan2018} reframed wave-particle
duality as a constraint on simultaneously performing two incompatible
discrimination tasks, in a game involving three parties 
Alice and Bob, who jointly play to win, and the House, which chooses which
of the two tasks is to be played.
A single quanton enters an $n$-port interferometer.
The House randomly decides between two games:\\
\emph{$\bullet$ Ways game} (Particle task): Alice measures the particle in the path basis.
Bob measures path detectors to guess which path the particle took.
They win if Bob's guess matches Alice’s outcome.
The success probability is represented by $P_{\text{way}}$, and
quantifies particle-like behavior.\\
\emph{$\bullet$ Phases game} (Wave task):
The House applies one of $n$ known sets of phases to the paths.
Alice must identify which phase set was applied, whereas Bob has no role to play here.
The success probability is represented by $P_{\text{ph}}$,
quantifies the wave-like behavior, and is directly related to the
$\ell_1$-norm coherence of the state.

The probability of winning the combined game is 
$P_{\text{win}} = \tfrac{1}{2}(P_{\text{ph}} + P_{\text{way}})$, and
is bounded \cite{Bagan2018} by
\begin{equation}
\frac{1}{2}(P_{\text{ph}} + P_{\text{way}}) \le \frac{1}{2}
+ \frac{1}{2\sqrt{n}} . \end{equation}
The upper bound of $P_{\text{ph}}$ depends on the $\ell_1$-norm coherence
of the state of the quanton, and $P_{\text{way}}$ is related to the
probability of discriminating between the $n$ path detector states using
minimum error discrimination.
Further analysis found that the normalized operational variables, defined as
$x = \frac{P_{\text{ph}}-1/n}{1-1/n}$ and
$y = \frac{P_{\text{way}}-1/n}{1-1/n}$, satisfy the bound
\begin{equation} \left(\frac{x+y-\tfrac{n-2}{n-1}}{\sqrt{n}/(n-1)}\right)^2
+\left(\frac{x-y}{\sqrt{n/(n-1}}\right)^2 \le 1 . 
\label{baganduality} \end{equation}
The inequality presented above improves upon the duality relation established by \citet{Bagan2016}. This bound is considered tight, although the inequality saturates only when the quanton entering the interferometer is in a maximally coherent state. It is pertinent to mention that the duality relations of \citet{Englert1996},
\citet{Bera2015} and \citet{Qureshi2017} saturate for any pure quanton state. When a duality relation like Eq.~(\ref{baganduality}) does not saturate for some pure states, it means there will be situations where a decrease in the \emph{measure of distinguishability} of paths does not lead to an increase in the quantum coherence of the quanton, something that violates the spirit of BCP.
Nevertheless, the strength of this work is that it replaces visibility-based
reasoning with an information-theoretic framework and provides a tight
bound for duality relations based on minimum error discrimination, for
multi-path interferometers. This approach was then extended to the presence of a quantum memory~\cite{Bu2018}.  
Other works also explored duality games in the context of wave-particle duality~\cite{Coles2014, Coles2016, Hillery2021}.

%---------------------------
\subsection{Wave-particle duality through entropic uncertainty relations}

Building on the duality game perspective introduced in the previous section, interferometric scenarios can be recast in terms of operational tasks, where particle-like and wave-like properties correspond, respectively, to the ability to predict which-path and which-phase information. In this setting, complementarity is framed as a trade-off between two incompatible guessing tasks, which can be naturally quantified using entropic measures, in particular the min- and max-entropies, admitting a clear operational interpretation in terms of guessing probabilities. Within this framework,~\citet{Coles2014, Coles2016} established a direct connection between wave-particle duality and entropic uncertainty relations. In this formulation, the impossibility of simultaneously optimizing both guessing tasks reflects a fundamental limitation on accessible information and provides a direct operational meaning to wave-particle trade-offs. In this sense, traditional duality relations are shown to be equivalent to entropic uncertainty relations~\cite{Coles2017}.

More specifically, \citet{Coles2014} demonstrated that the complementarity relations, particularly those given by Eqs.~\eqref{dualityG} and~\eqref{dualityE}, can be reinterpreted as instances of entropic uncertainty relations. These duality relations are equivalent to uncertainty relations expressed in terms of the min- and max-entropies, evaluated for complementary qubit observables. The min-entropy associated with a measurement $Z=|0\rangle\langle 0| - |1\rangle\langle 1|$ ($\{|0\rangle,|1\rangle\}$ are standard path basis states) is given by
\begin{align}
    H_{\min} = -\log P(Z),
\end{align}
where $P(Z)$ denotes the optimal probability of correctly guessing the outcome of the which-path observable $Z$ and throughout this article all logarithms are taken to base $2$. Similarly, the max-entropy for a measurement of the which-phase observable $W$ is
\begin{align}
    H_{\max}(W) = 2 \log \left(\sum_w \sqrt{p_w}\right),
\end{align}
where ${p_w}$ are the outcome probabilities of the observable $W$. These quantities admit a clear operational interpretation in terms of information-theoretic tasks. Within this formalism, Eq.~\eqref{dualityG} can be recast as the entropic uncertainty relation
\begin{align}
H_{\min}(Z) + \min_{W \in XY} H_{\max}(W) \geq 1,
\end{align}
where the minimization is taken over all observables $W$ lying in the $xy$-plane of the Bloch sphere, i.e., complementary to $Z$. Likewise, Eq.~\eqref{dualityE} corresponds to the conditional uncertainty relation
\begin{align}
    H_{\min}(Z|E) + \min_{W \in XY} H_{\max}(W) \geq 1,
\end{align}
where $H_{\min}(Z|E)$ denotes the conditional min-entropy, accounting for the presence of a quantum memory $E$.

This equivalence reveals that the trade-off between path information and interference is not merely a feature of specific experimental setups but rather a manifestation of fundamental constraints on the simultaneous knowledge of complementary observables. Consequently, complementarity can be understood as arising from information-theoretic limitations encoded in the structure of quantum theory. A recent experiment~\cite{Spegel-Lexne2024} provided a direct verification of this equivalence by implementing interferometric measurements using orbital angular momentum states of light. By accessing particle-like and wave-like quantities and comparing them with the corresponding min- and max-entropies, the authors confirm that wave-particle duality relations reproduce the same bounds as entropic uncertainty relations. Moreover, building on this connection, \citet{Raj2025} explored its implications in quantum cryptography. In a semi-device-independent framework, the authors showed that security witnesses can be expressed entirely in terms of interferometric quantities, such as visibility and distinguishability. This establishes a direct link between complementarity and cryptographic security, where the same constraints that limit simultaneous knowledge of complementary observables also bound an eavesdropper's information.~\citet{Lee2025} established a complementarity relation between quantum and classical uncertainties, quantified by distillable coherence and von Neumann entropy, which leads to tighter bounds than the entropic uncertainty relations enabling steering witnesses that quantify the additional distillable coherence unlocked by quantum steerability, even serving as a full entanglement measure for pure bipartite states. A different line of generalization considers multiple-measurement scenarios, where entropic uncertainty relations impose lower bounds on the total quantum coherence across different bases, as well as on quantum correlations such as thermal discord and conditional information, thereby establishing complementarity relations among these quantities~\cite{Fan2019}.

\subsection{Complementarity in many-body quantum systems}
\label{sec:manybody}

~\citet{Dittel2021} developed a general framework to extend wave-particle duality and complementarity relations to many-body quantum systems, addressing a fundamental gap in the standard formulation of these concepts. While wave-particle duality is well understood for single quantons---where interference and which-way information are quantitatively constrained---its generalization to systems of many identical quantons is highly nontrivial. In this context, particle-like behavior is no longer associated with the path of a single quanton but rather with information about which quanton took which path in a multi-system configuration. Conversely, wave-like behavior arises from the interference of many-quanton amplitudes, which can involve permutations of indistinguishable quantons.

An important feature of this framework is that the degree of distinguishability between quantons becomes a tunable parameter. By varying how distinguishable the quantons are---through internal states, interactions, or labeling---one can continuously interpolate between fully quantum and more classical regimes. In particular, indistinguishable bosons or fermions exhibit strong many-body interference effects, while increasing distinguishability suppresses interference and enhances classical behavior. The formalism is quite general and applies to both non-interacting and interacting systems, as well as to bosonic and fermionic statistics, as illustrated by paradigmatic examples such as Hong--Ou--Mandel-type interference and systems described by Bose--Hubbard dynamics.~\citet{Dittel2021} also shed light on the quantum-to-classical transition in complex systems, once classical behavior emerges as distinguishability increases and many-body coherence is lost.

%---------------------------%---------------------------
\subsection{Other approaches to duality and triality relations}
\label{sec:III_D}

With the advent of quantitative complementarity relations, a wide range of approaches has been developed. The derivation of such relations remains an active area of research. In this section, we review some of the recent advances that go beyond the standard framework of wave-particle duality quantification discussed so far. These contributions introduce new perspectives that extend the existing literature.

\citet{Luo2018} formulated Bohr's complementarity principle within the general framework of state-channel interactions, motivated by interferometric scenarios such as the Mach--Zehnder setup. Considering a quantum state $\rho$ undergoing a quantum operation $\Phi(\rho) = \sum_j K_j \rho K_j^\dagger$, the authors introduce two quantities that characterize complementary aspects of the state with respect to the channel. These are defined in terms of the symmetric (Jordan) and antisymmetric (Lie) combinations of $\sqrt{\rho}$ and the Kraus operators $K_j$, namely $J(\rho,\Phi) = \sum_j ||\{\sqrt{\rho},K_j\}||^2$ and
$I(\rho,\Phi) = \sum_j ||[\sqrt{\rho},K_j]||^2$,
where $||X||^2 = \mathrm{Tr}(X^\dagger  X)$ denotes the Hilbert--Schmidt norm. The quantity $I(\rho,\Phi)$ captures the coherence of the input state with respect to the channel and is associated with the decoherence induced by the environment, while $J(\rho,\Phi)$ quantifies the accessible information in the output state, and satisfies a trade-off relation of the form $J(\rho,\Phi) + I(\rho,\Phi) = \frac{1}{2}\mathrm{Tr}\big(\Phi(\rho) + \Phi^\dagger(\rho)\big)$, which reduces to the information conservation relation $J(\rho,\Phi) + I(\rho,\Phi) = 1$ for unital channels. By optimizing the phase shift in the Mach-Zehnder interferometer, these quantities capture the which-path information $\mathcal{D} (\rho,\Phi) = \min_\alpha I(\rho,\Phi)$ and the interferometric visibility $\mathcal{V}(\rho,\Phi) = \max_\alpha J(\rho,\Phi)$. These optimized quantities then satisfy the complementarity relation $P(\rho,\Phi) + W(\rho,\Phi) = 1$. This framework was further extended by \citet{WuZ2020}, who generalized the skew-information approach of \citet{Luo2018} by introducing a modified Wigner--Yanase--Dyson skew information. Within this generalized setting, they defined corresponding coherence measures and derived complementarity relations that preserve the same trade-off structure, reducing to a conservation law for unital channels.

\citet{Bagan2020} derived wave-particle duality relations based on entropic bounds for which-way information, employing the relative entropy of quantum coherence~\cite{Baumgratz2014}. In a multi-path interferometer with $N$-paths, the quanton becomes correlated with a path detector such that the global state can be written as $|\Psi\rangle = \sum_{j=1}^N \sqrt{p_j} |j\rangle_q |\eta_j\rangle_d$, where $|j\rangle_q$ denotes the state of the quanton associated with propagation through the $j$-th path, and $|\eta_j\rangle_d$ is the corresponding detector state. Tracing out the detector gives the reduced state of the quanton in the path basis $\rho_q = \operatorname{Tr}_d(|\Psi\rangle\langle\Psi|)$. The normalized relative entropy of coherence $\mathcal{C} \equiv \mathcal{C}_{\rm re}(\rho_p)/\log N,$ where $\mathcal{C}_{\rm re}(\rho_p) = S(\rho_p^{\rm diag}) - S(\rho_p)$ and $S$ denote the von Neumann entropy. The path distinguishability $\mathcal{D}$ is defined as the success probability $P_s = 1 - P_f$ of unambiguous (zero-error) discrimination of the detector states $|\eta_j\rangle_d$. Using Holevo's theorem~\cite{Nielsen2010}
for the mutual information between the path label $X$ and the measurement outcome $Y$ (which includes a failure outcome), the authors derived a tight entropic lower bound on the failure probability $P_f$, leading to the duality relation $\mathcal{D} + \mathcal{C} \leq 1.$ This bound becomes tight in the limit of large $N$ for specific symmetric and asymmetric configurations of detector states. The authors also extend the analysis to mixed discrimination strategies allowing both a nonzero error probability $P_e$ and failure probability $P_f$, obtaining a more general bound. In contrast to earlier approaches based on $\ell_1$-coherence and minimum-error distinguishability, this formulation relies on entropic bounds derived from Holevo's theorem and emphasizes the role of the chosen measurement strategy, providing an information-theoretic characterization of the wave-particle trade-off within the resource theory of coherence.

Further extensions of coherence-based approaches to complementarity have explored the relaxation of the orthogonality assumption on the reference basis.
\citet{Das2020} introduced a resource theory of quantum coherence for arbitrary (possibly non-orthogonal) linearly independent bases $\{|a_i\rangle\}$. They defined the free states as non-orthogonal incoherent states (NOIS) of the form $\chi = \sum_i p_i |a_i\rangle\langle a_i|$, where the free operations are non-orthogonal maximally incoherent operations (NOMIO)---channels that map every NOIS to another NOIS. A bona fide measure of non-orthogonal coherence was proposed via the trace distance, $\mathcal{C}_{\rm trace}^{\rm NO}(\rho) = \min_{\chi} \operatorname{Tr}|\rho - \chi|,$ with $|A|=\sqrt{A^\dagger A}$, where the minimization runs over all NOIS for the given basis. In the qubit case, this measure reveals striking differences from the standard (orthogonal) resource theory: for any fixed non-orthogonal basis there exists a unique maximally coherent pure state $|m\rangle$. Moreover, the geometry of the state space is significantly altered: below a threshold in the Bloch vector length, $\tilde{r} = \cos\alpha$ with $2\alpha$ being the angle between the basis vectors, every mixed state necessarily retains a strictly positive amount of non-orthogonal coherence. In other words, sufficiently mixed states cannot be written as convex combinations of the non-orthogonal basis states, implying that coherence cannot be completely eliminated in this regime. These features lead to tight complementarity relations between non-orthogonal coherence and mixedness (linear entropy) $\mathcal{C}_{\rm trace}^{\rm NO}(\rho) + \mathcal{E}_l(\rho) \le 1 + \cos\alpha$. The authors further demonstrated that non-orthogonal coherence furnishes a natural wave-particle duality in a leaky double-slit. The normalized non-orthogonal coherence $\tilde{C}$ of the quanton after the leaky channels quantifies residual wave nature, while the maximal probability $\tilde{\mathcal{D}}$ of unambiguous discrimination of the (non-orthogonal) detector states quantifies extracted path information, obeying the tight equality $\tilde{\mathcal{C}} + \tilde{\mathcal{D}} \le 1$ independent of the leakage parameter.

A resource-theoretic formulation of quantum coherence for systems with probabilistically non-distinguishable (linearly dependent) pointer states was introduced by \citet{Srivastava2021}, motivated by the question of how wave-particle duality should be formulated when the interferometric $d$-paths are not mutually orthogonal. In such scenarios, the usual notion of perfectly distinguishable slits is replaced by a general spanning set $\mathcal{B}=\{|\psi_i\rangle\}_{i=1}^n$ ($n\ge d$) of $\mathbb{C}^d$, which may be over-complete and linearly dependent. Within this framework, the incoherent or free states are identified with classical mixtures over the set of available pointer states, $\mathcal{F}_{\mathcal{B}} = \Bigl\{ \rho = \sum_{i=1}^n p_i |\psi_i\rangle\langle\psi_i| \;\Big|\; p_i \ge 0,\ \sum_i p_i = 1 \Bigr\}$, reflecting the fact that, in an interferometric setting, loss of coherence corresponds to classical ignorance about which non-orthogonal path was taken rather than perfect path distinguishability. The authors emphasize that this structure naturally generalizes wave-particle duality beyond the standard orthogonal double-slit paradigm. In particular, coherence with respect to the set $\mathcal{B}$ quantifies the ability of the system to exhibit interference even when the underlying paths are not perfectly distinguishable, while path information is limited by the intrinsic linear dependence of the detector states. To quantify this, they introduce coherence measures based on contractive distances to the free set, $\mathcal{C}_{\mathcal{B}}(\rho)=\min_{\sigma\in\mathcal{F}_{\mathcal{B}}}D(\rho,\sigma),$ with the trace-distance being an example $\tilde{\mathcal{C}}_{\mathcal{B}}(\rho)=\min_{\sigma\in\mathcal{F}_{\mathcal{B}}}\operatorname{Tr}|\rho-\sigma|$. In parallel, path distinguishability is defined via optimal state discrimination of the corresponding non-orthogonal pointer states. In an explicit interferometric implementation involving linearly dependent states $\{|0\rangle,|+\rangle,|1\rangle\}$, they show that coherence and distinguishability also satisfy a tight complementarity relation of the form $\tilde{\mathcal{C}}+\tilde{\mathcal{D}}\le1$.

Motivated by the triality relation first introduced by~\citet{Jakob2010} for two-qubit pure states,~\citet{Wu2020} proposed a triality-ignorance extension to mixed states. To this end, the local quantities for a bipartite system, each with dimension $d_k$, are defined in terms of the reduced state $\rho_k$. These quantities are predictability $\mathcal{P}_k = \sqrt{\frac{d_k}{d_k-1}\left(\sum_i \rho_{kk,ii}^2 - \frac{1}{d_k}\right)}$, visibility $\mathcal{V}_k = \sqrt{\frac{d_k}{d_k-1}\sum_{i\neq j} |\rho_{kk,ij}|^2} $, $Q_k = \sqrt{\frac{d_k}{d_k-1}\operatorname{Tr}(\rho_k^2) - \frac{1}{d_k-1}}$,
and mixedness $M_k = \sqrt{\frac{d_k}{d_k-1}[1 - \operatorname{Tr}(\rho_k^2)]}$. These quantities satisfy a closed set of local relations
\begin{align}
\mathcal{P}_k^2 + \mathcal{V}_k^2 = Q_k^2, \quad   Q_k^2 + M_k^2 = 1,
\end{align}
which combine into the identity 
\begin{align}
\mathcal{P}_k^2 + \mathcal{V}_k^2 + M_k^2 = 1.
\end{align}
For pure bipartite states, this relation extends to include entanglement, yielding the well known triality relation
\begin{align}
    P_k^2 + V_k^2 + C^2(\Psi)/\nu_k^2 = 1,
\end{align}
where $C(\Psi)$ is the concurrence and $\nu_k = \sqrt{2(n_k-1)/n_k}$~\cite{Jakob2010}.
For mixed states, however, this picture is no longer complete, as part of the uncertainty may arise from classical ignorance. This motivates the introduction of the tangle $\tau$, quantifying entanglement and defined as the convex-roof extension of the squared concurrence,
$\tau(\rho) = \min_{{p_i, \Psi_i}} \sum_i p_i C^2(\Psi_i),$ where the minimization is taken over all pure-state decompositions $\rho = \sum_i p_i |\Psi_i\rangle\langle\Psi_i|$. For pure states, this reduces to $\tau(\Psi)=C^2(\Psi)$. The separable uncertainty $B(\rho)$ is then introduced as the contribution of classical ignorance to the total uncertainty, defined through the triality balance as $B^2(\rho) = 1 - \sum_k \left(\mathcal{P}_k^2 + \mathcal{V}_k^2\right) - \tau(\rho)$, ensuring $B(\rho) \ge 0$ by construction. Equivalently, $B(\rho)$ quantifies the portion of the total uncertainty that cannot be attributed to either local coherence or quantum correlations, vanishing for pure separable states and becoming positive only in the presence of classical mixing. Within this extended framework, one obtains the central triality-ignorance relation 
\begin{align}
    \tilde{\mathcal{P}}_k^2 + \tilde{\mathcal{V}}_k^2 + (\tau(\rho) + B^2(\rho))/\nu_k^2 = 1.
\end{align}
As argued by the authors, this approach can be further extended to hierarchical bipartite systems (multipartite systems consisting of bipartite systems and to infinite-dimensional systems, where the weighting becomes uniform).

The quantum Fisher information (QFI) has recently been explored by~\citet{Niu2023} as a figure of merit to characterize wave-particle duality in a Mach-Zehnder interferometer equipped with a which-way detector. Let us recall that the Fisher information measures the amount of information one can estimate about an unknown random variable. The classical Fisher information for the estimation of an unknown parameter $\theta$, which is encoded in the probability distribution $\{p_x(\theta)\}_{x=1}^{N}$ conditioned on the value $\theta = \theta^*$, is $\mathcal{F}_c(\{p_x(\theta)\}) = \sum_{x=1}^{N}(\dot{p}_x(\theta))^2/p_x(\theta)$, where $\dot{p}_x(\theta) = \partial p_x(\theta)/\partial \theta.$ Then, the QFI can be as the maximum over all positive operator-valued measurements $\{M_x \ge 0, \sum_x M_x = I \}$ of the classical Fisher information given the state $\rho(\theta)$ with $p_x(\theta) = \Tr \rho(\theta) M_x$. In the eigenbasis $\{\lambda_j,|\phi_j\rangle\}$ of $\rho(\theta)$, the maximization yields  $\mathcal{F}_q(\rho(\theta)) = 2 \sum_{j,k}\,|\langle\phi_j|\dot{\rho}(\theta)|\phi_k\rangle|^2/( \lambda_j + \lambda_k)$. In their approach, the particle nature is quantified by the QFI $\mathcal{F}_{q}(\rho_{\rm D}(z))$, associated with estimating the which-path bias $z = \mathrm{Tr}(\rho \sigma_z)$ encoded in the detector state $\rho_{\rm D}$, while the wave nature is quantified by the QFI $\mathcal{F}_{q}(\rho_{\rm S}(\Phi))$, associated with the interferometric phase $\Phi$ retained in the state $\rho_{\rm S}$ of the system. This leads to a trade-off relation of the form 
\begin{align}
(1-\mathcal{P}^2)\mathcal{F}_{q}(\rho_{\rm D}(z))+\frac{\mathcal{F}_{q}(\rho_{\rm S}(\Phi))}{\mathcal{V}^2}\le 1   \end{align}
together with the bound $\mathcal{F}_{q}(\rho_{\rm D}(z)) \mathcal{F}_{q}(\rho_{\rm S}(\Phi)) \le 1/4$, which directly implies the~\citet{Englert1996} duality relation~\eqref{dualityE} as a special case. The formulation of wave-particle duality based on QFI has recently been placed within a broader framework of quantum uncertainty decomposition. In particular, building on the approach of~\citet{Luo2005}, the total uncertainty of a quantum state can be systematically decomposed into quantum uncertainty and classical uncertainty. In this framework, the quantum part of the uncertainty is associated with quantum coherence, while the classical part is linked to entanglement in a bipartite quantum system, provided that the global state is pure. From this decomposition,~\citet{Basso2021a} showed that complete complementarity relations can be interpreted either as a relation between quantum uncertainty, classical uncertainty, and predictability, or equivalently a relation between quantum coherence, entanglement, and local predictability. Building on this decomposition,~\citet{Sun2023} defined a measure of quantum coherence in multipath interference via quantum Fisher information associated with phase shifts encoded along each interferometric path. In this framework, the QFI captures the quantum contribution to the uncertainty, while the remaining part of the decomposition is associated with entanglement. The particle aspect is quantified by the predictability, defined through the purity of the corresponding classical probability distribution. These quantities satisfy a triality relation, in which the QFI explicitly represents the quantum part of the total uncertainty. This perspective was further explored by~\citet{Fu2022, Wang2025, Fu2025}.

%
%---------------------------
%
\section{Quantum complementarity relations from quantum mechanics %\textcolor{red}{Jonas}
}
\label{sec:QCP}

Heisenberg's uncertainty principle was formalized into quantum uncertainty relations, thus becoming a consequence of the postulates of quantum mechanics. On the other hand, although Bohr mentioned that his complementarity principle was a consequence of the quantum postulate, this assertion was not mathematically substantiated.
Over the years, several discussions have addressed the origin of BCP, questioning whether it follows from quantum uncertainty relations or whether it should be regarded as an additional principle to be incorporated into the formalism of quantum theory. In this section, we review how quantitative complementarity relations can be obtained directly from the mathematical structure of quantum theory. These relations emerge from fundamental properties of quantum states, such as positivity, normalization, and purity. It is worth noting that, in his seminal work,~\citet{Englert1996} already made use of the positivity of quantum states to derive his relation. However, this aspect was not explicitly emphasized at the time, and more recent work has taken it as a central motivation for deriving such relations directly from the formal structure of quantum theory.

As we have seen so far, Bohr's original notion of strictly mutually exclusive experimental arrangements required refinement in order to accommodate partial characterizations of wave-particle duality. This development began with the introduction of a quantitative framework by \citet{Wootters1979}, and was later solidified through the complementarity relations formulated by~\citet{Greenberger1988} and~\citet{Englert1996}, which enabled a systematic quantification of wave-particle duality. For a long time, interferometric visibility was regarded as a natural quantifier of wave-like behavior due to its operational appeal. However, as mentioned in the previous section, quantum coherence~\cite{Baumgratz2014} emerged as a suitable quantifier of wave-like behavior, consistent with the criteria for waveness quantifiers proposed by \citet{Durr2001} and \citet{Englert2008}. Experiments have been conducted demonstrating the use of quantum coherence as a quantifier of wave-like behavior~\cite{Gao2018,Yuan2018,Wang2021a,Pozzobom2021,Chrysosthemos2023a,Chrysosthemos2023b,YangX2025}. The similarities and discrepancies between interferometric visibility and quantum coherence, as well as the role of quantum coherence as the natural generalization of interferometric visibility, were analyzed in~\cite{Biswas2017,Qureshi2019a,Chrysosthemos2023a}.

By using quantum coherence measures, or similar functions that satisfy the \citet{Durr2001} and \citet{Englert2008} criteria, complementarity relations can be derived directly from the postulates of quantum mechanics. In this framework, a predictability quantifier naturally arises in the context of duality relations for each coherence function. By imposing the purity of a bipartite state, entanglement functions emerge, which can be shown to be entanglement monotones. Lastly, when the largest value of distinguishability is considered, the relation among predictability, entanglement, and distinguishability is established.

Finally, it is worth emphasizing that, although in the following we introduce complementarity relations using specific quantifiers of coherence, predictability, entanglement, and distinguishability, the central aspect of this approach is the method itself rather than the particular choice of measures. Indeed, several complementarity relations obtained through this procedure reproduce results that were already known in the literature, showing that they naturally emerge from the mathematical structure of quantum theory. Therefore, the importance of this framework lies in providing a systematic method to derive quantitative complementarity relations.

%---------------------------
\subsection{Duality from positivity}
\label{sec:CRpositivity}

Let us denote the set of density matrices by 
$\mathcal{D}(\mathcal{H}_A) \equiv \{\rho_A \in \mathcal{L}(\mathcal{H}_A) \, | \, \rho_A \ge 0,\; \Tr(\rho_A)=1\}$. They 
act on the Hilbert space $\mathcal{H}_A$ of the quantum system $A$, where $\mathcal{L}(\mathcal{H}_A)$ denotes the set of linear operators on $\mathcal{H}_A$. The procedure outlined by~\citet{Basso2020a} for obtaining complementarity relations starts by considering a given quantum coherence measure as a suitable quantifier for the wave-like behavior, e.g., the $\ell_2$-norm quantum coherence (or Hilbert--Schmidt quantum coherence)~\cite{Baumgratz2014}, given by 
\begin{align}
    \mathcal{C}_{\ell_2}(\rho_A) = \sum_{j\neq k}|\rho_{jk}|^2,
\end{align}
with $\rho_{jk}$ being the matrix elements of the density operator $\rho_A$ in a given basis. 

From the positivity of $\rho_A$, we have $|\rho_{jk}|^2 \le \rho_{jj}\rho_{kk}$ for all $j,k$ \cite{Zhang2011}, from which it is possible to write \cite{Basso2020a}
\begin{eqnarray}
\mathcal{C}_{\ell_2}(\rho_A) \le \sum_{j\neq k}\rho_{jj}\rho_{kk} \le \frac{d_A-1}{d_A},
\end{eqnarray}
where $d_A$ is the dimension of the quantum system $A$. 
Following the criterion P.1 of Sec.~\ref{sec:duality_multi-slits} by~\citet{Durr2001} and \citet{Englert2008} for a predictability measure, which must be a function only of the diagonal elements of the density matrix, and identifying the maximum value of the quantum coherence measure, one obtains
\begin{align}
    \mathcal{P}_{\ell_2}(\rho_A) & = \frac{d_A-1}{d_A} - \sum_{j\neq k}\rho_{jj}\rho_{kk}  = \sum_j (\rho_{jj})^2 - \frac{1}{d_A}.
\end{align}
Hence, the complementarity relation follows
\begin{eqnarray}
\mathcal{C}_{\ell_2}(\rho_A) + \mathcal{P}_{\ell_2}(\rho_A) \le \frac{d_A-1}{d_A}.
\label{eq:QCR_l2}
\end{eqnarray}

Following the same procedure, one can derive the complementarity relations
\begin{eqnarray}
& \mathcal{C}_{re}(\rho_A) + \mathcal{P}_{vn}(\rho_A) \le \log d_A, \\ \nonumber
& \mathcal{C}_{\ell_1}(\rho_A) + \mathcal{P}_{\ell_1}(\rho_A) \le d_A-1,
\label{eq:QCR_l1} \\
& \mathcal{C}_{wy}(\rho_A) + \mathcal{P}_{\ell_2}(\rho_A) \le \frac{d_A-1}{d_A}, \nonumber
\end{eqnarray}
where the quantum coherence measures correspond to the relative-entropy of quantum coherence, the $\ell_1$-norm quantum coherence, and the Wigner-Yanase quantum coherence~\cite{Baumgratz2014, Yu2017}, respectively. These are given by
\begin{align}
    & \mathcal{C}_{re}(\rho_A) = S_{vn}(\rho^A_{\text{diag}}) - S_{vn}(\rho_A), \\
    & \mathcal{C}_{\ell_1}(\rho_A) = \sum_{j\neq k}|\rho_{jk}|, \\
    & \mathcal{C}_{wy}(\rho_A) = \sum_{j\neq k}|(\sqrt{\rho_A})_{jk}|^2 ,
\end{align}
where $S_{vn}(\rho_A) = - \Tr(\rho_A \log \rho_A)$ is the von Neumann entropy, $\rho^A_{\text{diag}}$ is the diagonal part of the density operator $\rho^A$. The corresponding predictability measures are
\begin{align}
    & \mathcal{P}_{vn}(\rho_A) = \log d_A - S_{vn}(\rho^A_{\text{diag}}), \\
    & \mathcal{P}_{\ell_1}(\rho_A) = d_A-1 - \sum_{j\neq k}\sqrt{\rho_{jj}\rho_{kk}}, \\
    & \mathcal{P}_{\ell_2}(\rho_A) = \sum_j (\rho_{jj})^2 - \frac{1}{d_A}.
\end{align}

In general, by starting from a given quantum coherence measure and using the positivity of the density matrix $\rho_A$, a predictability quantifier naturally emerges in the context of duality relations for each coherence function. This allows one to obtain a quantum complementarity relation of the general form
\begin{eqnarray}
\mathcal{P}(\rho_A) + \mathcal{C}(\rho_A) \le \alpha(d_A),~\label{eq:QCR}
\end{eqnarray}
where $\alpha(d_A)$ is a constant that depends on the dimension of the system. It is worth noting that quantum complementarity relations expressed by Eq.~\eqref{eq:QCR} are saturated only for pure quantum states by the construction of the method. 
Besides, one can always normalize the coherence and predictability functions so that $\alpha(d_A)=1.$ 

Finally, \citet{Tsui2024} established a theorem stating that a valid path predictability measure $\mathcal{P}$ can be systematically constructed from any quantum coherence measure $\mathcal{C}$ that satisfies the criteria of \citet{Durr2001} and \citet{Englert2008}. The proof relies on restricting to pure states $|\psi\rangle = \sum_j c_j |j\rangle$, and noting that any quantum coherence becomes a function of the diagonal elements of the corresponding density operator, i.e., $\rho_{jj} = |c_j|^2$, once the inequality $|\rho_{jk}|^2 \le \rho_{jj}\rho_{kk}$ is saturated for pure states. The next step consists in defining a dual functional $\mathcal{P}$ from $\mathcal{C}$ such that it is symmetric under permutations of the paths and Schur-concave in the probability distribution. These properties ensure that $\mathcal{P}$ vanishes for the uniform distribution, attains its maximum for deterministic path occupation, and is a convex function of the density operator, thus fulfilling the criteria for a predictability measure.

%---------------------------
\subsection{Triality from purity}
\label{sec:CCRhs}

Triality relations arise as a way to fully characterize a quantum system by including quantum correlations with other quantum systems, as explored in Sec.~\ref{sec:triality}.  On the other hand, since it has already been shown that complementarity relations can be derived directly from the postulates of quantum mechanics, a natural question is whether Eq.~\eqref{eq:QCR_l2} can be completed by including an additional quantum correlation measure. This question was addressed by~\citet{Basso2020b}, which showed that the relation can indeed be completed by exploring the purity of the quantum state for bipartite pure quantum systems and further generalized the result to multipartite pure quantum systems. The key observation is that, when a subsystem becomes mixed, this mixedness reflects the presence of entanglement encoded in the global pure state. This observation allows us to derive complementarity relations linking local properties, such as predictability and coherence, with entanglement distributed among the parties.

By starting from a purification of $\rho_A$, i.e., a bipartite pure state 
$|\Psi\rangle_{AB}$ such that $\rho_A = \Tr_B(|\Psi\rangle_{AB}\langle\Psi|)$, 
where $\Tr_B$ denotes the partial trace over subsystem $B$, the complementarity 
relation above can be completed by using the purity condition 
$\Tr(\rho_{AB}^2)=1$, with $\rho_{AB}=|\Psi\rangle_{AB}\langle\Psi|$ to obtain 
the following triality relation (or complete complementarity relation) for subsystem $A$
\begin{eqnarray}
\mathcal{C}_{\ell_2}(\rho_A) + \mathcal{P}_{\ell_2}(\rho_A) + \mathcal{E}_l(\rho_A) 
= \frac{d_A - 1}{d_A},
\label{eq:CCRA}
\end{eqnarray}
where $\mathcal{E}_l(\rho_A) = S_l(\rho_A) = 1 - \Tr(\rho_A^2)$ is the linear entropy of subsystem $A$, 
which quantifies the entanglement between the subsystems $A$ and $B$ for globally pure states. It is worth noting that the same steps can be applied to subsystem $B$. The CCR given by Eq.~\eqref{eq:CCRA} is equivalent to the relation 
$\mathcal{P}_A^2 + \mathcal{V}_A^2 + (C_{AB}^{(d)})^2 = \frac{2(d_A - 1)}{d_A}$ 
reported by~\citet{Jakob2006,Jakob2007}, where  $C_{AB}^{(d)} = \sqrt{2S_l(\rho_A)}$ corresponds to a quantum concurrence measure, $\mathcal{P}_A = \sqrt{2\mathcal{P}_{\ell_2}(\rho_A)}$, and $\mathcal{C}_A = \sqrt{2\mathcal{C}_{\ell_2}(\rho_A)}$, which is also related to the triality relation presented in Eq.~\eqref{Eq:trialityl_2}  of Sec.~\ref{sec:triality}. For multipartite pure quantum systems,~\citet{Basso2020b} also showed that  complementarity is accompanied by the linear entropy of the subsystem under consideration, which is connected to the generalized quantum concurrence as introduced by~\citet{Bhaskara2017}. 

In addition,~\citet{Basso2020b} also derived a complete complementarity relation expressed in terms of the relative entropy of coherence, with the von Neumann entropy of the reduced states naturally appearing as the entanglement monotone, quantifying the correlations between subsystems. Because the global state is pure, the entropy of the subsystem $A$, $\mathcal{E}_{vn}(\rho_A) = S_{vn}(\rho_A) = - \Tr(\rho_A \log \rho_A)$, directly measures the amount of entanglement it shares with the remainder of the system.
As a consequence, the following complete complementarity relation is obtained:
\begin{eqnarray}
\mathcal{C}_{re}(\rho_A) + \mathcal{P}_{vn}(\rho_A) + \mathcal{E}_{vn}(\rho_A) 
= \log d_A,
\label{eq:CCRB}
\end{eqnarray}
which provides a closed equality relating local predictability $\mathcal{P}_{vn}(\rho_A) $, coherence $\mathcal{C}_{re}(\rho_A)$, and entanglement $\mathcal{E}_{vn}(\rho_A)$, thereby offering a complete account of how informational resources are distributed within the local quantum system. It is noteworthy that the above equation is equivalent to the complementarity relation proposed by~\citet{Angelo2015}, where the authors consider $ \mathcal{P}_{vn}(\rho_A) + \mathcal{E}_{vn}(\rho_A)$ as a measure of the particle-like property of system $A$. This is because the particle-like character can be associated with accessible which-path information and/or entanglement with another system. In fact, the sum  $ \mathcal{P}_{vn}(\rho_A) + \mathcal{E}_{vn}(\rho_A)$, as discussed in Sec.~\ref{sec:triality} and to be analyzed further below, can be regarded as a measure of distinguishability.

This framework has been further extended to the case of mixed states $\rho_{AB}$ by~\citet{Basso2021e}, where a generalized complementarity relation based on the von Neumann entropy is derived. In this scenario, the simple identification between subsystem entropy and entanglement no longer holds, and the total correlations must be more carefully quantified. In particular, this complete complementarity relation can be expressed as 
\begin{eqnarray}
\mathcal{C}_{re}(\rho_A) + \mathcal{P}_{vn}(\rho_A) + \mathcal{I}_{A:B}(\rho_{AB}) +  S_{A|B} (\rho_{AB}) 
= \log d_A,
\label{eq:CCRBmix}
\end{eqnarray}
where $ \mathcal{I}_{A:B}(\rho_{AB}) = S_{vn}(\rho_A) + S_{vn}(\rho_B) - S_{vn}(\rho_{AB})$ is the quantum mutual information and $S_{A|B}(\rho_{AB}) = S_{vn}(\rho_{AB}) - S_{vn}(\rho_B)$ is the quantum conditional entropy. This complete complementarity relation shows that the local properties of subsystem $A$ are constrained by its classical and quantum correlations with subsystem $B$, quantified by $\mathcal{I}_{A:B}(\rho_{AB})$, as well as by the remaining uncertainty about $A$ given access to $B$. Notably, the quantum conditional entropy can take negative values, although the sum $I_{A:B}(\rho_{AB}) + S_{A|B}(\rho_{AB})$ is always non-negative. Furthermore, recalling the definition of coherent information $S_{A>B}(\rho_{AB}) := -S_{A|B}(\rho_{AB})$, one sees that it quantifies the imbalance between global and local information~\cite{Wilde2013}. For instance, for a maximally entangled pure state, one has $S_{A>B}(\rho_{AB}) = S_{vn}(\rho_B) = \log 2$, indicating that the global state is more informative than its subsystems. Motivated by this interpretation and by the state merging protocol~\cite{Horodecki2005}, the conditional entropy $S_{A|B}(\rho_{AB})$ can be understood as quantifying how advantageous it is to access subsystem $A$ rather than the composite system. As a result, the trade-off between predictability, coherence, and correlations acquires a richer structure, reflecting the presence of both classical mixing and genuinely quantum features, while still preserving a closed balance between the different informational quantities.

Finally, these complete complementarity relations are not restricted to the traditional interpretation of wave-particle duality of a quanton in interferometric scenarios. More generally, they quantify the trade-off between local and global properties of a quantum system. In this broader perspective, predictability characterizes the amount of \textit{a priori} information about the state of the system in a given reference basis. For instance, when considering projectors onto the basis states $\Pi_j$, the statistical uncertainty associated with the measurement outcomes can be quantified through the variance of these projectors $\mathcal{V}(\rho_{A},\Pi_{j}) = \langle \Pi^2_{j}\rangle - \langle{\Pi_{j}}\rangle^2$. Summing these variances yields the linear entropy of the diagonal part of the density matrix $S_l(\rho_{A,\mathrm{diag}})$, which depends solely on the probability distribution associated with measurement outcomes in that basis and reflects the total uncertainty of the reference basis for the state $\rho_A$. Therefore, the predictability $\mathcal{P}_{\ell_2}(\rho_A) = \mathcal{E}_l^{\max} - \mathcal{E}_l(\rho_{A,\mathrm{diag}})$ measures how far this distribution deviates from a uniform one, reflecting our ability to anticipate the result of a measurement in the reference basis before it is performed. In contrast, the quantum coherence $\mathcal{C}_{\ell_2}(\rho_A)$ quantifies the presence of superpositions between basis states, while the entanglement monotone $\mathcal{E}_l(\rho_A)$ reflects the quantum correlations between the subsystem and the remaining parts of the composite pure system, capturing genuinely quantum features that cannot be described solely by classical probability distributions. It is worth noting that the same interpretation can be given for the complete complementarity relation from Eq.~\eqref{eq:CCRB}.

\begin{figure}
    \centering
    \includegraphics[width=1\linewidth]{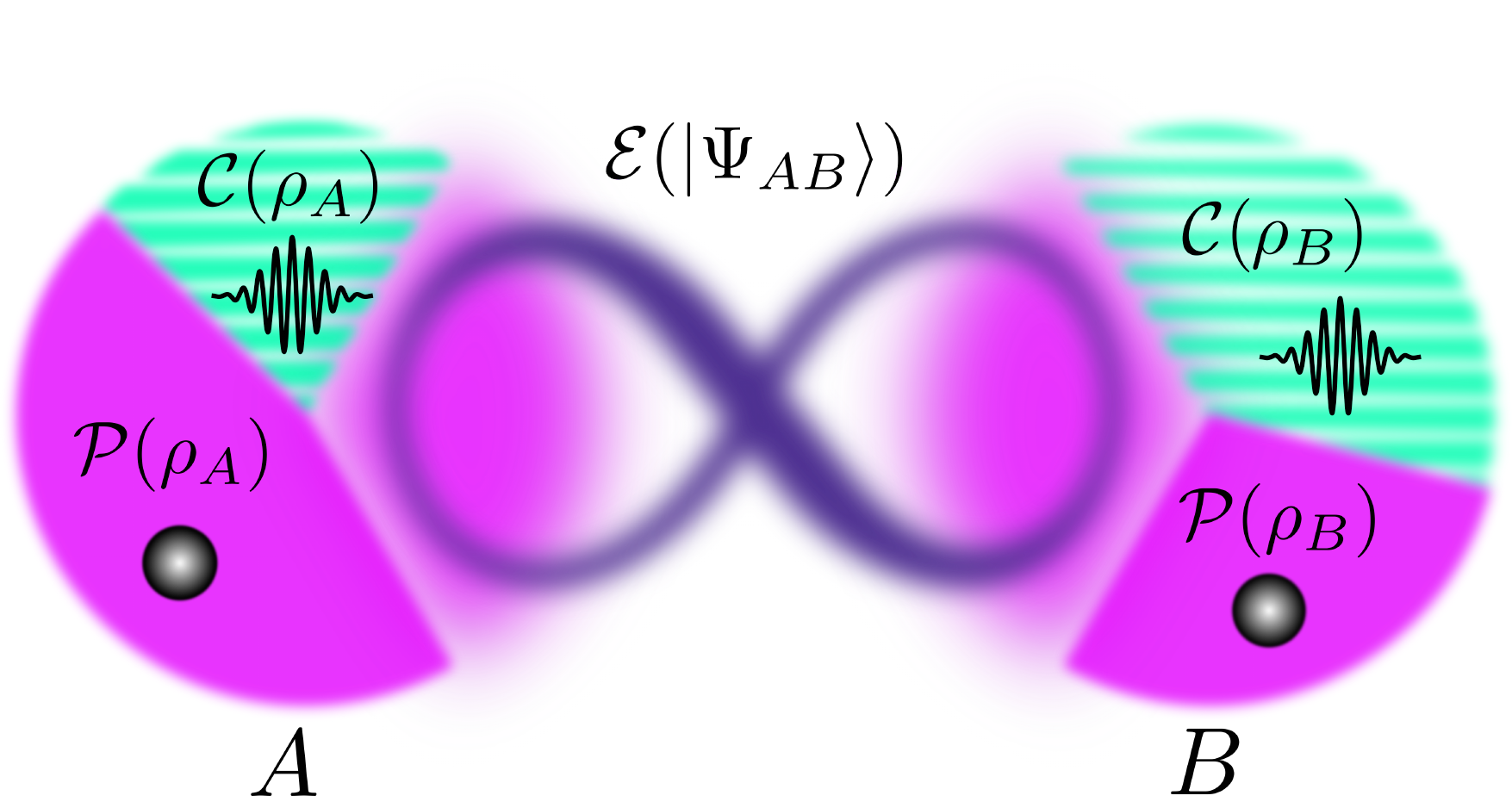}
    \caption{Schematic illustration of complete complementarity relations of two quantum systems, $A$ and $B$, highlighting their local features as well as their global feature. The local and global properties of the quantons $A$ and $B$ are described by the complete complementarity relations given in Eq.~\eqref{Eq:CCRg}, where $\rho_A$ and $\rho_B$ denote the reduced density matrices, $\mathcal{C}$ denotes the wave-like property quantified by quantum coherence, $\mathcal{P}$ is the predictability measure, and $\mathcal{E}(|\Psi_{AB}\rangle)$ is an entanglement monotone. The sum of these three quantities equals a constant $\alpha(d_A)$ that depends only on the dimension of the system. In general, the local quantities $\mathcal{P}$ and $\mathcal{C}$ may differ for the two quantons, whereas the global quantity $\mathcal{E}$ is identical for both subsystems $A$ and $B$.}
    \label{fig:CCR}
\end{figure}

%---------------------------
\subsection{Entanglement monotones from duality}
\label{sec:CCRtheorem}

The procedure described in Sec.~\ref{sec:CCRhs} allows one to obtain the complete complementarity relation based on the $\ell_2$-norm quantum coherence. This naturally raises the question of whether the other complementarity relations described in Sec.~\ref{sec:CRpositivity} can also be completed in a similar manner, and more generally whether the complementarity relation of the form given in Eq.~\eqref{eq:QCR} admits a corresponding complete version. This question is answered by the theorem established by~\citet{Basso2022a}, which shows that such a completion can indeed be achieved under suitable conditions.

The theorem can be stated as follows. Let Eq.~\eqref{eq:QCR} be a quantum complementarity relation for the state $\rho_A$ that is saturated only when $\rho_A$ is pure, with $\mathcal{P}(\rho_A)$ and $\mathcal{C}(\rho_A)$ being measures of predictability and visibility (or quantum coherence) that satisfy~\citet{Durr2001} and \citet{Englert2008} criteria, respectively, and $\alpha(d_A) \in \mathbb{R}^+$. Then the quantity
\begin{equation}
     \mathcal{E}(\ket{\Psi_{AB}}) := \alpha(d_A) - \mathcal{P}(\rho_A) - \mathcal{C}(\rho_A)
\label{eq:entmon_}
\end{equation}
defines an entanglement monotone when expressed in terms of the Schmidt coefficients with $\ket{\Psi_{AB}}$ being a purification of $\rho_A$. Hence, the complementarity relation is completed and takes the general form
\begin{eqnarray}
\mathcal{P}(\rho_A) + \mathcal{C}(\rho_A) + \mathcal{E}(\ket{\Psi_{AB}}) = \alpha(d_A). \label{Eq:CCRg}
\end{eqnarray}  
Figure~\ref{fig:CCR} provides a schematic illustration of two quantum systems, $A$ and $B$, highlighting the local features, predictability $\mathcal{P}$ and quantum coherence $\mathcal{C}$, as well as the global feature given by their entanglement $\mathcal{E}$.

The proof of this theorem is rather simple and here we outline the main steps. 
Any locally unitary invariant and real concave function $f$ on  $\mathcal{D}(\mathcal{H}_{AB})$ can be used to construct an entanglement monotone $\mathcal{E}(\ket{\Psi_{AB}})$ for the pure state 
$\ket{\Psi_{AB}}$~\cite{Zhu2017, Vidal2000}, defined as
\begin{align}
    \mathcal{E}(\ket{\Psi_{AB}}) := f(\Tr_B(|\Psi\rangle_{AB}\langle \Psi|)) 
    = f(\rho_A).
\label{eq:pure}
\end{align}
Since $\mathcal{C}$ and $\mathcal{P}$ are convex functions, the quantity
$\mathcal{E}(\ket{\Psi_{AB}}) = \alpha(d_A) - \mathcal{P}(\rho_A) - \mathcal{C}(\rho_A)$ is a concave function. Using the Schmidt decomposition $\ket{\Psi_{AB}} = \sum_k \lambda_k \ket{\phi_k}_A \otimes \ket{\Psi_k}_B$,
the reduced density matrix of subsystem $A$ can be written as $\rho_A = \sum_k \lambda_k^2 \ketbra{\phi_k}{\phi_k}$, which implies that $\mathcal{C}(\rho_A) = 0$ and $\mathcal{P}(\rho_A) \neq 0$. Furthermore, since $\mathcal{P}(\rho_A)$ is invariant under permutations of the components of the probability vector $\vec{\lambda} = (\lambda_0,\dots,\lambda_{d-1})$, the function $\mathcal{E}(\ket{\Psi_{AB}})=f(\rho_A)$ is likewise invariant under permutations of the entries of $\vec{\lambda}$, which completes the proof.
Recently, \citet{Ding2025} rediscovered this result.

From this theorem, it follows that the entanglement monotones obtained from the complementarity relations in Eq.~\eqref{eq:QCR_l1} are
\begin{align}
& \mathcal{E}_{vn}(\rho_A) := \log d_A - \mathcal{C}_{re}(\rho_A) - \mathcal{P}_{vn}(\rho_A), \\
& \mathcal{E}_{\ell_1}(\rho_A) := d_A - 1 - \mathcal{C}_{\ell_1}(\rho_A) - \mathcal{P}_{\ell_1}(\rho_A), \\
& \mathcal{E}_{wy}(\rho_A) := \frac{d_A - 1}{d_A} - \mathcal{C}_{wy}(\rho_A) - \mathcal{P}_{\ell_2}(\rho_A).
\end{align}
It is noteworthy that $\mathcal{E}_{vn}(\rho_A)$ reduces to the von Neumann entropy, $S_{vn}(\rho_A) = - \Tr(\rho_A \log \rho_A)$, which is a well-known entanglement monotone for globally pure states. Moreover, when restricted to the Schmidt
coefficients, $\mathcal{E}_{\ell_1}(\rho_A) = \sum_{j \neq k}\sqrt{\lambda_j\lambda_k}$ coincides with the robustness of entanglement for globally pure states~\cite{Vidal1999}, while $\mathcal{E}_{wy}(\rho_A) = 1 - \sum_j \lambda_j^2$ corresponds to the linear entropy of the state $\rho_A$. However, it is important to emphasize that, although $\mathcal{E}_{\ell_1}$ and $\mathcal{E}_{wy}$ reduce to these well-known measures when expressed in terms of the Schmidt coefficients (for which the coherence vanishes), their general expressions differ whenever coherence is nonzero, since they are defined in terms of the corresponding predictability and coherence quantifiers appearing in the complementarity relations. In particular, even though the predictability quantifier is the same in both cases, the resulting entanglement functions $\mathcal{E}_{\ell_2}$ and $\mathcal{E}_{wy}$ differ due to the distinct coherence measures employed in their definitions~\cite{Basso2021f}.

Hence, entanglement monotones for globally pure states can be derived from complementarity relations and subsequently extended to mixed states through the convex-roof construction \cite{Rothlisberger2012}. This result relies only on the assumption that the predictability and visibility (or quantum coherence) quantifiers satisfy the criteria established in the literature by~\citet{Durr2001} and \citet{Englert2008}, and that the corresponding complementarity relation saturates only for pure states. In this way, the framework provides a systematic method for constructing new entanglement measures while formally connecting two fundamental features of quantum mechanics.

%---------------------------
\subsection{Entanglement, distinguishability, and predictability}

So far we have seen that two kinds of complementarity relations have been discussed in the literature, namely predictability-based and distinguishability-based complementarity relations. As described in Sec.~\ref{sec:III_D},~\citet{Qureshi2021b} and \citet{Roy2022} investigated the connection between predictability, distinguishability, and entanglement in the context of multipath interference. Their results show that entanglement quantitatively links distinguishability and predictability, indicating that given duality relation can be interpreted as being part of a broader triality relation involving predictability, coherence, and entanglement.

Moreover, it is worth noting that \citet{Englert2000a} introduced a hierarchy relating predictability and distinguishability by using the following argument. The distinguishability $\mathcal{D}$ represents the information that an experimenter can obtain from the path-detector device (or, equivalently, from measuring an observable of the environment), which the authors refer to as which-alternative knowledge. On the other hand, the maximum possible value of this quantity, $\mathcal{D}_{max}$, characterizes Nature's information about the actual alternative. This situation corresponds to the ideal case in which the quanton becomes coupled to a path-detector such that the detector states become mutually orthogonal, as in an ideal von Neumann measurement. In addition, the predictability $\mathcal{P}$ quantifies the knowledge available to the experimentalist prior to measuring the path-detector device. From these considerations, the authors established the hierarchy $\mathcal{P}\le \mathcal{D}\le\mathcal{D}_{max}$.

As already discussed in Sec.~\ref{sec:duality_multi-slits}, in a $d$-slit interferometer, let $\ket{j}_A$ denote the state corresponding to the quanton $A$ taking the $j$-th path, so that a general pure state can be written as $\ket{\psi}_A=\sum_j a_j\ket{j}_A$, where ${\ket{j}_A}$ forms an orthonormal path basis. To obtain which-path information, one can introduce a path detector initially in state $\ket{d_0}_B$, which interacts unitarily with the quanton according to $U(\ket{j}_A\otimes\ket{d_0}_B)=\ket{j}_A\otimes\ket{d_j}_B$. After the interaction, the joint state becomes $\ket{\Psi}_{AB}=\sum_j a_j \ket{j}_A\otimes\ket{d_j}_B.$ Measuring a detector observable $\mathcal{O}_B$ with eigenstates ${\ket{d_j}_B}$ yields outcomes $o_k$ with probabilities $p_k = \Tr(\mathbb{I}_A \otimes \Pi^k_B \rho_{AB})$, producing conditional states $\rho_A^{(k)}$ of the quanton $A$. A non-selective measurement leads to the ensemble $\rho_A=\sum_k p_k\rho_A^{(k)}$, sorted into sub-ensembles according to the detector outcomes. For each sub-ensemble, one can define a distinguishability measure. Let us consider here $\mathcal{D}_{vn}^k(\rho_A^{(k)})=\log d_A - S_{vn}(\rho^{(k)}_{A,\text{diag}})$ and the quantum coherence $\mathcal{C}_{re}^k(\rho_A^{(k)})=S_{vn}(\rho^{(k)}_{A,\text{diag}})-S_{vn}(\rho^{(k)}_A).$ Averaging over all outcomes gives the total distinguishability $\mathcal{D}^{av}_{vn}(\rho_A) = \sum_k p_k \mathcal{D}^k_{vn}(\rho^{(k)}_A)$ and the average coherence $\mathcal{C}^{av}_{re}(\rho_A) = \sum_k p_k \mathcal{C}^k_{re}(\rho^{(k)}_A)$, which satisfy the complementarity relation
\begin{align}
\mathcal{D}^{av}_{vn}(\rho_A)+\mathcal{C}^{av}_{re}(\rho_A)\le \log(d_A).    
\end{align}
Moreover, by noting that the largest value $\mathcal{D}^{max}_{vn}$ of the averaged distinguishability $\mathcal{D}^{av}_{vn}$ is $\mathcal{D}^{max}_{vn} = \log(d_A)$, given that the maximum is reached when the states $\ket{d_j}_B$ are orthogonal and therefore $S_{vn}(\rho^{(k)}_{A diag}) = 0$. In this case, one can show that $\mathcal{D}^{max}_{vn}(\rho_A) = \mathcal{P}_{vn}(\rho_A) + S_{vn}(\rho_A)$.

This construction can be carried out independently of the specific function used to define quantum coherence or distinguishability. Therefore, we can generally write
\begin{align}
\mathcal{D}^{av}(\rho_A)+\mathcal{C}^{av}(\rho_A)\le \alpha(d_A), \label{eq:avCR}    
\end{align}
for any underlying measure that satisfies the criteria of~\citet{Durr2001} and \citet{Englert2008}. This allows us to state the following theorem by~\cite{Basso2022b}. Let $\mathcal{D}^{max}$ be the maximum value of a distinguishability measure appearing in a complementarity relation given by Eq.~\eqref{eq:avCR} together with a corresponding predictability
measure $\mathcal{P}$ used in complementarity relations of the type~\eqref{eq:QCR}, which saturates only for pure states and also satisfies the criteria of~\citet{Durr2001} and \citet{Englert2008}. Then the quantity
\begin{equation}
\mathcal{E}(\ket{\Psi_{AB}}) := \mathcal{D}^{max} - \mathcal{P}(\rho_A) 
\label{eq:entmon_dis}
\end{equation}
defines an entanglement monotone. The proof of this theorem is similar to that presented in Sec.~\ref{sec:CCRtheorem}. For instance, let us consider the entanglement quantity, introduced by~\citet{Qureshi2021b}, defined as the difference between the distinguishability measure $\mathcal{D}_Q$ given by Eq.~\eqref{DQn} and the corresponding predictability $\mathcal{P}_Q$ given by Eq.~\eqref{PQ}, leading to Eq.~\eqref{eq:E_Q}. The distinguishability $\mathcal{D}_Q$ reaches its maximum when the detector states are mutually orthogonal. In this limit,
\begin{align}
\mathcal{E}_Q=\frac{1}{d_A-1}\sum_{j\neq k}\sqrt{\lambda_j\lambda_k},
\end{align}
where $\lambda_j$ are the Schmidt coefficients. In this case, $\mathcal{E}_Q$ is proportional to the robustness of entanglement for pure bipartite states~\cite{Vidal1999}.

\section{Retro-inference experiments}
\label{sec:retro}

Many experiments have demonstrated the manifestation of wave-particle duality, as discussed so far. We now turn to a class of experiments in which this duality is examined through a more subtle issue: the inference of wave-like or particle-like behavior from measurement arrangements and outcomes that are fixed only at a later stage of the experiment. The family of experiments commonly associated with delayed choice, quantum erasure, and related tests of complementarity has given rise to a vast and diverse literature. This body of work encompasses foundational proposals, experimental realizations, interpretational debates, information-theoretic analyses, causal-model approaches, as well as more recent developments in quantum simulation and platform-specific implementations.

The Wheeler delayed-choice experiment, first introduced in Refs.~\cite{Wheeler1978,Wheeler1984}, involves an interferometric setup where the late insertion or removal of the second beam splitter in a Mach--Zehnder interferometer dictates whether the final detection statistics exhibit interference patterns or reveal which-way information. This has often been described as a situation in which a later experimental choice appears to determine whether the quanton behaved as a wave or as a particle in the past. More precisely, however, the experiment raises a question about retroinference rather than retrocausation. What is fixed by the later choice is the experimental context in which the earlier behavior may be meaningfully described. This distinction is central to the analysis below.

For the theoretical and experimental work on the delayed-choice experiments, we refer the reader to a review by \citet{Ma2016}. Early and modern implementations of WDCE include single-photon interferometric experiments~\cite{Baldzuhn1989,Hellmuth1987,Jacques2007,Jacques2008a}, followed by realizations using atoms and matter waves~\cite{Lawson-Daku1996,Manning2015}, as well as temporal and memory-based versions~\cite{Dong2020,Hong2023}. More recent developments include implementations in nanophotonic and integrated architectures~\cite{Chen2021}, as well as in nuclear magnetic resonance (NMR) and superconducting platforms~\cite{LiuX2017,Roy2012}. Additional variants explore quantum-walk dynamics and decoherence effects~\cite{Jeong2013,Lee2014}, satellite-ground interferometer~\cite{Vedovato2017}, and schemes involving dual-selection or structured photons~\cite{YangY2025,YangZ2024}.

A second closely related family is formed by quantum eraser experiments. In these setups, which-way information is first made available by correlating the quanton with an additional degree of freedom and is then erased by measuring that degree of freedom in a complementary basis. The original proposals and early demonstrations~\cite{Scully1982,Kwiat1994,Zajonc1991,Walborn2002} led naturally to delayed-choice quantum eraser variants, where the erasure or retrieval of which-way information may occur after the quanton has already been detected. These experiments have generated extensive discussions concerning post-selection, conditional interference, the role of correlations, and the status of delayed-choice reasoning~\cite{Englert2000a,Kim2000,Ma2016,Qureshi2020,Qureshi2021a,Qureshi2025}. Extensions involving entanglement and entanglement swapping push the same conceptual structure into explicitly nonlocal scenarios~\cite{Peres2000,Sciarrino2002,Kaiser2012,Ma2012,Ma2013,Xin2015}.

A third line of experimental setups consists of variations in two-slit interferometric arrangements that have been claimed to challenge BCP by allowing simultaneous access to wave-like and particle-like behavior in its full extent. Afshar-type experiments are the most prominent examples~\cite{Afshar2006,Afshar2007}. In these proposals, wave-like behavior is inferred from the presence of an interference pattern, or from a minimally disturbing probe placed at interference minima, while particle-like behavior is inferred retrospectively from correlations between the particle's path and the detector outcomes. The analyses of~\citet{Unruh2004,Unruh2007} provides an important counterpoint, emphasizing that the relevant inferences depend on the full experimental configuration and that conclusions obtained in one arrangement cannot generally be transferred unchanged to the other one.

A novel idea of a \emph{quantum switch} was presented in which the causal order of two operations becomes indefinite in the sense of quantum superposition \cite{Chiribella2013}. Recently, complementarity has been examined through the lens of this quantum switch, where the sequence of operations, involving observing wave characteristics and acquiring path information of a quanton, may be indefinite \cite{Siddiqui2026}. It may be noted that in Afshar-type experiments, the interference is probed first, and the path information is inferred later. In conventional which-way experiments it is the other way round.

In the previous paragraphs, we referred to several existing theoretical and experimental studies on this broad topic. However, the aim of this section is not to deliver a comprehensive literature review, but to highlight those complementary aspects that are specifically relevant to the present work. To this end, we analyze these configurations---Wheeler's delayed-choice experiment, its quantum-controlled version, quantum eraser and partial-erasure schemes, and Afshar--Unruh-type arrangements---using the complementarity relations introduced in Secs.~\ref{sec:1st} and~\ref{sec:QCP}. This perspective will be used to clarify that any apparent tension with BCP arises from limitations of the specific measures used to quantify it, rather than from the BCP itself.

%---------------------------
\subsection{Wheeler's delayed-choice experiment}
\label{sec:WDCE}

\begin{figure}
    \centering
    \includegraphics[width=1\linewidth]{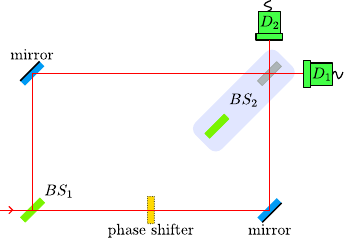}
    \caption{Wheeler’s delayed-choice experiment is implemented using a Mach--Zehnder interferometer, where $BS_{1(2)}$ denotes the first (second) beam splitter and $D_{1(2)}$ denotes one of the two detectors. The insertion of $BS_2$ is delayed until after the quanton has already passed through $BS_1$. When $BS_2$ is present, the photon exhibits wave-like interference, whereas when $BS_2$ is removed, the photon displays particle-like behavior, as judged by the sensitivity of the detection probabilities to phase variations inside the interferometer.}
    \label{fig:mzi_WDCE}
\end{figure}

Within Bohr's original concept of complementarity, according to which a photon's behavior is determined by the experimental setup taken as a whole, \citet{Wheeler1978, Wheeler1984} introduces the delayed-choice experiment. 
The Wheeler's delayed-choice experiment (WDCE) was first introduced as a modified double-slit experiment and simplified by a two-level system in a Mach--Zehnder interferometer (MZI), as depicted in Fig.~\ref{fig:mzi_WDCE}. 

The state of a quanton entering a MZI through the first beam splitter ($BS_1$), assuming a 50:50 beam splitter, is represented in the path degree of freedom as a coherent superposition between the paths representing the inner arms of the interferometer. After the phase shifter (which adds the phase factor $e^{i\phi_1}$ in the lower state) and the reflections from the mirrors, the state can be written as
\begin{equation}
\ket{\psi_{\text{out}}} = \frac{1}{\sqrt{2}}(ie^{i\phi_1}\ket{\psi_{\text{lower}}} -\ket{\psi_{\text{upper}}}),
\label{eq:psi_bs1}
\end{equation}
where $\ket{\psi_{\text{lower}}}$ and $\ket{\psi_{\text{upper}}}$ denote the quantum states of the quanton propagating inside the MZI along the lower and upper paths, respectively, as illustrated in Fig.~\ref{fig:mzi_WDCE}.
The term $i$ and the minus sign are introduced by the reflection term $e^{i\pi/2}=i$, which occurs in the beam splitter and the mirrors. If the $BS_2$ is present, the state after $BS_2$ is given by
\begin{equation}
\ket{\psi}_{\text{in}} = -\frac{1}{2}\left( (e^{i\phi_1} + 1)\ket{D_1} - i(e^{i\phi_1} - 1)\ket{D_2}\right),
\label{eq:MZI_BS2}
\end{equation}
where $\ket{D_1}$ and $\ket{D_2}$ are the quantum states after the $BS_2$ that lead to detector $D_1$ and $D_2$, respectively.
For the moment, we consider two distinct experimental configurations, each analyzed separately, i.e., mutually exclusive setups. The first corresponds to the configuration in which the second beam splitter ($BS_2$) is present, as described by the state above. The second corresponds to the configuration in which $BS_2$ is absent, Eq.~\eqref{eq:psi_bs1}, so that the relevant state is the one obtained after $BS_1$ but before reaching $BS_2$. The labels ``in'' and ``out'' assigned to the states were chosen deliberately in anticipation of the subsequent discussion.

We now proceed to analyze these configurations in terms of complementarity relations. The experiment that reveals wave-like behavior corresponds to the configuration in which $BS_2$ is inserted and is, essentially, the two-path discrete version of the setup considered by~\citet{Greenberger1988}. In this case, the complementarity relation is given by the equality in Eq.~\eqref{dualityG}, since a pure state is injected into the MZI. The wave-like behavior can be quantified through the interferometric visibility defined in Eq.~\eqref{eq:vis}, computed from the detection probabilities associated with the state after $BS_2$, given in Eq.~\eqref{eq:MZI_BS2}. According to the Born rule, the probability of detection at detector $D_1$ is $\Pr(D_1|\psi_{\text{in}}) = (1+\cos\phi_1)/2$. By determining the maximum and minimum values of this probability as a function of $\phi_1$, one obtains $\mathcal{V}_1 = 1$. A similar analysis applies to detector $D_2$, for which $\Pr(D_2|\psi_{\text{in}}) = (1-\cos\phi_1)/2$, also leading to $\mathcal{V}_2 = 1$.
From Eq.~\eqref{dualityG}, the complementarity relation is trivially satisfied. Since $BS_1$ is a 50:50 beam splitter, it follows that $\mathcal{P} = 0$ and $\mathcal{V}_j = 1$, with $j = 1,2$ labeling the visibilities associated with each detector, obtained from the extremal values of the corresponding detection probabilities. It is worth noting that, in this case, the detection probabilities are sensitive to phase variations.

On the other hand, particle-like behavior is observed in the configuration where $BS_2$ is removed, as argued by \citet{Ionicioiu2011}, since the detection probabilities become insensitive to phase variations. In this case, the probabilities described by Eq.~\eqref{eq:psi_bs1} provide information about the path followed by the quanton. More specifically, a detection at $D_1$ indicates that the quanton took the upper path, while a detection at $D_2$ indicates that it took the lower path. However, note that, in terms of the complementarity relation given by Eq.~\eqref{dualityG}, both predictability and visibility vanish, i.e., $\mathcal{P} = 0$ and $\mathcal{V} = 0$. In contrast, when expressed in terms of the complementarity relations given by Eq.~\eqref{eq:QCR}, the situation is different: the predictability still vanishes, while the coherence attains its maximum value, capturing the underlying superposition established within the interferometer.~\citet{Dieguez2022} further investigates this aspect by showing that the wave-like element of reality remains maximal in the quantum version of WDCE.

Nevertheless, according to the proposal of postponing the choice of whether to insert or not $BS_2$ considers the situation in which the quanton has already passed through $BS_1$. According to \citet{Wheeler1984}, ``\textit{In this sense, we have a strange inversion of the normal order of time. We, now, by moving the mirror in or out have an unavoidable effect on what we have a right to say about the already past history of that photon}'' and ``\textit{of all the features of the \emph{act of creation} that is the elementary quantum phenomenon, the most startling is that seen in the delayed-choice experiment. It reaches back into the past in apparent opposition to the normal order of time.}''
This feature arises from the role of the beam splitter~\cite{Grangier1986} and from the fact that, following Bohr's viewpoint, the experimental arrangement must be analyzed as a whole. When $BS_2$ is inserted, the observed interference---constructive at detector $D_1$ and destructive at detector $D_2$---indicates that the quanton behaves as if it propagates along both paths. In contrast, when $BS_2$ is absent, detections occur at only one detector at a time, revealing path information in a fundamentally random manner. Since each configuration is associated with a well-defined type of behavior, postponing the choice of which configuration to implement until after the quanton has passed through $BS_1$ gives rise to this apparent retroactive inference about its past evolution.

The apparent ``back-in-time'' causation arises from the specific way wave- and particle-like behaviors are defined in terms of detection probabilities, particularly through their sensitivity or insensitivity to phase variations. In this framework, the absence of phase dependence in the detection probabilities (when $BS_2$ is removed) leads to a particle-like interpretation, while its presence (when $BS_2$ is inserted) leads to a wave-like description, giving the impression that a later choice influences the past behavior of the quanton. However, this apparent retrocausal feature is avoided when one adopts complementarity relations of Eq.~\eqref{eq:QCR} based on quantum coherence, or alternative formulations based on realism~\cite{Dieguez2022}. In these approaches, the characterization of wave- and particle-like properties does not rely solely on detection probabilities, thereby removing the need for such back-in-time interpretations. Although \citet{Qureshi2020} does not formulate the discussion in terms of coherence-based complementarity relations, his analysis is consistent with the view that phase-insensitive detection statistics do not exhaust the quantum features present within the MZI. \citet{Araujo2025} proposed an experimental setup in which entanglement serves as a quantifier of wave-like behavior, building on the ideas introduced by~\citet{Angelo2015}. In this scheme, wave properties are characterized through the presence of tripartite entanglement involving the path degree of freedom and an additional qubit placed in each arm of the MZI. The underlying idea is that only a delocalized system---i.e., one exhibiting wave-like behavior---can generate such entanglement. From this perspective, the identification of wave-like behavior via quantum coherence after the first beam splitter is supported by the corresponding observation of entanglement.

\citet{Chrysosthemos2023a} used a more general MZI, considering that both $BS_1$ and $BS_2$ could have controllable transmissibility $T$ and reflectivity $R$. For example, consider that $BS_1$ is an ordinary 50:50 beam splitter. Then both WDCE experimental configurations are obtained by controlling $T_2$ and $R_2$ in $BS_2$. In the more general case, allowing the transmission and reflection coefficients of both beam splitters to be controlled, the state in Eq.~\eqref{eq:MZI_BS2} after $BS_2$ is modified to
\begin{equation}
  \hspace{-5pt}  \ket{\psi_{\text{in}}^\prime} \! = \! -\big(e^{i \phi_1} T_1 R_2 + R_1 T_2 \big)\ket{D_1} %\nonumber
%    \\ 
    %&
    + i\big(e^{i \phi_1}T_1 T_2 - R_1 R_2 \big) \ket{D_2}.
    \label{eq:psi3}
\end{equation}
By maximizing and minimizing the detection probabilities with respect to $\phi_1$, Eq.~\eqref{eq:vis} yields two expressions for the interferometric visibility, one associated with each detector:
\begin{align}
\mathcal{V}_1 & := \frac{\Pr(D_1)_{\max}-\Pr(D_1)_{\min}}{\Pr(D_1)_{\max}+\Pr(D_1)_{\min}} = \frac{2T_{1}R_{1}T_{2}R_{2}}{T_{1}^{2}R_{2}^{2}+R_{1}^{2}T_{2}^{2}}, \label{eq:vis0} \\
\mathcal{V}_2 & := \frac{\Pr(D_2)_{\max}-\Pr(D_2)_{\min}}{\Pr(D_2)_{\max}+\Pr(D_2)_{\min}} = \frac{2T_{1}R_{1}T_{2}R_{2}}{T_{1}^{2}T_{2}^{2}+R_{1}^{2}R_{2}^{2}}.
\end{align}
The complementarity relations are then analyzed employing these expressions for the visibilities, and scenarios are identified in which such relations appear to be violated when combined with the standard expression for predictability of~\citet{Greenberger1988}.
To address this issue, an alternative expression for predictability is derived by explicitly imposing the validity of the complementarity relation. However, this modified quantity does not correspond to the conventional definition of~\emph{a priori} which-way information, highlighting a conceptual tension between enforcing the complementarity bound and preserving its standard interpretation. This experimental configuration enables an intermediate regime when $T_2 \neq R_2$, which, to our knowledge, is properly addressed only by the state-dependent approach of~\citet{Starke2024}, discussed in Sec.~\ref{sec:QCP}, and by the realism-based framework of~\citet{Bilobran15}, which will be analyzed in Sec.~\ref{sec:realism}. Consequently, interferometric visibility should not be regarded as the unique measure of wave-like behavior, nor as a universally applicable one.

%---------------------------
%\subsubsection{Quantum control in WDCE}
%\label{sec:qWDCE}

\vspace{0.25cm}

\textit{Quantum control in WDCE:} A step forward in the WDCE was made by \citet{Ionicioiu2011}, who proposed a quantum version of the delayed-choice experiment by placing $BS_2$ in a coherent superposition of being present and absent. The central idea is to introduce an ancilla---prepared, for instance, in a biased state $\cos\alpha \ket{0} + \sin\alpha \ket{1}$---to control which type of behavior is observed.

They defined the particle state $\ket{\text{particle}}$ as the state given in Eq.~\eqref{eq:psi_bs1}, corresponding to the configuration without $BS_2$, and the wave state $\ket{\text{wave}}$ as the state in Eq.~\eqref{eq:MZI_BS2}, corresponding to the configuration with $BS_2$ in place, following the characterization of particle- and wave-like behavior based on detection probabilities. The joint state of the quanton and the ancilla can then be written as
\begin{eqnarray}
\ket{\psi} = \cos\alpha\ket{\text{particle}}\ket{0} + \sin\alpha\ket{\text{wave}}\ket{1}.
\label{eq:psi_QDC}
\end{eqnarray}
\citet{Ionicioiu2011} found that the quanton manifests morphing behavior between particle-like and wave-like, where the detector $D_1$ measures $I_1(\phi,\alpha) = I_p(\phi)\cos^2\alpha + I_w(\phi)\sin^2\alpha$, with the corresponding visibility $\mathcal{V} = \sin^2\alpha$. Here, $I_p (\phi) = \frac{1}{2}$ and $I_w = \cos^2\frac{\phi}{2}$, which are obtained from the case of a 50:50 beam splitter. The joint measurements of the photon and ancilla for the state in Eq.~\eqref{eq:psi_QDC} are given by $\Pr(a, b) = (\frac{1}{2}\cos^2\alpha,\sin^2\alpha\cos^2\frac{\alpha}{2},\frac{1}{2}\cos^2\alpha,\sin^2\alpha\sin^2\frac{\phi}{2})$, with $a$ being the photon path and $b$ the ancilla.

The authors state that ``\textit{Our result suggests a reinterpretation of the complementarity principle; instead of the complementarity of experimental setups (Bohr's view), we have the complementarity of experimental data.}''
They conclude their work with the following remark about the order of time: ``\textit{There is no inversion of the normal order of time-in our case, we measure the photon before the ancilla deciding the experimental setup (open or closed interferometer). It is only after we interpret the photon data, by correlating them with the results of the ancilla, that either a particle-like or wave-like behavior emerges: behavior is in the eye of the observer.}''
This was experimentally verified through NMR techniques \cite{Auccaise2012, Peruzzo2012} and through photonic experiments \cite{Roy2012, Wang2019}. A further step in this direction is provided by~\citet{Wang2022}, where quantum-controlled schemes are extended by incorporating entanglement as a resource to coherently control the manifestation of wave- and particle-like behavior, allowing for a more general manipulation of complementarity so that any notion of an objective wave-particle character for the individual system becomes ill-defined when the interferometric setting is not well defined.

\citet{Dieguez2022} revisited the quantum delayed-choice experiment from the perspective of path coherence. They argue that, even in the absence of $BS_2$, the quanton still possesses non-vanishing path coherence inside the interferometer. Therefore, the standard interpretation, based on detection probabilities and their insensitivity to phase, misidentifies this configuration as particle-like. According to a coherence-based definition the state remains wave-like, and thus no genuine morphing behavior occurs between wave and particle regimes in the usual quantum delayed-choice setup. To address this issue, the authors propose a modification of the experiment in which the first beam splitter is placed under quantum control. As a result, true morphing behavior emerges when wave and particle properties are defined in terms of path coherence, rather than detection probabilities.
They used a liquid-state NMR experiment with two spin-1/2 qubits to test and verify the physical realism predictions proposed in their quantum-controlled interferometric framework. 

\citet{Qureshi2013b} analyzed the scenario of two-slit interference with a path detector in a quantum superposition of being present and absent. An inequality was established that limits the visibility of interference and the extent of which-way information that can be acquired within the framework of such modified experiments. Given that the wave nature can only be demonstrated through a collection of detections, it was contended that in these experiments an individual detection can be interpreted as pertaining only to one of the two sub-ensembles, which are associated with either the wave nature or the particle nature. Consequently, each detected quanton behaves either as a particle or as a wave, but never simultaneously as both, thereby fully honoring BCP.
\citet{Siddiqui2021} carried out a more general and detailed analysis of a multipath interference experiment that has a path detector in a quantum superposition of being present and absent. They demonstrated that a tight multipath wave-particle duality relationship is upheld in all these circumstances, and complementarity is fully respected. The perceived violation of complementarity mostly stems from an inaccurate assessment of path distinguishability in these contexts. More interestingly, they argued that the underlying physics of the interference experiments with a quantum device \cite{Ionicioiu2011} is the same as that of a MZI with a \emph{biased} second beam splitter which is completely classical.

~\citet{Chaves2018} introduced a novel perspective on delayed-choice experiments by analyzing them within the framework of causal models~\citet{Pearl2013}. Rather than focusing on wave-particle duality in the traditional sense, the authors reformulate both Wheeler’s delayed-choice experiment and its quantum-controlled variant (QDCE) in terms of causal structures and statistical dependencies. The central idea is to treat these experiments as instances of a prepare-and-measure scenario, where a system is prepared in a certain state and later measured under different configurations. Within this framework, one can represent the experiment using causal diagrams, where hidden variables encode underlying properties of the system, and observable outcomes arise from conditional probability distributions. This approach allows one to systematically investigate whether the observed statistics can be reproduced by classical causal models subject to specific constraints. An interesting result is that standard delayed-choice experiments, including the quantum delayed-choice version in which the second beam splitter is coherently controlled, can be reproduced by a classical causal model without invoking retrocausality. In other words, the seemingly paradoxical features of these experiments---often interpreted as requiring influences from the future or a breakdown of classical causality---can instead be explained within a causal framework that preserves a forward-in-time causal structure. However, the authors go further by identifying conditions under which classical explanations fail. By slightly modifying the experimental scenario, they construct a situation in which the observed statistics cannot be reproduced by any non-retrocausal hidden-variable model with the same dimensionality as the quantum system. The main conceptual insight is that what makes delayed-choice experiments counterintuitive is not the apparent need for retrocausality or wave-particle duality per se, but rather the fact that any classical explanation must either use a hidden variable with higher dimensionality than the corresponding quantum system or invoke retrocausal influences.

%---------------------------
\subsection{The quantum eraser}
\label{sec:qeraser}

An interesting experiment was proposed by \citet{Scully1982}, in which one could choose to observe the particle nature of a quanton by entangling the path degree of freedom with a path detector. Alternatively, one could decide to erase the path labels by measuring the path-detector in a different way. It was shown that the interference would return if the which-way information is erased. The basic idea is as follows. When a two-state path-detector is entangled with the particle, the state is given by Eq.~\eqref{entstate}. This state does not yield any interference because the path-detector carries the potential which-way information of the quanton. Looking at an observable of the path-detector whose eigenstates are $|d_{1,2}\rangle$ would provide information about which of the two slits the quanton went through. If one finds the path-detector in the state $|d_{1}\rangle$ ($|d_{2}\rangle$), it would imply that the wavefunction of the quanton was $\psi_1(x)$ ($\psi_2(x)$). This would correspond to the particles forming one of the two slightly shifted  distributions on the screen, without any fringes. This is in full agreement with the BCP, showing that when full path-information is present, there is no interference. On the other hand, one could also write the same state as Eq.~(\ref{ent2}). Suppose now one measures an observable of the path detector whose eigenstates are $|d_{\pm}\rangle$, and finds that the state is (say) $|d_-\rangle$, this changes the state of the quanton to
\begin{eqnarray}
\Psi_f(x) &=& \tfrac{1}{2} [ \psi_1(x)+ \psi_2(x)]\langle d_-|d_+\rangle + \tfrac{1}{2} [ \psi_1(x) - \psi_2(x)]\langle d_-|d_-\rangle \nonumber\\
 &=& \tfrac{1}{\sqrt{2}} [ \psi_1(x) - \psi_2(x)] ,
\label{erased}
\end{eqnarray}
with a suitably modified normalization. This yields an interference with maximum visibility. When the path-detector falls into the state $|d_-\rangle$, the potential information about whether the state could be $|d_1\rangle$ or $|d_2\rangle$ is lost, and the \emph{path information is erased}. In such a situation, BCP allows the appearance of interference. On the other hand, if one looks only at those quantons for whom the path-detector state is found to be $|d_+\rangle$, they will show an interference corresponding to the state $\tfrac{1}{\sqrt{2}} [ \psi_1(x) + \psi_2(x)]$. The two interference patterns will be complementary to each other, in the sense that the maxima of one will be the positions of the minima of the other, such that the two add up to give no interference. The physical picture that one draws from this is that if one looks at the states $|d_1\rangle,|d_2\rangle$, the quanton randomly passes through one of the slits, but if one looks at the states $|d_{\pm}\rangle$, the path information is erased, and the quanton passes through both slits, consequently giving interference. \citet{Kwiat1992} were the first to experimentally implement the quantum eraser, although similar experiments, without this context, had been carried out earlier \cite{Zajonc1991}. As expected, if one chooses to erase the which-way information partially, one would recover an unsharp interference \cite{Englert2000a}.

\begin{figure}
\centerline{\resizebox{8.5cm}{!}{\includegraphics{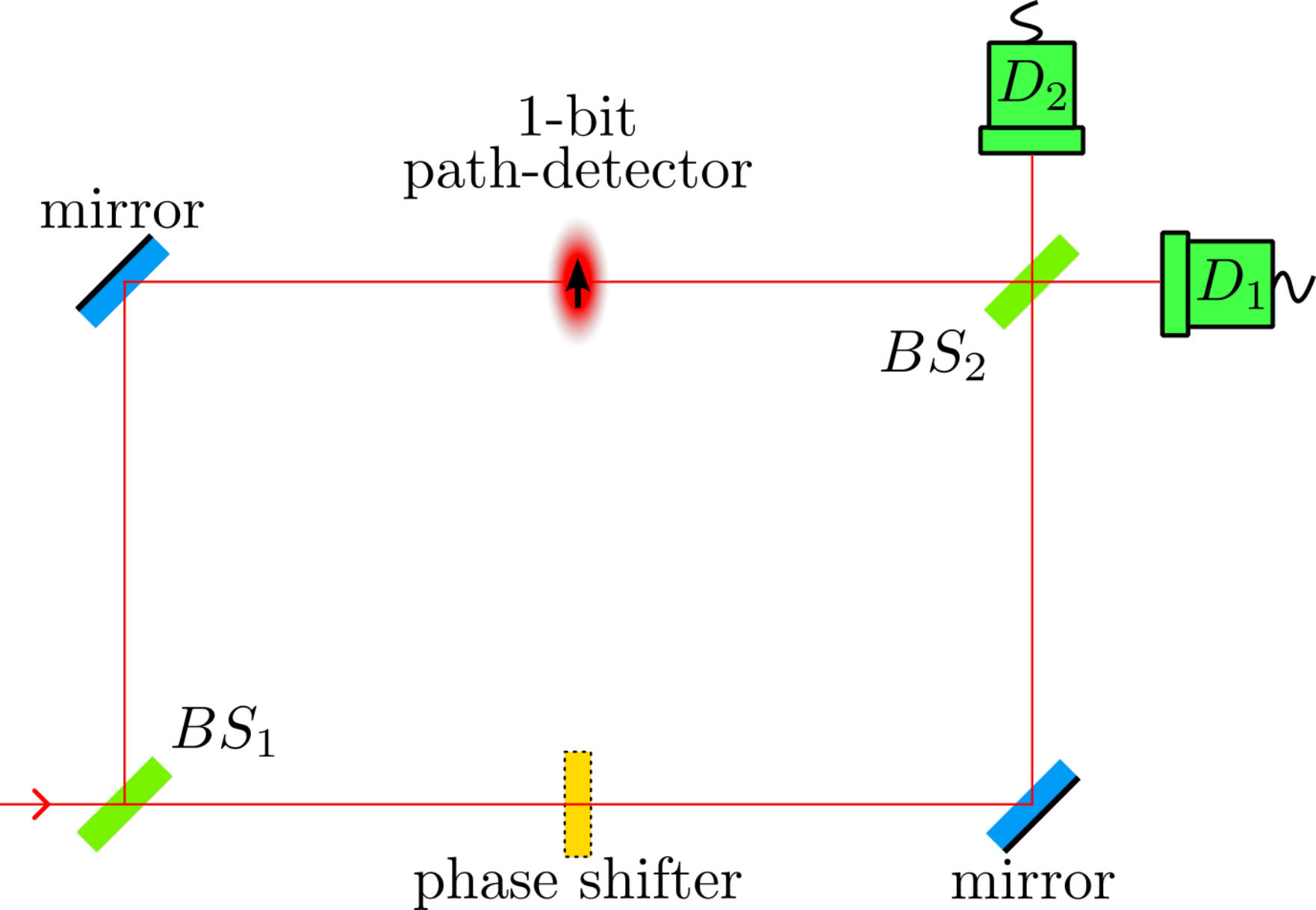}}}
\caption{Schematic diagram of a Mach--Zehnder interferometer in the presence of a which-way (or path) detector.}
\label{delayedmz}
\end{figure}

A more surprising implication of the quantum eraser is that quantum erasing only requires correlating the position of the quanton on the screen to the measured state of the which-way detector, irrespective of whether the which-way detector is measured first or the quanton lands on the screen and the which-way detector is measured later. This latter case is called the delayed-choice quantum eraser and has been a subject of lively debates over several decades~\cite{Walborn2003}.
When people interpret the delayed-choice quantum eraser as allowing the experimenter to either erase the path information or keep it, even after the quanton has been recorded on the screen, some may perceive this as having an impact on the particle's past. The question, how one can retain the choice to make the quanton pass through both the slits or just one slit, after it has already traversed the double-slit and landed on the screen, has been a subject of debate for a long time. 

The prevalent view on the interpretation of the delayed-choice quantum eraser has been due to \citet{Englert1994,Englert1999}. Recently it has been shown that this view is untenable, and that in the delayed mode the which-way information is always erased \cite{Qureshi2020,Qureshi2021a,Qureshi2025}. This can be seen most easily if one uses a MZI to set up a quantum eraser, as seen in Fig.~\ref{delayedmz}. Without the which-way detector, the state of a quanton inside the interferometer is $\tfrac{1}{\sqrt{2}}(|\psi_1\rangle+|\psi_2\rangle)$, where $|\psi_1\rangle,|\psi_2\rangle$ represent the state of the particle if it follows the upper or lower path, respectively. A quanton in this state always arrives at the detector $D_1$, and never reaches $D_2$. However, if the phase difference between the two paths is tuned, using the phase shifter, in such a way that the state is $\tfrac{1}{\sqrt{2}}(|\psi_1\rangle - |\psi_2\rangle)$, the quanton always arrived at $D_2$, and never reaches $D_1$. The detector receiving all the quantons represents bright fringe of the conventional two-slit interference, and the one receiving no quanton represents a dark fringe. Now in the presence of the which-way detector, the state of the quanton and the which-way detector can be written as 
\begin{eqnarray}
    |\psi\rangle &=& \tfrac{1}{\sqrt{2}}(|\psi_1\rangle|d_1\rangle + |\psi_2\rangle|d_2\rangle) \nonumber\\
    &=& \tfrac{1}{2}(|\psi_1\rangle+|\psi_2\rangle)|d_+\rangle + \tfrac{1}{2}(|\psi_1\rangle-|\psi_2\rangle)|d_-\rangle,
    \label{mzstate}
\end{eqnarray}
where the path-detector states are the same as introduced in the preceding discussion. The state from~Eq.~(\ref{mzstate}) implies that finding the which-way detector in the state $|d_+\rangle$ tells one that the quanton passed through both paths, and its state was $\tfrac{1}{\sqrt{2}}(|\psi_1\rangle+|\psi_2\rangle)$. Finding the which-way detector in the state $|d_-\rangle$ also tells one that the quanton passed through both paths, but its state was $\tfrac{1}{\sqrt{2}}(|\psi_1\rangle-|\psi_2\rangle)$. Now, we know that the state $\tfrac{1}{\sqrt{2}}(|\psi_1\rangle+|\psi_2\rangle)$ always reaches detector $D_1$, and $\tfrac{1}{\sqrt{2}}(|\psi_1\rangle-|\psi_2\rangle)$ always reaches $D_2$, which means that every quanton registered at the final detectors tells one whether the path-detector state was $|d_+\rangle$ or $|d_-\rangle$. This consequently indicates that, in the delayed mode, each quanton detected at the final detectors unequivocally reveals that it traversed both paths, as well as the phase difference that existed between the two paths. It is obvious that once the path-detector state is revealed to be (say) $|d_+\rangle$, trying to measure an observable of the path-detector, whose eigenstates are $|d_1\rangle,|d_2\rangle$, yields no information about the path of the quanton. The bottom-line is that in the delayed mode, the which-way information is always erased, and the quanton always follows both paths, like a wave. 

The situation is a bit more complex in the two-slit delayed choice quantum eraser, but the resolution is the same as in the case of the MZI. The key is to recognize that $|d_{\pm}\rangle$ are not the only states that can be used to erase the path-information and recover interference. One can use any of the infinite number of mutually unbiased bases $|d^{\theta}_{\pm}\rangle = \tfrac{1}{\sqrt{2}}(|e^{i\theta}d_1\rangle \pm |d_2\rangle)$, and each measured $|d^{\theta}_{+}\rangle$ state will lead to a recovered interference which will not be centered at $x=0$, but at another position which depends on the phase $\theta$. Conversely, for a quanton landing at any random position $x$ on the screen, there always exists a phase $\theta_x$ such that the recovered interference corresponding to the path detector state $|d^{\theta_x}_{+}\rangle$ has a \emph{maximum} at that position. The probability of $|d^{\theta_x}_{-}\rangle$ at that position, by definition, is zero. Hence, when a quanton lands at any random position $x$, the which-way detector changes to a \emph{definite} state $|d^{\theta_x}_{+}\rangle$, which corresponds to the particle following both paths with a phase difference $\theta_x=\tfrac{2\pi xd}{\lambda D}$, where $\lambda$ is the wavelength of the quanton, and $d, D$ are as in Fig.~\ref{twoslit} \cite{Qureshi2025}. So in the two-slit delayed-choice quantum eraser too, in the delayed mode, the which-way information is \emph{always} erased. There is no choice left for the experimenter. \citet{YangW2025} have performed this experiment using angular orbital momentum as the path marker.

%---------------------------
%\subsubsection{Partial quantum eraser}
%\label{sec:qeraser}

\begin{figure}
    \centering
    \includegraphics[width=1\linewidth]{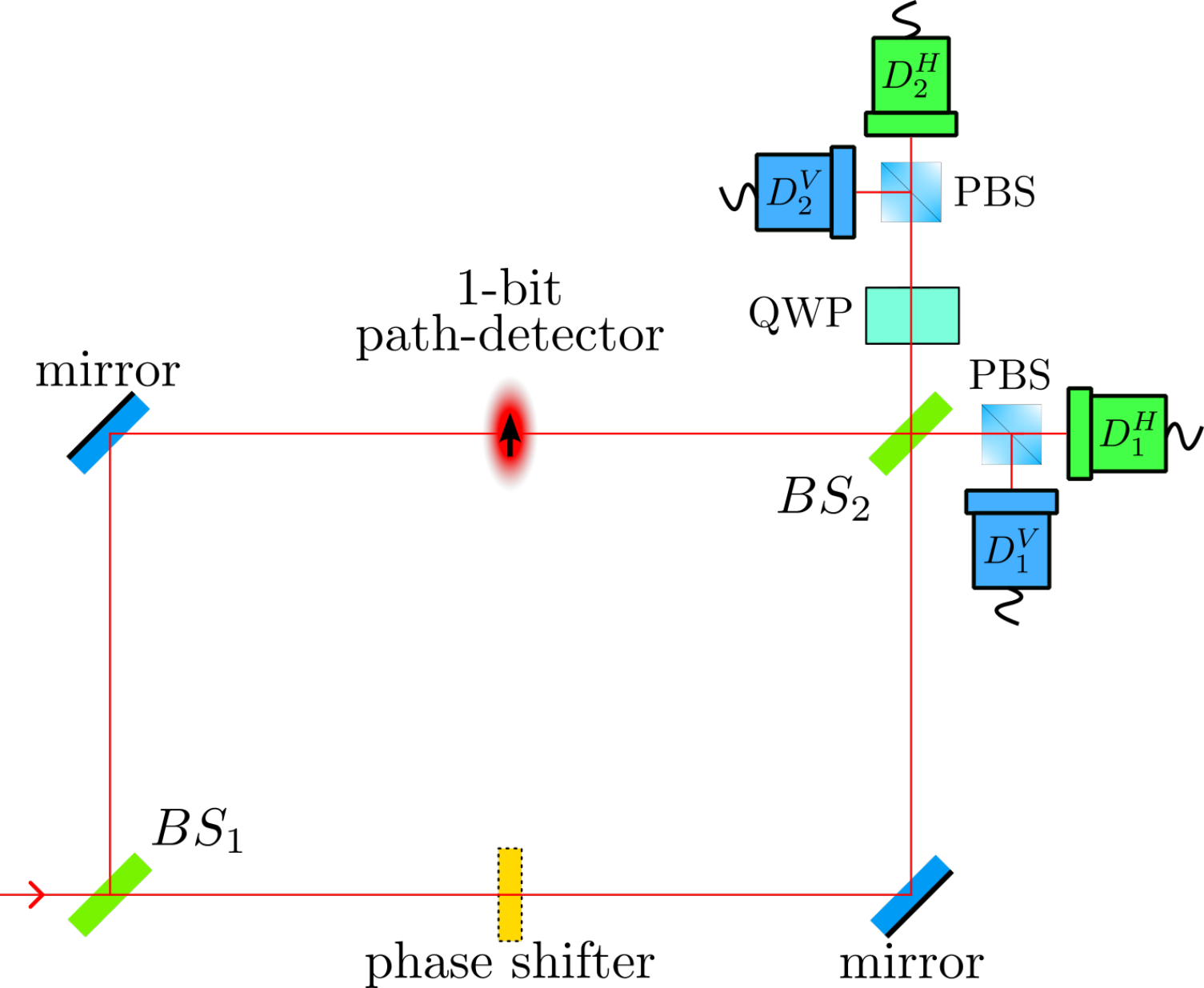}
    \caption{Partial quantum eraser is a configuration that modifies the standart verion of the quantum eraser presented in Fig.~\ref{delayedmz}. QWP stands for quarter-wave plate and PBS stands for polirazing beam splitter.}
    \label{fig:PQE}
\end{figure}

\vspace{0.25cm}

\textit{The partial quantum eraser:} This experiment is a variation of the quantum eraser discussed above, which was reported in Refs.~\cite{PessoaJr2000,PessoaJr2013}. In this setup, each output port of the second beam splitter ($BS_2$) is followed by a quarter-wave plate (QWP) and a polarizing beam splitter (PBS). The QWP converts linear polarization into circular polarization, effectively changing the polarization basis according to the transformations $\ket{d_1} \to \ket{d_+}$ and $\ket{d_2} \to \ket{d_-}$. Subsequently, the PBS separates the polarization components by transmitting, for instance, the horizontal polarization, denoted by $\ket{d_1}$, and reflecting the vertical polarization, denoted by $\ket{d_2}$.

If no QWP is inserted at the outputs of the MZI, detections at the corresponding detectors provide direct which-path information about the quanton. By starting with the horizontal polarization, lets suppose that the $1$-bit path detector changes the horizontal to vertical polarization, effectively marking each path. In particular, at the output, a click at $D_j^V$ or $D_j^H$ indicates that the quanton has taken the upper or lower path within the MZI, respectively, with $j=1,2$. Conversely, when a QWP is placed in both output arms, the which-path information is effectively erased, as in the standard quantum eraser scenario. The situation becomes especially interesting when the QWP is inserted in only one of the outputs, as illustrated in Fig.~\ref{fig:PQE}.
In this case, we have the following quantum state after the PBSs
\begin{align}
\begin{aligned}
\ket{\psi}_{\text{PBS}} &=  \frac{1}{2}(e^{i\phi}\ket{D_1^{H}}\ket{d_1}+\ket{D_1^{V}}\ket{d_2}),
\\
& \hspace{-9.8pt} +\frac{1}{2\sqrt{2}}\left[(1-e^{i\phi})\ket{D_2^{H}}\ket{d_1} - (1+e^{i\phi})\ket{D_2^{V}}\ket{d_2}\right]
\end{aligned}
\end{align}
The detection probabilities at the detectors without the QWP are given by $\Pr(D_1^H) = \Pr(D_1^V) = 1/4$, which are no sensitive to phase variation, while the probabilities at the detectors where the quanton first passes through the QWP are given by
\begin{equation}
\Pr(D_2^H) = \frac{1}{4}(1-\cos\phi),\ \Pr(D_2^V) = \frac{1}{4}(1+\cos\phi),
\end{equation}
explicitly dependent of phase variation.
Within this experimental configuration, the behavior exhibited by the quantons depends on which detector is triggered. In contrast to the standard quantum eraser scenario, wave-like behavior can be reintroduced through the phase dependence of the detection probabilities. Notably, when the QWP is inserted in only one of the outputs of $BS_2$, the behavior is not predetermined prior to detection. If one of the detectors $D_1^k$ clicks, $k=H,V$ provides which-path information. Conversely, if one of the detectors $D_2^k$ clicks, the which-path information is erased, and wave-like behavior is manifested, as indicated by the phase sensitivity of the corresponding probabilities.

\citet{Starke2024} analyzed this experiment with the framework of Sec.~\ref{sec:QCP}. Considering that the system is initially prepared in a pure state entering only one input mode of the first beam splitter and with horizontal polarization $\ket{d_1}$, after the $BS_1$ and before the quanton has interacted with $1$-bit path-detector, the state is a quantum superposition of paths with a separable state in the polarization degree of freedom
\begin{eqnarray}
\ket{\psi_{BS_1}} = \frac{1}{\sqrt{2}} \left( \ket{\psi_{\text{lower}}} + i\ket{\psi_{\text{upper}}} \right) \ket{d_1}.
\end{eqnarray}
The reduced density matrix describing the path degree of freedom after $BS_1$ is written as $\rho_{BS_1,a} = \Tr_A\left(\ketbra{\psi_{BS_1}}{\psi_{BS_1}}\right)$, where the lowercase index $a$ labels the path and the uppercase index $A$ labels the polarization. Then, the complete complementarity relation from Eq.~\eqref{Eq:CCRg} reads $\mathcal{C}(\rho_{BS_1,\ a}) = 1$ and $\mathcal{P}(\rho_{BS_1,\ a}) = \mathcal{E}(\rho_{BS_1,\ a}) = 0$, characterizing maximal wave-like behavior.
After the interaction of the quanton with the $1$-bit path-detector phase shifter and mirrors, the quanton state is changed to
\begin{eqnarray}
\ket{\psi_{d}} = \frac{1}{\sqrt{2}} \left( ie^{i\phi}\ket{\psi_{\text{lower}}}\ket{d_1} -\ket{\psi_{\text{upper}}}\ket{d_2}\right). 
\end{eqnarray}
So, now the complete complementarity relation from Eq.~\eqref{Eq:CCRg} reads $\mathcal{C}(\rho_{d,\ a}) = \mathcal{P}(\rho_{d,\ a}) = 0$ and $\mathcal{E}(\rho_{d,\ a}) = 1$, which indicates maximal entanglement, characterizing maximal particle-like behavior through quantum correlation, where $\rho_{d, a} = \Tr_A \left(\ketbra{\psi_{d}} \right)$.
This highlights one of the features brought about by the proposed update to the principle of complementarity explored by \citet{Starke2024} in Sec.~\ref{sec:QCP}. The wave-particle duality needs to be updated as the quantum states evolve.
Although we are free to evaluate these complementarity relations for the states after $BS_2$ or after any later optical elements, the crucial point is that the optical components placed after $BS_2$ cannot alter the dual behavior occurring inside the MZI.
These operations merely apply transformations that modify the quantum state, thereby preventing the experimenter from accessing the wave- and particle-like behavior.

The partial quantum eraser highlights an important conceptual shift in the understanding of complementarity. While Bohr’s original formulation associates complementary properties with mutually exclusive experimental arrangements, partial erasure demonstrates that phase sensitivity-based wave and particle-like features can depend on the detector used. On the other hand, this experiment is most naturally understood in terms of the quantum state of the system, where coherence and correlations determine the observed interference. Such experiments, therefore, support a state-dependent perspective on complementarity, in which the trade-offs between different physical properties are encoded in the structure of the state rather than solely in the global experimental configuration.

%---------------------------
\subsection{Afshar paradox}
\label{sec:afshar}

\begin{figure}
    \centering
    \includegraphics[width=1\linewidth]{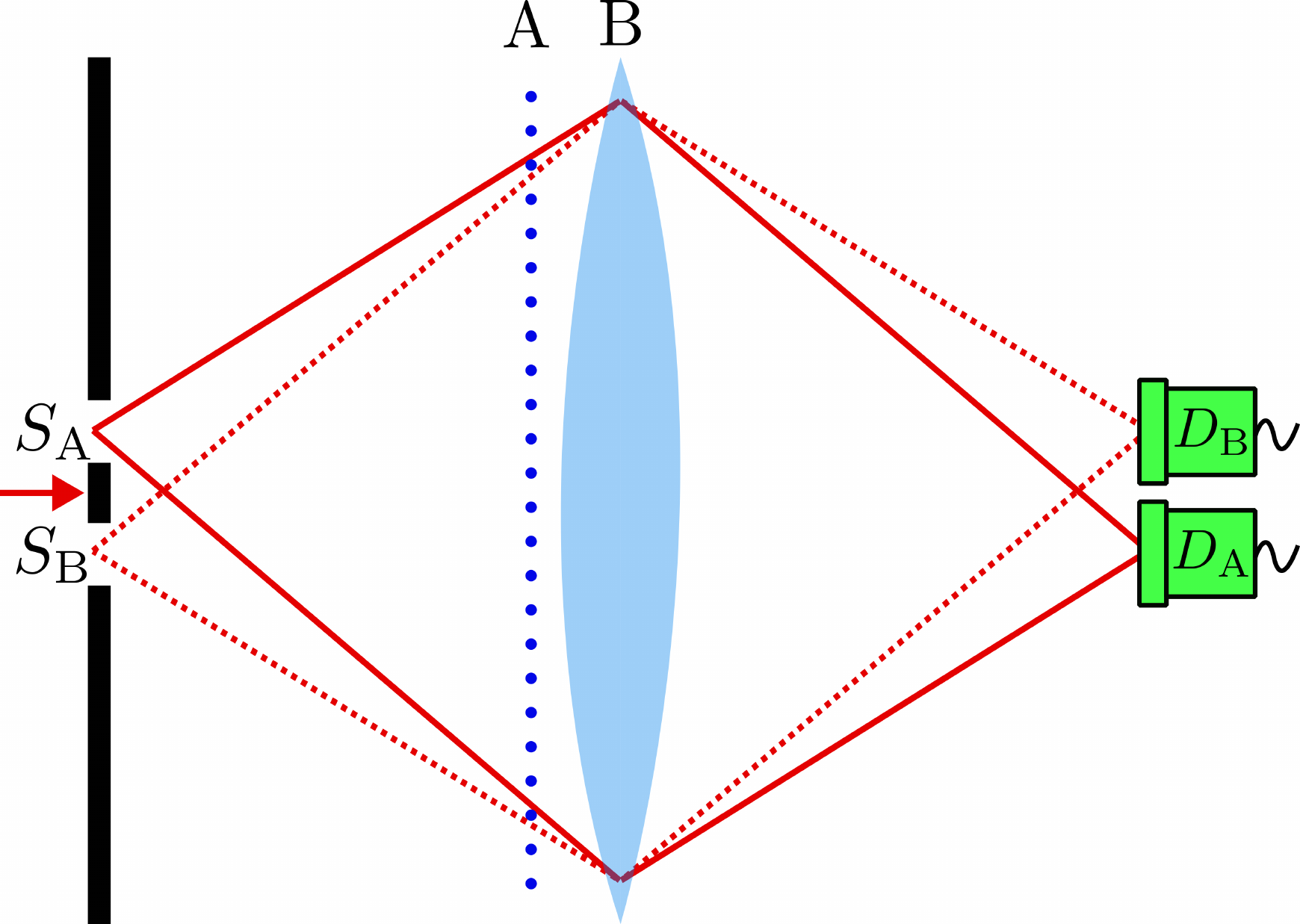}
    \caption{Schematic representation of the Afshar experiment. The experiment consists in a modification of the double-slit experiment with the aim of quantifying the wave-particle duality through a non-perturbative measurement process of the wave behavior. The photon hits the double slit from left to right and reaches the slits $S_\text{A}$ and $S_\text{B}$. In region $\text{A}$ a grid of extremely thin wires is positioned at the minima of the interference pattern in order to quantify the wave behavior non-destructively. In region $\text{B}$ a lens is positioned that redirects the photons to two very specific regions, where the detectors $D_\text{A}$ and $D_\text{B}$ are positioned.}
    \label{fig:afshar}
\end{figure}

\citet{Afshar2005,Afshar2006} proposed a modified version of Young’s double-slit experiment that was claimed to challenge Bohr’s complementarity principle. The central argument is that full wave-like behavior can be inferred indirectly, while still allowing one to determine, after detection, which slit the photon passed through, thus suggesting a violation of complementarity.

The experiment, as depicted in Fig.~\ref{fig:afshar}, is implemented as follows. A quanton passes through two slits, $S_A$ and $S_B$, and a lens placed in region $B$ redirects the emerging beams toward two well-defined detection spots, where detectors $D_A$ and $D_B$ are positioned. The procedure can be divided into three main steps. First, the lens is adjusted in order to identify the precise locations where the detectors should be placed. Second, a correspondence between slits and detectors is established by closing one slit at a time and verifying that only one detector clicks: detector $D_A$ fires when slit $S_B$ is closed, and detector $D_B$ fires when slit $S_A$ is closed. According to Afshar, this path-detector correspondence remains valid even when both slits are open.

Third, with both slits open, a very thin wire grid is inserted at the positions corresponding to the minima of the interference pattern. The purpose of this grid is to probe wave-like behavior in a minimally disturbing way since ideally no particles should pass through the interference minima. To support this claim, a series of experiments were performed to verify that the presence of the grid does not significantly reduce the detection counts. Additional runs confirm that when one slit is closed, detections remain correlated with the corresponding detector, reinforcing the slit-detector association. With these elements in place, the experiment is interpreted as follows: the detection outcome determines which slit the quanton traversed via the established slit-detector correspondence, thus indicating particle-like behavior, while the negligible disturbance introduced by the wire grid is taken as evidence of a well-defined interference pattern, i.e., wave-like behavior. On this basis, Afshar argues that both aspects are simultaneously observed within a single experimental arrangement, suggesting a violation of Bohr’s complementarity principle.

\begin{figure}
    \centering
    \includegraphics[width=1\linewidth]{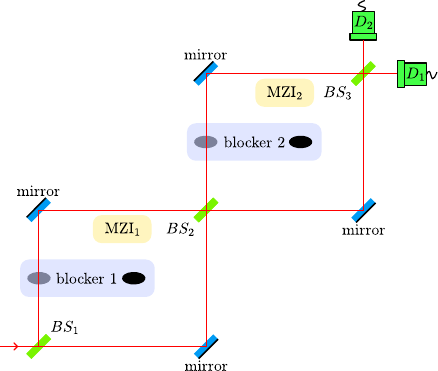}
    \caption{Interferometric setup proposed by Unruh as an analogue of Afshar’s experiment.}
    \label{fig:unruh}
\end{figure}

One of the main objections to this experiment concerns the assumed correspondence between the slits and the detectors. When both slits are open, there is no guarantee that an individual detection event can be unambiguously associated with a specific slit~\cite{Steuernagel2007}. It was also argued that the which-way information is completely destroyed once interference is established \cite{Qureshi2012}. A more transparent criticism was presented by \citet{Unruh2004}, who analyzed an earlier version of Afshar's proposal. In that work, Unruh introduced a simplified two-level system analogous to the Afshar setup, as illustrated in Fig.~\ref{fig:unruh}, in order to examine the validity of the underlying assumptions. The central point of this critique is that the slit-detector relationship is established without simultaneously including the wire grid at the interference minima. Since the presence of the grid can modify the detection outcomes, determining the path-detector correspondence independently may lead to misleading conclusions. This issue motivated further experimental refinements aimed at addressing these concerns~\cite{Afshar2006,Afshar2007,Jacques2008b}.
For a more extensive discussion, we refer the reader to Refs.~\cite{Avner2021,Drezet2005,Drezet2010,Erol2022,Ferrari2010,Flores2007,Flores2009,Flores2010,Georgiev2007a,Georgiev2007b, Georgiev2012,Gergely2022,Jacques2008b,Knight2020,Kastner2005,Kastner2009,Lai2009,LiuZ2024,Navia2017,PessoaJr2013,Qureshi2010,Qureshi2012,Reitzner2007,Starke2024,Steuernagel2007,Tang2013,Unruh2004,Unruh2007,Wechsler2019}.

Recently a modified form of Afshar's two-slit experiment was theoretically analyzed by introducing polarization of the photon as a path marker \cite{Qureshi2023}.
The photons at the two detectors may be sorted according to their polarization. The analysis showed that the photons whose polarization indicates that they passed through one of the two slits do not show interference. For all the photons that \emph{do} show interference, the polarization indicates that they passed through both slits, although they randomly land at one of the two detectors \cite{Qureshi2023}. This contradicts Afshar's inference that the photons showing interference and landing at a particular detector pass through only one of the two slits.

%------------------
%\subsubsection{Unruh's experiment}
%------------------

\vspace{0.25cm}

\textit{Unruh's experiment:} We now turn to the interferometric arrangement introduced by Unruh and analyze how both particle- and wave-like behaviors can be inferred within this framework. The discrete version of the Afshar experiment proposed by \citet{Unruh2004}, shown in Fig.~\ref{fig:unruh}, is implemented using two Mach--Zehnder interferometers in series. The first interferometer (MZI$_1$) represents the double-slit configuration, where the upper and lower paths correspond to slits $S_A$ and $S_B$, respectively. The second interferometer (MZI$_2$) plays the role of region $A$ in Afshar's setup, where the wire grid is placed, while the beam splitter $BS_3$ mimics the function of the lens in Fig.~\ref{fig:afshar}. When no blockers are inserted, both detectors register clicks, and the upper arm of MZI$_2$ corresponds to a dark fringe, meaning that no quantons propagate along that path. The introduction of blocker 1 allows one to establish a path-detector correspondence in MZI$_1$, analogous to the procedure in Afshar's experiment. By performing a series of runs, one verifies that blocking the lower path leads to detections at $D_1$, associating the upper path with this detector, while blocking the upper path leads to detections at $D_2$, associating the lower path with that detector. When both paths in MZI$_1$ are open, Afshar's reasoning would suggest that inserting blocker 2 in the upper arm of MZI$_2$ corresponds to a situation of maximal wave-like behavior, since that path remains unoccupied. At the same time, the detection outcomes are taken to retroactively indicate which path the quanton followed in MZI$_1$.

Unruh's argument can be summarized as follows. Suppose that blocker 2 is kept in the upper arm of MZI$_2$ while the path-detector correlation test is repeated. In this situation, one finds that the previously established path-detector correspondence is no longer preserved. Based on this observation, Unruh concludes that Afshar's experiment does not violate the principle of complementarity. As discussed earlier, Afshar responded by repeating the experiment while taking this issue into account. His results indicate a reduction in photon counts, as well as the appearance of counts in the opposite detector. However, the path-detector correspondence is not completely destroyed, in contrast to what is predicted in Unruh's idealized scenario. Afshar therefore argues that the original conclusions remain essentially unaffected. In turn, Unruh proposed a refinement of this argument by suggesting that, instead of a complete blocker 2, one could introduce an attenuator in the corresponding arm. This would partially degrade the path-detector correspondence while enhancing the wave-like character, thereby restoring consistency with the complementarity principle.

It becomes evident that two distinct assumptions underlie the interpretation of these experiments. The first is the acceptance that Bohr’s complementarity principle can be quantified across the entire experimental setup, even when wave-like and particle-like behaviors are evaluated in different regions. In the Afshar experiment, for instance, particle-like behavior is inferred through retrodiction, by associating a detection event at a given detector with a specific slit. In contrast, wave-like behavior is quantified independently in the region of the wire grid, typically through interferometric visibility. The second assumption is that conclusions obtained in one configuration remain valid in another. More specifically, the inference that each detector click corresponds to a definite path---established when one path is blocked---is assumed to hold even when both paths are open.
However, these correspond to distinct experimental configurations, a point that has already been criticized by~\citet{Unruh2004}.

From the perspective of the complementarity relation given by Eq.~\eqref{eq:QCR}, as argued by~\citet{Starke2024}, this issue becomes transparent. Turning to Unruh's experiment, when both paths are open, the state exhibits maximal path superposition in the MZI$_1$ region while displaying maximal path predictability in MZI$_2$. Conversely, when one path is blocked, the state shows maximal predictability in MZI$_1$ and maximal superposition in MZI$_2$. As a result, the complementarity relation is satisfied in each configuration individually, and no violation arises when the experiments are analyzed consistently within their respective setups. This is another kind of issue that is resolved by applying the quantum state dependent approach to complementarity. Related interferometric variations and delayed-choice extensions of this setup were further explored by~\cite{PessoaJr2013,Brasil2025}, further analyzing the role of phase shifts, retroinference, and wave-particle interpretations in this class of experiments, while reinforcing the conclusion that the complementarity relations given by Eq.~\eqref{eq:QCR} provide a consistent description across a wide range of scenarios, thereby ensuring that Bohr's complementarity principle remains valid.

%
%---------------------
%
\section{Broader aspects of quantum complementarity}
\label{sec:furthertopics}
In this section, we explore the broader aspects of quantum complementarity. We begin by examining the interplay between complementarity relations and contextuality, as well as their role within generalized notions of realism. We then discuss recent developments that cast complementarity within resource-theoretic frameworks, including approaches based on predictability, coherence, and purity. Further connections are established with quantum information protocols, such as entanglement swapping in partially entangled systems, and with thermodynamic settings, where wave-particle duality has been linked to the performance of quantum heat engines. We also analyze the relationship between complementarity and incompatible observables, emphasizing its deep roots in the non-commutative structure of quantum mechanics. Finally, we consider extensions of complementarity beyond non-relativistic quantum theory, highlighting perspectives that emerge in relativistic and quantum field-theoretic settings.

%---------------------
\subsection{Complementarity relations and contextuality}
The phenomenology of quantum interference is often regarded as one of the clearest manifestations of non-classical behavior. However, not every feature of interference is intrinsically non-classical in a strict foundational sense.~\citet{Catani2023a} showed that some interference features can actually be reproduced by noncontextual ontological models~\cite{Budroni2022}, that is, classical hidden-variable models that assign fixed outcomes to measurements regardless of context. This naturally raises the question of which interference phenomena genuinely resist any classical explanation?

In order to tackle this question,~\citet{Catani2023b} introduced the framework to discuss noncontextuality for prepare-measure scenarios and what it means for an interference experiment to be ``classical'' or ``non-classical''. Starting from an operational theory that describes physical systems in terms of laboratory procedures: preparations (ways of preparing a system) and measurements (ways of probing it), one considers a set of preparation procedures $P \in \mathcal{P}$ and a set of measurement procedures $M \in \mathcal{M}$. For each preparation $P$, each measurement $M$, and each outcome $k$ of $M$, the theory assigns a probability $p(k|M,P)$. These probabilities are the only empirically accessible quantities. At this level, the theory does not posit any underlying description of reality, it only encodes input-output statistics. An ontological model attempts to explain these operational statistics in terms of underlying physical states, called ontic states, denoted by $\lambda \in \Lambda$. In such a model, each preparation $P$ is represented by a probability distribution $\mu(\lambda|P)$ over the ontic state space $\Lambda$. Each measurement $M$ is represented by response functions $\xi(k|M,\lambda)$, which give the probability of obtaining outcome $k$ given ontic state $\lambda$. The ontological model reproduces the operational predictions if
\begin{equation}
  p(k|M,P) = \sum_{\lambda \in \Lambda}
\xi(k|M,\lambda)\mu(\lambda|P).  
\end{equation}
Thus, the observed probabilities arise from classical ignorance about the underlying ontic state. Moreover, another central notion is that of operational equivalence. Two preparations $P$ and $P'$ are operationally equivalent if they yield identical statistics for all measurements: $
p(k|M,P) = p(k|M,P') \quad \forall M \in \mathcal{M},  \forall k.$ The principle of generalized non-contextuality states that if two preparations are operationally equivalent, then they must be represented identically at the ontological level: $\mu(\lambda|P) = \mu(\lambda|P') \quad \forall \lambda \in \Lambda.$ The core idea is that an ontological model should not introduce distinctions that have no operational counterpart. If it does, the model is contextual.

Using a MZI as the operational scenario and adopting the definitions of predictability $\mathcal{P}$ and visibility $\mathcal{V}$ introduced by~\citet{Greenberger1988}, the authors \citet{Catani2023b} analyzed the constraints imposed by generalized non-contextuality on wave-particle duality and proved that an operational theory admits a non-contextual ontological model if, and only if, the predictability $\mathcal{P}$ and visibility $\mathcal{V}$ satisfy the linear bound $\mathcal{V} + \mathcal{P} \le 1$. This inequality therefore characterizes the full set of duality relations compatible with generalized noncontextuality. Quantum mechanics, however, predicts the strictly stronger quadratic relation $\mathcal{V}^2 + \mathcal{P}^2 \le 1$, which allows combinations of $\mathcal{V}$ and $\mathcal{P}$ that violate the linear bound. Consequently, quantum theory violates $\mathcal{V} + \mathcal{P} \le 1$, demonstrating that there exist quantum states realized in a MZI that are incompatible with any noncontextual ontological model and thus constitute a witness of contextuality. It is important to emphasize, however, that the appearance of linear complementarity relations---such as those introduced in Sec.~\ref{sec:CRpositivity}---does not by itself imply compatibility with a noncontextual model. The fact that a complementarity relation can be written in linear form does not guarantee that it corresponds to the noncontextual bound. In many cases, such linear relations can be written in quadratic form and properly normalized, such as those introduced in Sec.~\ref{sec:CRpositivity}.

%---------------------------
\subsection{Complementarity in the realism framework}
\label{sec:realism}

Drawing inspiration from the concept of elements of reality introduced by Einstein, Podolsky, and Rosen~\cite{EPR1935}, which states that ``\textit{if, without disturbing the system, it is possible to predict with certainty (i.e., with probability equal to 1) the value of a physical observable, then there exists an element of reality corresponding to this physical quantity}'',~\citet{Bilobran15} and \citet{Dieguez18} developed an operational framework to characterize (ir)realism by considering a protocol that prepares a bipartite quantum system in a given quantum state $\rho_{AB}$.

The task is to determine, via state tomography, the most complete description of the prepared state. This task can be accomplished by repeating the preparation procedure as many times as necessary to obtain ideal tomography. At the end of the protocol, one knows that every time the procedure is repeated, the system is described by the density operator $\rho_{AB}$. Now consider the same situation, but suppose that between preparation and quantum state tomography a measurement of the observable
$\mathcal{O} = \sum_j o_j \Pi^{\mathcal{O}}_j$, with projectors $\Pi^{\mathcal{O}}_j = \ketbra{o_j}{o_j}$ acting on subsystem $A$, is secretly performed each time the procedure is repeated. After a measurement of the observable $\mathcal{O}$, the state of the system becomes
\begin{align}
\frac{1}{p_{o_j}}(\Pi^{\mathcal{O}}_j \otimes I_B)\rho_{AB}(\Pi^{\mathcal{O}}_j \otimes I_B)
= \Pi^{\mathcal{O}}_j \otimes \rho_{B|o_j}
\end{align}
with probability $p_{o_j} = \Tr[(\Pi^{\mathcal{O}}_j \otimes I_B)\rho_{AB}]$, where
\begin{align}
\rho_{B|o_j} = \frac{1}{p_{o_j}} \Tr_A \Big[(\Pi^{\mathcal{O}}_j \otimes I_B)\rho_{AB}(\Pi^{\mathcal{O}}_j \otimes I_B)\Big]
\end{align}
is the conditional state of subsystem $B$ given the outcome $o_j$. Since the measurement of the observable is performed secretly (either by another agent or simply by the environment), after repeating the same procedure many times and performing state tomography, the best description of the system is given by the weighted average over all possible outcomes of the observable $\mathcal{O}$, namely
\begin{align}
\Phi_{\mathcal{O}}(\rho_{AB}) &
= \sum_j p_{o_j} \Pi^{\mathcal{O}}_j \otimes \rho_{B|o_j} = \sum_j (\Pi^{\mathcal{O}}_j \otimes I_B)\rho_{AB}(\Pi^{\mathcal{O}}_j \otimes I_B).
\end{align}
Thus, the description of the system given above is epistemic with respect to the observable $\mathcal{O}$, since the probabilities $p_{o_j}$ reflect our ignorance about the actual value of $\mathcal{O}$ after each measurement.

The final step of the protocol consists in comparing the situations with and without a secret measurement of the observable $\mathcal{O}$. When $\Phi_{\mathcal{O}}(\rho_{AB}) = \rho_{AB}$, the element of reality associated with the observable $\mathcal{O}$ was already implicit in the preparation of $\rho_{AB}$. In this case, the unrevealed measurements did not create an element of reality (i.e., did not render the observable definite), but merely revealed that the observable was already definite beforehand. Hence, an observable $\mathcal{O}$ possesses an element of reality with respect to the preparation $\rho_{AB}$ if, and only if, $\Phi_{\mathcal{O}}(\rho_{AB}) = \rho_{AB}$. One can see that this criterion reproduces the Einstein--Podolsky--Rosen notion of reality when the preparation is an eigenstate of the observable $\mathcal{O}$ and also assigns an element of reality to mixtures of eigenstates. Moreover, the criterion automatically implements the idea that a measurement preserves a preexisting element of reality associated with $\mathcal{O}$, since $\Phi_{\mathcal{O}}\big(\Phi_{\mathcal{O}}(\rho_{AB})\big) = \Phi_{\mathcal{O}}(\rho_{AB}).$ The definition also suggests a measure of how far a given state $\rho_{AB}$ is from a state in which $\mathcal{O}$ is real (or definite), i.e., 
\begin{align}
\mathcal{I}_\mathcal{O}(\rho_{AB})
:= S_{vn}(\Phi_{\mathcal{O}}(\rho_{AB})) - S_{vn}(\rho_{AB}),
\end{align}
defines a measure of irrealism (or indefiniteness) of the observable $\mathcal{O}$ given the preparation $\rho_{AB}$. The realism (or definiteness) of the observable $\mathcal{O}$ given the preparation $\rho_{AB}$ is then defined as
\begin{align}
\mathcal{R}_\mathcal{O}(\rho_{AB})
= \log d - \mathcal{I}_\mathcal{O}(\rho_{AB}),
\end{align}
where $d$ is the dimension of the bipartite quantum system.

Within this approach, it is possible to define the elements of reality associated with wave and particle behavior of the subsystem $A$ inside two-path quantum-controlled interferometric setups~\cite{Dieguez2022}, with $B$ playing the role of the control subsystem. The wave and particle observables can be defined as $\mathfrak{W} = \sigma_\perp$ and $\mathfrak{P} = \sigma_z$ with eigenstates $\{\frac{1}{\sqrt{2}}\big(e^{i\theta}|0\rangle \pm |1\rangle\big)\}$ and $\{|0\rangle, |1\rangle\}$, respectively, where $|0\rangle$ and $|1\rangle$ labels each arm of the interferometer. The elements of reality of $\mathfrak{W}$ and $\mathfrak{P}$ inside the interferometer are simultaneously prevented by the following complementarity relation
\begin{align}
\mathcal{R}_{\mathfrak{W}}(\rho_{AB}) + \mathcal{R}_{\mathfrak{P}}(\rho_{AB})
\le 2 \log d_A - I(\rho_A) - I_{A:B}(\rho_{AB}),
\end{align}
where $I(\rho_A) = \log d_A - S_{vn}(\rho_A)$ is a purity measure for the subsystem $A$ and $I_{A:B}(\rho_{AB})$ is the mutual information between the subsystem $A$ and $B$. It tell us that the purity of the subsystem $A$ and the correlations it shares with subsystem $B$ prevents the observables $\mathfrak{W}$ and $\mathfrak{P}$ from possessing simultaneous elements of reality.

A connection between local elements of reality and complete complementarity relations can also be established~\cite{Basso2021e} by noting that the local realism of an arbitrary observable $\mathcal{O}$, given the local state $\rho_A$, can be expressed as
\begin{align}
     \mathfrak{R}_{\mathcal{O}}(\rho_A) = \mathcal{P}_{vn}(\rho_A) + S_{vn}(\rho_A),
\end{align}
which captures the idea that the local realism of an observable $\mathcal{O}$ is tied to its predictability prior to a projective measurement---namely, its ``pre-existing'' reality---as well as to the possible generation of entanglement with an informer, that is, a degree of freedom capable of recording information about the state of the system. While, the local irrealism of an observable $\mathcal{O}$ given the local state $\rho_A$ can be expressed as 
\begin{equation}
\mathfrak{I}_{\mathcal{O}}(\rho_A) = \mathcal{C}_{re}(\rho_A),
\end{equation}
which means that the local irrealism is directly related to the quantum coherence of $\rho_A$ in the eigenbasis of $\mathcal{O}$.

%---------------------------
\subsection{Wave-particle duality in optics}

For optical fields, correlations between fluctuating electric field components evaluated at distinct space-time points are described as coherence, whereas correlations between orthogonal field components evaluated at the same space-time point are identified as polarization. From this perspective, it is natural to ask how these two notions---coherence and polarization---are quantitatively related. A significant advance in addressing this question was achieved with the introduction of the polarization coherence theorem by~\citet{Eberly2017}, later experimentally verified by~\citet{Kanseri2019} and further explored in subsequent optical implementations~\cite{Sanchez2019}. This theorem establishes an interesting relation between the degree of polarization $\mathfrak{P}$, the interferometric visibility $\mathcal{V}$, which is a measure of coherence, and the degree of path distinguishability $\mathcal{D}$.

By considering an optical field emerging from a two-path interferometric arrangement, such as a Young-type setup, the optical field can be written as
\begin{align}
    F(\mathbf{r}_\perp,z) = u_a(\mathbf{r}_\perp,z)\varphi_a(q) + u_b(\mathbf{r}_\perp,z)\varphi_b(q),
\end{align}
where $u_a$ and $u_b$ are orthogonal spatial modes associated with the two paths, and $\varphi_a(q)$ and $\varphi_b(q)$ represent complex amplitudes depending on additional degrees of freedom $q$ (such as polarization or time). The detected intensity after superposition is
\begin{align}
    I = I_a + I_b + 2\mathrm{Re}\left(\langle \varphi_a^* \varphi_b \rangle\right),
\end{align}
where $I_a = \langle |\varphi_a|^2 \rangle$ and $I_b = \langle |\varphi_b|^2 \rangle$, and angular brackets denote ensemble averaging. The interferometric visibility is defined as usually, i.e.,
\begin{align}
    \mathcal{V} = \frac{I_{\max}-I_{\min}}{I_{\max}+I_{\min}} = \frac{2|\langle \varphi_a^* \varphi_b \rangle|}{I_a+I_b},
\end{align}
while the degree of distinguishability is 
\begin{align}
\mathcal{D} = \frac{|I_a - I_b|}{I_a + I_b},    
\end{align}
being closely related to the measure of predictability, as introduced by~\citet{Greenberger1988}.
From the Cauchy-Schwarz inequality, one obtains the well-known duality bound $\mathcal{V}^2 + \mathcal{D}^2 \le 1$.

To connect these quantities with polarization, one introduces the $2\times2$ coherence (or polarization) matrix
\begin{equation}
\mathcal{W} = 
\begin{pmatrix}
\langle \varphi_a^* \varphi_a \rangle & \langle \varphi_b^* \varphi_a \rangle \\
\langle \varphi_a^* \varphi_b \rangle & \langle \varphi_b^* \varphi_b \rangle
\end{pmatrix},
\end{equation}
with the degree of polarization associated with this field being defined as
\begin{align}
    \mathfrak{P} = \sqrt{1 - \frac{4\det \mathcal{W}}{(\operatorname{Tr}\mathcal{W})^2}}.
\end{align}
Evaluating the determinant and trace explicitly, one finds
\begin{align}
\mathfrak{P}^2 = \frac{(I_a - I_b)^2 + 4|\langle \varphi_a^* \varphi_b \rangle|^2}{(I_a + I_b)^2}.    
\end{align}
Recognizing that the first term corresponds to $\mathcal{D}^2$ and the second to $\mathcal{V}^2$, one obtains the polarization coherence theorem, i.e., $\mathfrak{P}^2 = \mathcal{D}^2 + \mathcal{V}^2$. Before the formulation of the polarization coherence theorem,~\citet{Mandel1991} showed that the degree of indistinguishability equals the degree of coherence, while~\citet{Lahiri2011} obtained an inequality between the degree of polarization and the degree of distinguishability, namely, $\mathfrak{P} \ge \mathcal{D}$. This inequality can in fact be obtained directly from the polarization coherence theorem itself.

Further developments have connected the degree of polarization with entanglement~\cite{DeZela2014, DeZela2018, Qian2023}, establishing a direct quantitative relation between concurrence and the degree of polarization, leading to complementarity-type identities between polarization and entanglement. In addition, complementarity relations explicitly incorporating the vectorial nature of the electric field have been derived~\cite{Norrman2017, Norrman2020}, while~\citet{Qian2020b} investigated the wave-particle duality of a photon emitted by genuine point sources, namely a pair of entangled two-level atoms, showing that the complementarity relation is a conditional feature determined by the purity of the atomic state. More recently,~\citet{Pillinen2025} derived a complementarity relation linking which-path information to the geometric phase of the photon, identifying the latter as a fundamental indicator of wave-like behavior. Finally,~\citet{Khatiwada2025} proposed a compact wave-particle duality elliptical relation for a qubit, establishing a quantitative connection between the visibility $\mathcal{V}$, the degree of path distinguishability $\mathcal{D}$, and the degree of coherence $\gamma$, namely, $ \frac{\mathcal{V}^2}{\gamma^2} + \mathcal{D}^2 = 1$. The authors further applied this relation to quantum imaging with undetected photons, a protocol in which spatial information about an object is inferred through measurements performed on entangled photons that never directly interact with the object. In this context, wave-particle duality plays a functional role in the extraction of imaging information, and the duality ellipse explicitly shows how the trade-off between wave- and particle-like behavior---governed by the degree of coherence---constrains and guides the imaging performance.

%---------------------
\subsection{Resource theories of predictability, complementarity and purity}
\label{sec:Presource}

A noteworthy aspect of complete complementarity relations is that quantum coherence and quantum entanglement are features of quantum states that may be regarded as resources enabling specific tasks in quantum information and quantum computation. This perspective has motivated the systematic formulation of quantum resource theories~\cite{Chitambar2019}. Such theories provide a rigorous framework based on the idea that, under constrained classes of operations, certain physical properties of quantum systems acquire operational value as resources for accomplishing particular tasks. After the resource theories of quantum entanglement~\cite{Horodecki2009} and quantum coherence~\cite{Streltsov2017} had been developed and studied for some time, a careful examination of CCRs naturally leads to the question of whether predictability can be regarded as a quantum resource.

Recently, there have been developments aimed at establishing predictability as a quantum resource~\cite{Das2025, Basso2022d}. In particular, it has been shown that the von-Neumann predictability of a state $\rho$ with respect to a discrete non-degenerate observable $X=\sum_j x_j |x_j\rangle \langle x_j|$, denoted by $P_{vn}^X(\rho)$, is equal to the relative entropy of quantum coherence, defined relative to observable $Y = \sum_k y_k Y_k = \sum_k y_k \ketbra{y_k}$ that is mutually unbiased with respect to $X$, of the state $\Phi_X(\rho)$, which is obtained from a nonselective projective measurement of $X$, namely $\mathcal{C}_{re}^Y(\Phi_X(\rho))$, i.e.,
\begin{align}
    \mathcal{P}_{vn}^X(\rho) = \mathcal{C}_{re}^Y(\Phi_X(\rho)),
\end{align}
with $[X,Y] \neq 0$ such that $\abs{\langle{x_j}|{y_k}\rangle}^2 = 1/d \ \forall j,k$. This result captures the idea that if a quantum state is completely predictable in a given basis---hence being an eigenstate of the corresponding observable---then it necessarily exhibits maximal coherence with respect to a mutually unbiased basis. On the other hand, if the observables $X$ and $Y$ are not mutually unbiased, then
\begin{align}
    \mathcal{C}^Y_{re}(\Phi_X(\rho)) +  \mathcal{P}^Y_{vn}(\Phi_X(\rho)) =  \mathcal{P}^X_{vn}(\rho) \le \log d.
\end{align}
This shows that the complementarity relation associated with the observable $Y$, evaluated for the state $\Phi_X(\rho)$, is governed by the predictability of the observable $X$ prior to performing any non-selective measurement on the state $\rho$. A similar relation was obtained by~\citet{LiuZh2025}, where both quantities are treated operationally within the framework of resource theories. In particular, it was shown that, for an ensemble of mutually orthogonal pure states, the sum of ``co-bits'', which quantify the coherence preserved under incoherent free operations, and classical bits, representing the distinguishability extracted through quantum state discrimination, is upper bounded by $\log d$. Moreover, predictability has been shown to be useful for efficiently detecting and characterizing phase transitions in quantum systems, such as Anderson and many-body localization~\cite{Pernambuco2025}. Additionally, in the context of entanglement swapping from partially entangled pure states, predictability has been operationally linked to the entanglement properties of the post-measurement state~\cite{Basso2022c}.

The resource theory of predictability is defined by identifying the free states, the free operations, and resource states together with suitable predictability monotones~\cite{Basso2022d}. The free states are those whose outcome probabilities are uniform in the eigenbasis of a reference observable $X$ and have the general form
\begin{align}
    \rho_{free} = & \frac{1}{d}\Big(\sum_j \ketbra{x_j} + \sum_{j \neq k} \epsilon_{jk} \ketbra{x_j}{x_k}\Big)\\
    & = (1 - p)\frac{I}{d}  + p |\psi_d\rangle \langle \psi_d |, \nonumber
\end{align}
with $\epsilon_{jk} = p e^{i (\phi_j - \phi_k)} \in \mathbb{C}$ such that $|\epsilon_{jk}| \le 1$ , $p \in [0,1]$, and $\ket{\psi_d} = \frac{1}{\sqrt{d}}\sum_j e^{i \phi_j} \ket{x_j}$. These states exhibit vanishing predictability according to any bona fide measure, forming a compact and convex set known as the set of unpredictable states. Free operations are completely positive and trace-preserving maps that do not generate predictability from free states and do not increase it for arbitrary states, such as monitoring maps that interpolate between the identity and a resource-destroying map composed of nonselective measurements of maximally incompatible observables, i.e.,
\begin{align}
    \Lambda_{\epsilon}(\rho) := (1 - \epsilon) \rho + \epsilon \Phi_{XY}(\rho), \label{eq:lamb}
\end{align}
where $\Phi_{XY}$ denotes the composition of nonselective projective measurements of two maximally incompatible observables $X$,$Y$ and $\epsilon \in [0,1]$. On the other hand, if $\Phi_{XY}$ is replaced by $\Phi_X$ in Eq.~\eqref{eq:lamb}, we obtain an example of a free operation that always preserves predictability. Moreover, all states whose probability distributions in the $X$ basis are nonuniform are resource states such that predictability can be quantified by monotones that depend solely on the diagonal elements of the state in the $X$ basis. A convenient and operationally simple way to construct such predictability monotones relies on the commuting condition between free operations and the resource-destroying map~\cite{LiuH2017}, i.e., $\Lambda_{\epsilon} \circ \Phi_{XY} = \Phi_{XY} \circ \Lambda_{\epsilon} $. In this framework, it suffices to consider a contractive function $D(\rho,\sigma)$, not necessarily a metric, satisfying $D(\Lambda_{\epsilon}(\rho),\Lambda_{\epsilon}(\sigma))\le D(\rho,\sigma).$ As a consequence, the quantity $\mathcal{D}(\rho):=D\left(\Phi_X(\rho),\Phi_{XY}(\rho)\right)$ defines a valid predictability monotone, since it is monotonically non-increasing under free operations. Typical examples include those introduced in Sec.~\ref{sec:CRpositivity}.

The relationship among the resource theories of predictability, coherence, and purity, within the framework of resource-destroying maps, is also analyzed by~\citet{Basso2022d}. Coherence, defined with respect to an observable $X$, is characterized by the set of incoherent states obtained via the dephasing map $\Phi_X(\rho) = \sum_j X_j \rho X_j$, with $X_j = |x_j \rangle \langle x_j|$. Purity, in turn, is formulated as a resource theory whose only free state (with respect to $X$) is the maximally mixed state $I/d$, obtained through a maximally destroying map built from mutually unbiased observables $X$ and $Y$. The corresponding information (or purity) measure is $I(\rho) = \log d - S_{vn}(\rho)$, which can also be written as a relative entropy distance to the maximally mixed state~\cite{Horodecki2003}. An interesting result is that the set of maximally incoherent states coincides with the intersection between the set of incoherent states and the set of unpredictable states, and that the sum of coherence and predictability measures equals the total information content, i.e.,
\begin{align}
  \mathcal{C}^X_{re}(\rho) + \mathcal{P}^X_{vn}(\rho) = I(\rho) \le \log d.
\end{align}
Operationally, successive non-selective measurements of mutually unbiased observables progressively erase resources: measuring $X$ destroys coherence but preserves predictability, while measuring $Y$ further eliminates predictability, driving the state to maximal mixedness and zero information. Besides, complementarity relations between predictability and quantum coherence, which saturates if, and
only if, the state of the system is pure and vanish for maximally mixed states, can be used as purity measures, as notice by~\citet{Pozzobom2021}. In addition,~\citet{Ganardi2025} explored the connection between quantum entanglement and local purity extraction within quantum resource theories, introducing the framework of Gibbs-preserving local operations and classical communication to model local cooling processes, where two parties aim to cool their subsystems toward ground states using only one copy of a shared quantum state. In the case of fully degenerate local Hamiltonians, optimal local cooling reduces to local purity extraction, thereby establishing a purity-entanglement complementarity: the amount of locally extractable purity is fundamentally constrained by the entanglement present in the state. This result is in agreement with earlier findings by~\citet{Oppenheim2003} that established an operational complementarity between locally accessible information and nonlocal (entanglement-based) information under LOCC, showing that extracting one necessarily limits the other, which has also been experimentally verified using entangled photons in linear-optical interferometric setups~\cite{Fedrizzi2011}.

\subsection{Complementarity and the entanglement swapping for partially entangled states}

Within the framework of CCR discussed in Sec.~\ref{sec:QCP},~\citet{Basso2022c} demonstrated that entanglement swapping protocol~\cite{Zukowski1993} exhibits a counterintuitive behavior when applied to partially entangled pure two-qubit states. In particular, even when the initial entanglement shared by the parties is arbitrarily small, the protocol still allows Alice and Bob to obtain a maximally entangled state with a small but strictly nonzero probability. Remarkably, this occurs despite starting from vanishing local quantum coherence in the subsystems.

As usual, in the entanglement swapping protocol, we have three laboratories operated by Alice, Bob, and Charlie. In addition, suppose that Darwin, in a separate laboratory, prepares two pairs of qubits in partially entangled pure states given by
\begin{align}
& |\xi\rangle_{AC} = \sqrt{p}|0\rangle_A|0\rangle_{C} + \sqrt{1-p}|1\rangle_A|1\rangle_{C}, \\
& |\eta\rangle_{C'B} = \sqrt{q}|0\rangle_{C'}|0\rangle_{B} + \sqrt{1-q}|1\rangle_{C'}|1\rangle_{B},
\end{align}
with $p,q\in[0,1]$. The qubits $C$ and $C'$ are sent to Charlie, who performs a selective Bell-basis measurement on them, while the qubit $A$ is sent to Alice and the qubit $B$ to Bob.

After Charlie makes the Bell-basis measurement, the possible post-measurement states are 
\begin{align}
\hspace*{-6pt} \ket{\phi_{\pm}}_{AB} &= \frac{1}{N_{\phi}}\big(\sqrt{pq}|0\rangle_A|0\rangle_{B}\pm\sqrt{(1-p)(1-q)}|1\rangle_A|1\rangle_{B}\big),
\label{eq:phipm}
\\
\hspace*{-6pt} \ket{\psi_{\pm}}_{AB} &= \frac{1}{N_{\psi}} \big(\sqrt{p(1-q)}|0\rangle_A|1\rangle_{B}\pm\sqrt{(1-p)q}|1\rangle_A|0\rangle_{B}\big),
\label{eq:psipm}
\end{align}
with $N_{\phi} = \sqrt{pq+(1-p)(1-q)}$ and $N_{\psi} = \sqrt{p(1-q)+(1-p)q}$. These states are obtained with probabilities 
\begin{align}
& \Pr(\phi_{\pm}) = \frac{1}{2}\Big(pq + (1-p)(1-q)\Big),
\\ 
& \Pr(\psi_{\pm}) = \frac{1}{2}\Big(p(1-q) + (1-p)q\Big),
\end{align} 
which are the same as the probabilities for Charlie to obtain the Bell states $\ket{\Phi_{\pm}}_{CC'}$ and $\ket{\Psi_{\pm}}_{CC'}$ respectively.
Further analysis shows that, for any fixed value of $q$, the quantities $S_{vn}(\rho^{\phi}_A)$ and $S_{vn}(\rho^{\psi}_A)$ vary continuously as functions of $p$. In particular, as $p$ ranges over the interval $[0,1]$, both entropies start at $0$, increase until reaching a maximum value of $1$, and then decrease back to $0$. Consequently, for every fixed $q$, there always exists a value of $p$ for which $S_{vn}(\rho^{\phi}_A)$ and $S_{vn}(\rho^{\psi}_A)$ attain their maximal values. This can be verified by analyzing the first and second derivatives of these functions with respect to $p$:
\begin{align}
& \frac{d}{dp}S_{vn}(\rho^{\phi}_A) = 0 \ \land \ \frac{d^2}{dp^2}S_{vn}(\rho^{\phi}_A) < 0 \iff p = 1 - q, \nonumber \\
& \frac{d}{dp}S_{vn}(\rho^{\psi}_A) = 0 \ \land \ \frac{d^2}{dp^2}S_{vn}(\rho^{\psi}_A) < 0 \iff p = q. \nonumber
\end{align}

On the other hand, when $p \neq q$ and $p \neq 1-q$, the states $\ket{\phi_{\pm}}_{AB}$ and $\ket{\psi_{\pm}}_{AB}$ produced by the protocol are partially entangled. As an illustrative example, consider $p = 0.1$ and $q = 0.75$. In this case, the initial states $|\xi\rangle_{AC}$ and $|\eta\rangle_{C'B}$ have entropies $S_{vn}(\rho^{\xi}_A) = S_{vn}(\rho^{\xi}_C) \approx 0.4689$ and $S_{vn}(\rho^{\eta}_{C'}) = S_{vn}(\rho^{\eta}_B) \approx 0.8112$, respectively. After Charlie’s measurement, Alice and Bob obtain the states $\ket{\phi_{\pm}}_{AB}$ and $\ket{\psi_{\pm}}_{AB}$ with entropies $S_{vn}(\rho^{\phi}_A) = S_{vn}(\rho^{\phi}_B) \approx 0.8112$ and $S_{vn}(\rho^{\psi}_A) = S_{vn}(\rho^{\psi}_B) \approx 0.2222$, occurring with probabilities $\Pr(\phi{\pm}) = 0.15$ and $\Pr(\psi_{\pm}) = 0.35$, respectively. Therefore, the outcomes associated with lower entanglement occur more frequently than those with higher entanglement. Nevertheless, if a sufficiently large ensemble of partially entangled quantons is available at the beginning of the protocol, it is possible---by means of local operations---to concentrate this entanglement into a smaller number of maximally entangled pairs.

The analysis also showed that the structure imposed by the CCR formalism constrains how quantum resources are distributed and redistributed during the swapping process, revealing that entanglement amplification can arise probabilistically from states that are nearly separable yet exhibit nonzero predictability. They established a linear relationship between the local predictability of the pre-measurement states and the probability of obtaining a maximally entangled state after the Bell-basis measurement. This implies that even when the initially shared entanglement is vanishingly small, a maximally entangled state can still be obtained with a strictly nonzero probability, provided that there is nonzero predictability available to be consumed.

This result was further generalized by~\citet{Maziero2022} to include arbitrary pure quantum states. Using the Jacob-Bergou triality relation given by Eq.~\eqref{JBrelation}, it was possible to formalize the resource-conversion underlying the protocol by providing explicit transformation equations showing how the distributed entanglement after the Bell-basis measurement is strictly constrained by the pre-measurement values of the predictability and visibility (or quantum coherence). The analysis thus makes explicit the resource-conversion mechanism underlying entanglement swapping: local quantum features that characterize the initial state are transformed into quantum correlations after the measurement. To validate these findings, the authors performed simulations using IBM's quantum computers, where, despite hardware noise, the results showed strong agreement with the theoretical prediction that local features are converted into nonlocal correlations through the Bell-state measurement.

These ideas were further extended to high-dimensional systems (qudits) by~\citet{Starke2025}, where the entanglement swapping protocol is analyzed for partially entangled qudit pairs. An interesting result of this generalization is a no-free-lunch bound for entanglement distribution: while it remains possible to obtain a final state with greater entanglement in a single run of the entanglement swapping protocol, the average entanglement that Alice and Bob can distribute via ESP is fundamentally bounded above by the initial entanglement of one of the input pairs, and also by the product of the initial entanglements of both pairs. This means that, on average over many runs, the protocol cannot increase entanglement beyond what was initially present. Together, these works establish an operational picture in which entanglement swapping, coherence, and predictability are quantitatively linked, i.e., local quantum resources constrain the probabilistic generation of entanglement, which may have important implications for quantum communication and distributed quantum information processing.

\subsection{Wave-particle duality and quantum thermal machines}

The analysis of wave-particle duality in the context of quantum thermal machines was carried out by~\citet{Janovitch2023}, where the authors investigated how this fundamental quantum feature manifests in the performance of a bosonic quantum heat engine. Their approach consists of comparing the quantum engine with two distinct classical counterparts: a classical wave heat engine, based on coherent classical fields, and a classical particle heat engine, described in terms of stochastic particle transport. In thermodynamics, a heat engine converts heat absorbed from a hot reservoir into useful work delivered to an external load. The average output power is the mean rate at which work is extracted from the engine, typically defined as the average work performed per unit time in the steady-state regime. It is a central thermodynamic quantity characterizing the engine's performance at the level of expectation values. The analysis showed that both classical models can reproduce the average output power of the quantum engine under certain conditions. In this sense, mean energetic performance alone does not necessarily reveal uniquely quantum behavior. However, a different picture emerges when one examines power fluctuations, which quantify the statistical variance of the output power around its mean value. In quantum thermodynamics, fluctuations play a crucial role because work and heat are stochastic quantities at the microscopic level, and their distributions encode genuinely quantum features such as coherence and quantum statistics. The investigation showed that neither classical model reproduces the full fluctuation statistics of the quantum engine: the classical wave model fails to capture vacuum fluctuations, which arise from zero-point quantum noise, while the particle model cannot correctly describe the bosonic bunching effects inherent to quantum transport. These results demonstrate that, although average thermodynamic quantities may admit effective classical descriptions, higher-order statistical properties---such as fluctuations---reveal intrinsic quantum signatures that can be linked to wave-particle duality. This indicates that the non-equilibrium operation of the quantum heat engine, particularly at the level of fluctuation statistics, is governed by the simultaneous presence of wave-like coherence and particle-like quantum statistics, embodying wave-particle duality within a genuinely thermodynamic setting.

In addition, another thermodynamic perspective on wave-particle duality has been proposed, in which a single quantum system is treated as a quantum battery and its dual character is analyzed in terms of energy storage and extraction processes~\cite{YangL2025}. Within this framework, the maximal extractable energy---quantified by the so-called energy capacity, a quantity determined solely by the eigenvalues of the quantum state---provides a direct link to operational measures traditionally associated with wave- and particle-like behavior, such as interferometric visibility and path distinguishability. This approach leads to a novel formulation of duality in terms of a squared energy-coherence equality. In this way, the distinct roles of wave and particle attributes are reinterpreted in terms of their contribution to the performance of quantum batteries, offering a physically grounded and device-independent characterization of complementarity. The theoretical predictions are further supported by experimental results obtained with single photons, reinforcing the viability of this thermodynamic viewpoint.

%---------------------------
\subsection{Complementarity and incompatible observables}
\label{sec:Cincomp}
While complementarity, incompatibility, and uncertainty  are terms often used interchangeably in heuristic discussions, contemporary research establishes a hierarchical and functional relationship among them. The analysis developed by \citet{Busch1997,Bush2006,Busch2007} provides a refined understanding of the relationship between complementarity, incompatibility, and uncertainty within the standard quantum formalism. In contrast to views that either identify complementarity with incompatibility or treat it as an independent principle, Busch \textit{et al.} argue that these notions are ultimately rooted in the structure of quantum mechanics, particularly in the non-commutativity of observables. In Bohr's original formulation, complementarity expresses the impossibility of simultaneously attributing properties such as wave-like and particle-like behavior to a quantum system. This impossibility is often interpreted as a mutual exclusivity between experimental arrangements: an experiment designed to reveal interference cannot, at the same time, provide which-path information. However, this qualitative formulation leaves open a fundamental question: is this exclusivity absolute, or merely a practical limitation associated with the type of measurement being performed?

The modern approach reformulates this question in terms of the \textit{incompatibility} of observables. Instead of referring to experimental arrangements, one considers whether two observables can be measured jointly within a single measurement scheme. In this approach, incompatibility is defined broadly as the impossibility of measuring two or more observables simultaneously using a single device. Operationally, two observables $X$ and $Y$ are incompatible if there exists no joint observable $G$ such that the marginals of $G$ recover the statistics of $X$ and $Y$ for all possible system preparations. \textit{Complementarity}, in the sense of Bohr and refined by~\citet{Bush2006}, represents an extreme form of this incompatibility. It is characterized by the disjointness of effects: two observables are complementary if the information gathered about one precludes any deterministic knowledge of the other, often formalized through the condition that their corresponding effects have no common ``tests'' or overlap in the Hilbert space structure.

If the analysis is restricted to observables represented by self-adjoint operators with spectral measures (PVMs), one recovers the well-known result that non-commuting observables do not admit a joint exact measurement. In this restricted setting, complementarity is often identified with non-commutativity. However, this identification is overly restrictive. The general theory of quantum measurements allows observables to be described by positive operator-valued measures (POVMs), which naturally incorporate imperfections and unavoidable noise present in any realistic experimental procedure.
At this point,~\citet{Bush2006} analysis becomes essential. By introducing POVMs, it becomes possible to define unsharp versions of incompatible observables. These correspond to measurements that do not perfectly distinguish ideal outcomes but still retain relevant information about the system. The central result is that incompatible observables may admit an approximate joint measurement, provided one accepts a controlled degradation of precision. This observation profoundly changes the interpretation of complementarity: it ceases to be an absolute prohibition and instead becomes a quantitative restriction. In this generalized framework, complementarity acquires a quantitative formulation through trade-off relations that constrain the simultaneous accessibility of different observables. A paradigmatic example is provided by interferometric scenarios, where one considers the balance between which-path information and interferometric visibility. When these quantities are described within the POVM formalism, they can be jointly measured only at the cost of reduced sharpness, leading to complementarity relations, such as the ones expressed by Eqs.~\eqref{dualityG} and~\eqref{dualityE}, or analogous expressions depending on the specific definitions adopted. These relations capture the essence of complementarity by establishing precise limits on how much information about complementary properties can be obtained in a single experimental arrangement. Moreover, using standard measures of uncertainty, it can be shown that, in the context of Mach--Zehnder interferometry experiments, quantitative duality relations are equivalent to the uncertainty relation corresponding to a suitable pair of associated observables~\cite{Bjork1999, Durr2000, Liu2012, Bosyk2013}. Despite this close connection, complementarity and uncertainty are not conceptually equivalent~\cite{Bush2006}. Uncertainty relations can be formulated without explicit reference to measurement contexts or mutually exclusive experimental arrangements, whereas complementarity emphasizes the structure of accessible information and the role of the measurement setup~\cite{Lahti1980}. In this more general setting, it has also been shown that complementarity itself may depend on the chosen measure of uncertainty and need not be a symmetric or universally well-defined relation between observables, further highlighting its operational and context-dependent character~\cite{Luis2002}.

In addition,~\citet{Kiukas2019} provided a comprehensive review and a unified mathematical characterization of complementarity, largely following the operational legacy of~\citet{Bush2006, Busch2007}. The authors formalize complementarity through the lens of quantum effects and positive operator-valued measures, utilizing a fundamental factorization lemma for quantum effects to derive various characterizations of the concept. In this operational approach, two POVMs $A_a$ and $B_b$ are called complementary if all their measurements are mutually exclusive.  More precisely, for every pair of outcomes $a$ and $b$, the only POVM $O$ satisfying $\Tr(\rho O) \le \Tr(\rho A_a)$ and $\Tr(\rho O) \le \Tr \rho B_b$, for all states $\rho$, is the null operator, i.e., $O = 0$. If there existed a nonzero operator $O$ fulfilling these inequalities, then the two-outcome $\{O,\mathbb{I}-O\}$ would provide nontrivial information about both outcome probabilities $\Tr(\rho A_a)$ and $\Tr(\rho B_b)$ in every state for which $\Tr(\rho O) \neq 0$. In this framework, canonical pairs such as position-momentum, number-phase, or energy-time in the harmonic oscillator provide standard examples of complementary observables, since no nonzero positive operator can simultaneously bound their effects. In addition,~\citet{Kiukas2019} also distinguishes between sharp and unsharp complementarity. For sharp observables, complementarity is often synonymous with the disjointness of spectral projections. However, the authors extend this to unsharp (noisy) observables, demonstrating that even when measurements are fuzzy, they can maintain a form of complementarity. More recently, complementarity has also been formulated in information-theoretic terms through information exclusion relations~\cite{Huang2024}, providing a general framework that captures wave-particle duality as a particular instance and expresses complementarity as a constraint on the accessible information in quantum measurements.

On the other hand,~\citet{Saha20} proposed a paradigm shift by establishing a theory-independent, purely operational framework for complementarity and its relationship with preparation uncertainty. Unlike traditional approaches that rely on Hilbert space formalism (such as non-commuting operators), this work defines complementarity and uncertainty solely through measurement statistics.  They also postulate that complementarity should be viewed as the fundamental cause of uncertainty. Specifically, they propose that measures of complementarity should constitute the right-hand side of uncertainty relations. This means that the more complementary two observables are, the higher the minimum joint uncertainty must be, regardless of the system's preparation. Interestingly, this approach seems to frame uncertainty not as a special quantum mechanical feature but as a structural consequence of the operational concept of complementarity. Furthermore,~\citet{Saha20} applied this operational approach of complementarity and uncertainty relations in quantum information tasks. Particularly, it derives the Tsirelson bound for the CHSH inequality~\cite{CHSH1969} and links uncertainty relations to the Information Content Principle~\cite{Czekaj2017} (a variant of Information Causality~\cite{Pawlowski2009}). This suggests that complementarity and incompatibility are not just abstract properties of quantum systems, but fundamental constraints on the flow and storage of information in any physical theory that exhibits quantum-like features. It is worth noting that~\citet{Guhne2023} provided a comprehensive overview of observable incompatibility and its relationships to notions such as joint measurability, contextuality, and the operational definitions of complementarity.

%---------------------------
\subsection{Complementarity beyond non-relativistic quantum mechanics}

%---------------------
\subsubsection{Complementarity and relativity}
\label{sec:Rel_comp}

\begin{figure*}[t]
    \centering
    \includegraphics[width=1\linewidth]{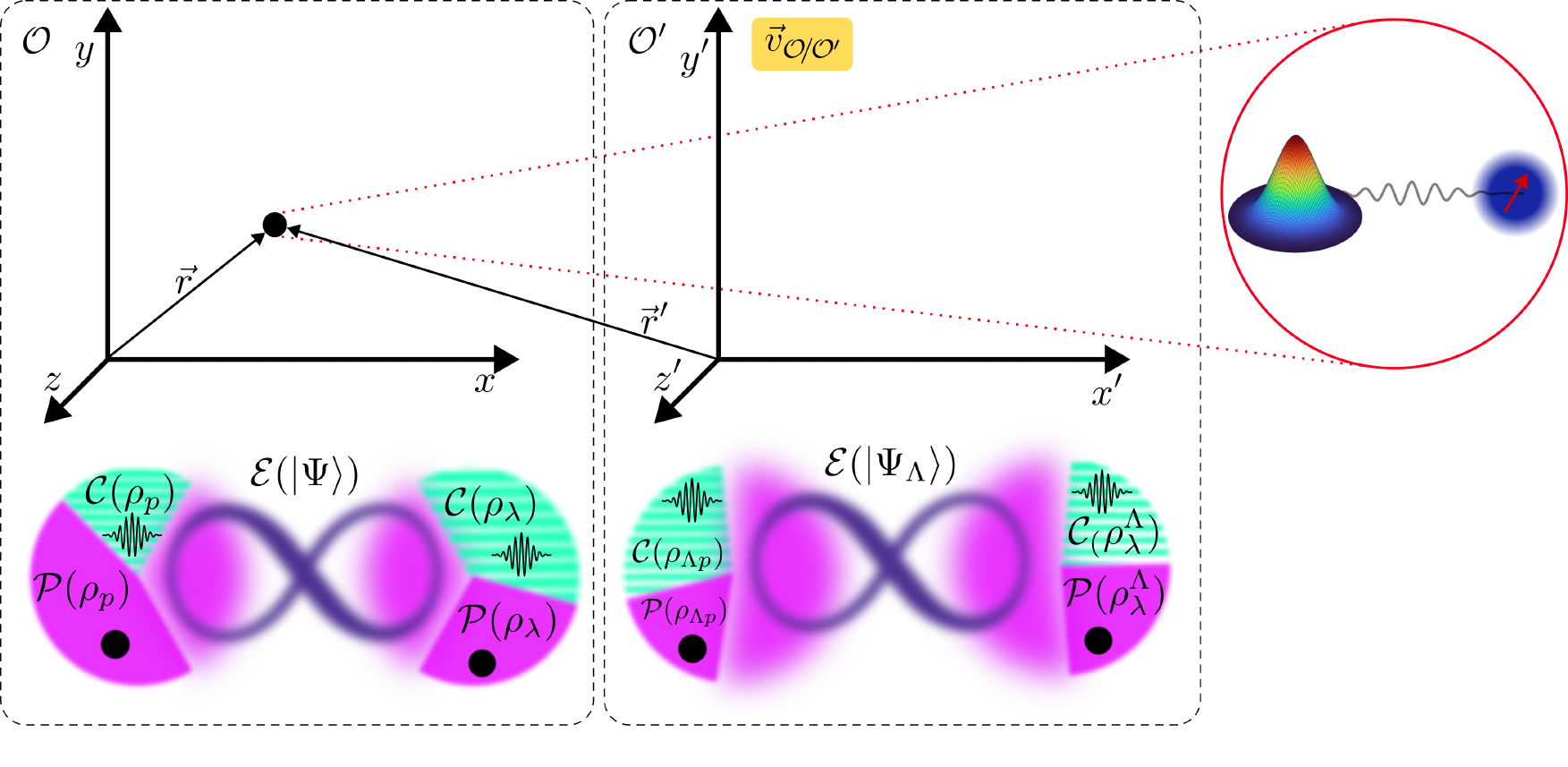}
    \caption{Illustration of the Lorentz invariance of the complete complementarity relations, considering two inertial observers, $\mathcal{O}$ and $\mathcal{O'}$, moving with relative velocity $\vec{v}_{\mathcal{O}/\mathcal{O}'}$, and describing a spin-$1/2$ particle with discrete momentum spectrum. Although the individual quantities---quantum coherence, predictability, and entanglement---are not Lorentz invariant and may change under transformations between inertial frames, their combined relation remains invariant, ensuring that the complementarity bound is independent of the observer's reference frame.}
    \label{fig:CCR_LI}
\end{figure*}

From a relativistic standpoint, quantities such as predictability, coherence, and entanglement are not a priori guaranteed to be invariant under Lorentz transformations, especially when relativistic degrees of freedom---such as spin---are taken into account. This raises the natural question of whether complementarity, when expressed through triality relations, represents an observer-independent feature of quantum systems.

For instance,~\citet{Czachor1997} analyzed a relativistic version of the Einstein--Podolsky--Rosen experiment with massive spin-$1/2$ particles and argued that the degree of Bell inequality violation depends on the particles' velocities. Shortly thereafter,~\citet{Gingrich2002} showed that entanglement itself can become frame dependent: while spin-spin entanglement between two particles may decrease under a Lorentz boost, the entanglement between spin and momentum degrees of freedom can increase. This behavior reflects a redistribution of entanglement among different subsystems induced by Lorentz transformations. In the same year,~\citet{Peres2002a} demonstrated that the entanglement entropy of a single massive spin-$1/2$ particle is not Lorentz invariant. These findings motivated extensive investigations into the behavior of entanglement under Lorentz transformations~\cite{Ahn2003, Gingrich2003, Terashima2004, Caban2006, Peres2004a, Fuentes2005, Lamata2006, Alsing2006, Jordan2007, Dunningham2009, Fuentes2010, Friis2010, Palge2015, Bittencourt2018, Zambianco2019, Caban2023}, leading to the conclusion that the behavior of entanglement under Lorentz boosts is highly dependent on the physical settings considered, including the structure of the particle states and the geometry of the Lorentz transformations involved.

Let us consider an inertial frame $\mathcal{O}$ that describes the quantum state of a massive spin-$1/2$ particle as
\begin{align}
    |\Psi\rangle = \sum_{p, \lambda} \psi_{\lambda}(p) |p\rangle|\lambda\rangle,
\end{align}
where the basis vectors $|p\rangle|\lambda\rangle \equiv |p\rangle \otimes |\lambda\rangle$ span the composite Hilbert space, with $p$ labeling the momentum eigenstates and $\lambda$ denoting the spin states. Let us consider another inertial observer $\mathcal{O}'$ moving with respect to $\mathcal{O}$ $(\vec{v}_{\mathcal{O}/\mathcal{O}^\prime})$ and related to it by a Lorentz boost $\Lambda$. The observer $\mathcal{O}'$ assigns the state $|\Psi_{\Lambda}\rangle = U(\Lambda)|\Psi\rangle$ to the same physical system, where $U(\Lambda)$ denotes the unitary representation of the Lorentz transformation. The action of this operator on the basis states reads
\begin{align}
    U(\Lambda) |p\rangle |\lambda\rangle = |\Lambda p\rangle \otimes D(W(\Lambda,p)) |\lambda\rangle,
\end{align}
with $W(\Lambda,p)$ being the associated Wigner rotation. The collection of such rotations forms the Wigner's little group, which is a subgroup of the Poincar\'e group~\cite{Wigner1939}. Thus, under a general Lorentz boost $\Lambda$, the momentum degrees of freedom transform according to $p \mapsto \Lambda p$, while the spin undergoes a momentum-dependent unitary transformation $D(W(\Lambda,p))$. The Wigner's little group is isomorphic to the three-dimensional rotation group $\mathrm{SO(3)}$, which, for massive spin-$1/2$ particles, is homomorphic to the special unitary group $\mathrm{SU(2)}$~\cite{Weinberg1995}.

Following the reasoning established by~\citet{Basso2020b} and discussed in Sec.~\ref{sec:CCRhs}, let us consider the spin reduced density operator, $\rho_{\lambda} = \mathrm{Tr}_p (|\Psi\rangle \langle \Psi|)$. From the purity of the composite quantum system, $1 - \mathrm{Tr}(\rho^2) = 0$, with $\rho = |\Psi\rangle\langle\Psi|$, one can derive the following CCR in the inertial reference frame $\mathcal{O}$:
\begin{align}
  \mathcal{C}_{\ell_2}(\rho_\lambda) +   \mathcal{P}_{\ell_2}(\rho_\lambda) + \mathcal{E}_l(\rho_\lambda)
  = \frac{d_\lambda - 1}{d_\lambda}, \label{eq:CCRhs}
\end{align}
where $d_\lambda = 2$ is the dimension of the spin-$1/2$ Hilbert space. Given that unitary transformations preserve purity, we have $\Tr (\rho_{\Lambda}^{2}) = \Tr (\rho^{2})$, where $\rho_{\Lambda} = |\Psi_{\Lambda}\rangle \langle\Psi_{\Lambda}|$, and since the derivation of Eq.~\eqref{eq:CCRhs} relies solely on the purity of the global state, the condition $1-\Tr(\rho_{\Lambda}^{2})=0$ guarantees that the CCR for the spin subsystem remains valid in the boosted frame $\mathcal{O}'$, namely
\begin{align}
\mathcal{C}_{\ell_2}(\rho^\Lambda_\lambda) + \mathcal{P}_{\ell_2}(\rho^\Lambda_\lambda) +
\mathcal{E}_l(\rho^\Lambda_\lambda)
  = \frac{d_\lambda - 1}{d_\lambda}, 
\end{align}
where $\rho^\Lambda_{\lambda} =  \Tr_{\Lambda p}(|\Psi_{\Lambda}\rangle \langle \Psi_{\Lambda}|)$. The same reasoning, based on the purity of the composite quantum system, may be employed to show that the other CCRs described in Sec.~\ref{sec:CCRtheorem} also remain invariant under Lorentz transformations.

This result demonstrates that CCR for spin degrees of freedom can be consistently formulated in Minkowski spacetime and remain invariant under Lorentz transformations, independently of whether the momentum is treated as a continuous or a discrete variable. The same conclusion holds for CCRs applied to the reduced momentum state, in which momentum is treated as a discrete subsystem and the spin degrees of freedom are traced out. This approximation can be justified by considering narrow momentum wave packets centered at distinct values, which can be represented by orthogonal states~\cite{Palmer2012}. Figure~\ref{fig:CCR_LI} illustrates how the CCR can be displayed in each inertial reference frame. Even though the individual measures of coherence, predictability, and entanglement may not each be Lorentz invariant, the CCR itself is invariant, so the complementarity bound does not depend on the observer’s reference frame. Moreover, this result can be easily extended to systems composed of an arbitrary number of particles with arbitrary spin~\cite{Basso2021a}. Furthermore, local frames of reference, defined at each point in curved spacetime through a tetrad field, have been employed to extend Wigner rotations to curved spacetimes~\cite{Terashima2004} and to investigate the invariance of CCRs in this more general setting~\cite{Basso2021b, Basso2024}. This framework has also been used to analyze interferometric setups in curved spacetime for massive spin-$1/2$ particles~\cite{Basso2021c, Basso2021d, Basso2024}.

Another interesting application of complementarity relations is to probe general relativistic effects in quantum particles.~\citet{Zych2011} proposed a novel experimental framework to test the effects of general relativity on quantum systems. Unlike traditional matter-wave interferometry, which typically measures phase shifts---effects that can often be mimicked by effective potentials in flat spacetime---this proposal focuses on a uniquely relativistic phenomenon: gravitational time dilation. The core of the proposal involves an interferometer where the interfering particle is not a simple point mass, but a ``clock''---a system with evolving internal degrees of freedom. As the particle travels through two different paths in a gravitational field, it experiences different rates of proper time flow due to general relativity. Because proper time flows at different rates along the paths, the internal state of the clock evolves differently. By the time the paths recombine, the clock's state labels the path it took. If the gravitational time dilation is significant enough that the internal states of the clock corresponding to each path become distinguishable, path information becomes available in principle. According to the principle of complementarity, this which-path information necessarily leads to a reduction in the visibility of the interference pattern.

Finally,~\citet{Cepollaro2025} showed that the sum of entanglement and local coherence remains invariant under quantum reference frame transformations~\cite{Giacomini2019, Ahmad2022}. This invariance is demonstrated both when entanglement is quantified by the von Neumann entropy and coherence by the relative entropy of coherence, and when entanglement is quantified by the linear entropy and coherence by the Hilbert--Schmidt (or 
$l_2$-norm) measure. This work naturally leads to the question of whether the CCRs, such as the one given by Eq. \eqref{eq:CCRhs}, also remain invariant under such transformations. It is worth noting that one of the key points in their demonstration is that the sum of entanglement and local coherence can be expressed as a function of the diagonal elements of the reduced density operator. By rewriting Eq.~\eqref{eq:CCRhs} as $\mathcal{C}_{\ell_2}(\rho_\lambda) + \mathcal{E}_l(\rho_\lambda) = \mathcal{E}_l^{\text{max}} - \mathcal{P}_{\ell_2}(\rho_\lambda)$ and observing that $\mathcal{P}_l(\rho_\lambda)$ depends only on the diagonal elements of the reduced density matrix, which plays a crucial role in establishing invariance, the results obtained here seem to contain the essential ingredients to address this question. The same line of reasoning may be applied to the other CCRs described in Sec.~\ref{sec:CCRtheorem}.

%---------------------------
\subsubsection{Complementarity in quantum field theory}
\label{sec:QFT}

Motivated by the Lorentz invariance of complete complementarity relations discussed in the previous section, recent works have studied the CCR framework to quantum field theory settings. For instance, \citet{Bittencourt2022} investigated the application of CCRs to neutrino oscillation phenomena, treating two-flavor neutrino states as bipartite states, i.e., a two flavor-neutrino state can be represented as
\begin{equation}
    |\nu_\alpha(t)\rangle = a_{\alpha\alpha}(t)|\nu_\alpha\rangle + a_{\alpha\beta}(t)|\nu_\beta\rangle ,
\end{equation}
where $\alpha, \beta = e, \mu$ denote flavor indices. Using the following correspondence
\begin{eqnarray}
|\nu_\alpha\rangle = |1\rangle_\alpha |0\rangle_\beta, \quad |\nu_\beta\rangle = |0\rangle_\alpha  |1\rangle_\beta,
\end{eqnarray}
the composite nature of neutrino flavor states becomes explicit. For instance, an initially electronic neutrino reads $|\nu_e(t)\rangle = a_{ee}(t)|1\rangle|0\rangle + a_{e\mu}(t)|0\rangle|1\rangle$, where $\mu$ denotes the muonic neutrino. Both plane-wave (pure-state) and wave-packet (mixed-state) regimes were analyzed, allowing the study of the interplay between predictability, quantum coherence, and quantum correlations during neutrino propagation. Later, \citet{Bittencourt2024} extended the analysis for three-flavor neutrino oscillations by considering the three-flavor neutrino states as tripartite states, while~\citet{Blasone2024} further extended this framework by incorporating the effects of CP violation. The study reveals that bipartite CCRs capture the redistribution of quantum correlations among neutrino flavor modes, but are not sufficient to uniquely characterize genuine tripartite correlations, due to ambiguities associated with residual correlation measures.

Complete complementarity relations were also applied to analyze the generation and redistribution of quantum correlations in tree-level quantum electrodynamics scattering processes~\cite{Blasone2025a}, using Bhabha scattering as a paradigmatic example, namely the elastic scattering process between an electron and a positron. For an initially factorized fermionic state of the form $ |i\rangle = |p_1,a\rangle_A \otimes |p_2,b\rangle_B,$ where $p_1$ and $p_2$ denote the incoming momenta and $a$ and $b$ label the spin degrees of freedom of the two fermions, the corresponding final state after the interaction is given by
\begin{align}
    |f\rangle = & |i\rangle +
 i \sum_{r,s} \int \frac{d^3p_3 d^3p_4}{(2\pi)^6 2E_{p_3} 2E_{p_4}} 
  \\ & \times \delta^{(4)}(p_1+p_2-p_3-p_4) 
  \mathcal{M}(a,b; r,s)
  |p_3,r\rangle_A  |p_4,s\rangle_B, \nonumber
\end{align}
where $ \delta^{(4)}(p_1+p_2-p_3-p_4)$ enforces $4$-momentum conservation and $\mathcal{M}(a,b; r,s)$ denotes the scattering amplitude. Within this framework, CCRs for the spin degrees of freedom provides a quantitative description of the interplay between predictability, local coherence, and entanglement in the outgoing state. The analysis shows that scattering processes can generate nontrivial spin entanglement from initially separable states, while for initially correlated states CCRs characterize the redistribution of quantum correlations. In particular, maximally entangled initial states were found to preserve maximal entanglement under scattering.~\citet{ChenD2026} investigated quantum correlations in top-antitop production by employing complete complementarity relations to connect quantum coherence and mutual information in the spin degrees of freedom, thus providing a natural and consistent description of nonclassical features in high-energy processes. In addition to applications in relativistic scattering of elementary particles, complementarity relations have also been explored in the context of resonant Auger scattering~\cite{LiuG2024} and interferometric setups that can be used to witness the quantum nature of the gravity~\cite{Maleki2022, Sugiyama2022, ChenD2025}.

Moreover, a new perspective on wave-particle duality in quantum mechanics was considered by~\citet{Aiello2023}. Employing methods from quantum field theory, the author proposed a form of complementarity between the continuous wave-like degrees of freedom of the field, associated with field quadratures, and the discrete particle-like degrees of freedom, associated with particle number measurements, rather than the usual trade-off between path distinguishability and fringe visibility. This approach reveals a nontrivial complementarity between the continuous and discrete aspects of the electromagnetic field when prepared in a single-photon state. In particular, it is operationally impossible to simultaneously observe a nonzero field quadrature amplitude beyond vacuum fluctuations and to detect a single photon in two distinct regions of the same light beam.

%---------------------
\section{Perspectives}
\label{sec:persp}

In this section, we outline a number of open questions and directions for future research that naturally emerge from the discussions presented throughout this review. Rather than aiming at a comprehensive account, we briefly highlight selected issues that remain unresolved, including both conceptual and technical challenges associated with the formulation and generalization of complementarity relations.

Throughout this review, several connections between complementarity and uncertainty have appeared in different contexts, including the discussion of momentum kicks, wave-particle duality formulated through entropic uncertainty relations, and the relation between incompatible observables and interferometric duality relations in two-slit (or Mach-Zehnder) interferometers. While uncertainty relations traditionally quantify limitations on the simultaneous sharpness of incompatible observables, complementarity relations characterize trade-offs between mutually (or partially) exclusive physical manifestations, such as wave- and particle-like behavior. Nevertheless, modern approaches increasingly suggest that these two notions may represent different aspects of a common underlying structure of quantum theory, as advocated by~\citet{Bush2006, Busch2007}. In particular, complementarity relations formulated in interferometric scenarios are often closely connected to entropic, variance-based, or information-theoretic measures, suggesting a more general equivalence between uncertainty relations and complementarity relations extending beyond conventional two-path interferometers, in the sense that wave-particle duality relations may be derived from uncertainty relations and vice versa. Along this direction,~\citet{Coles2016} generalized the equivalence between entropic uncertainty relations and wave-particle duality relations to multipath interferometers, reinforcing the idea that these connections are not restricted to simple two-path scenarios. From this perspective, an interesting line of investigation is the development of a unified framework capable of encompassing different types of uncertainty relations together with the different forms of complementarity relations discussed throughout this review.

Beyond the complementarity relations directly connected to wave-particle duality and Bohr's original formulation, the literature has also seen the development of a broad class of trade-off relations involving different quantum resources. Although many of these works are referred to as complementarity relations, their connection with Bohr's notion of complementarity is often not explicit and still calls for further conceptual investigation. In this broader sense,~\citet{Singh2015} derived a trade-off relation between the $\ell_1$-norm coherence and the degree of mixedness in $d$-dimensional systems, identifying the class of states that maximizes coherence for a fixed amount of noise. Within the resource-theoretic framework of quantum coherence,~\cite{Das2020, Srivastava2021} have extended complementarity relations to generalized notions of coherence and superposition in nonorthogonal bases~\cite{Theurer2017}. More recently,~\citet{Chen2024} established analogous trade-off relations involving quantum imaginarity~\cite{Wu2021}, linear entropy, and the Brukner–Zeilinger invariant information measure, while~\citet{Liu2026} explored further relations formulated directly in terms of imaginarity-based quantities. The exploration of trade-off-type relations involving other quantum resources and generalized resource theories, such as  magic~\cite{Veitch2014} and texture~\cite{Parisio2024}, represents a promising research perspective that may help reveal broader structures underlying wave-particle duality and complementarity.

In view of the recent development of dynamical resource theories of quantum coherence \cite{Xu2019, Saxena2020, Fan2026} and of quantum entanglement \cite{Gour2020, Zhou2022}, a natural and largely unexplored direction concerns the extension of complementarity relations from quantum states to quantum channels. In particular, \citet{Luo2018} formulated Bohr's complementarity principle within a general state--channel interaction framework, where coherence and accessible information are quantified through skew-information-like quantities associated with the Kraus representation of a quantum operation. Their approach establishes trade-off relations directly at the dynamical level and allows interferometric quantities such as visibility and which-path information to be characterized in terms of the action of the channel itself. However, since there also exist other types of complementarity relations involving quantities such as predictability, coherence, and entanglement formulated at the level of quantum states, it is conceivable that analogous constraints may emerge at the dynamical level. In this context, a key challenge is to define a meaningful notion of predictability for quantum channels.
This perspective naturally connects with existing resource-theoretic frameworks, where coherence and entanglement have been extensively characterized as resources under suitable sets of free operations. Extending these ideas, one may envision a theory of complementarity at the level of channels, where different operational tasks---such as information transmission, interferometric visibility, and channel discrimination---are constrained by generalized duality or triality relations. Establishing such a framework could provide new insights into the role of quantum processes in complementarity, with potential implications for quantum communication, metrology, and the theory of open quantum systems.

Complementarity relations have been formulated along different, yet closely related, lines. On the one hand, duality relations involving predictability and coherence have been extensively studied as a way to quantify particle- and wave-like behaviors. On the other hand, alternative formulations based on distinguishability and coherence provide a complementary perspective, particularly in interferometric scenarios where which-path information is explicitly accessible. In addition, triality relations incorporating predictability, coherence, and entanglement have further enriched this framework, revealing the role of quantum correlations in shaping complementarity.
Moreover, connections between predictability, entanglement, and distinguishability have been established in specific contexts, suggesting that these quantities are not independent, but rather different facets of a more general structure~\cite{Qureshi2021b,Roy2022, Basso2022b}.
In this direction, it is natural to consider measures that interpolate between predictability and distinguishability, providing a unified description of particle-like behavior across different regimes. This naturally leads to an open question: whether it is possible to formulate a general complementarity relation that is obtained from the quantum postulates and that consistently combines predictability, distinguishability, and entanglement within a single framework, and under which conditions such a unification can be achieved.

Within the standard Copenhagen interpretation, BCP is introduced as an independent conceptual principle. It is therefore natural to ask how this principle manifests in alternative interpretations of the theory. For instance, in the de Broglie--Bohm interpretation, a given initial condition gives rise to well-defined particle trajectories, even though fundamental uncertainty remains in the initial positions. However, from the modern perspective developed throughout this review, complementarity can be formulated independently of any particular interpretation, either in terms of the non-commutative structure of quantum observables and within a theory-independent operational framework, as discussed in Sec.~\ref{sec:Cincomp}, or through the derivation of quantitative complementarity relations directly from the mathematical structure of quantum mechanics, as presented in Sec.~\ref{sec:QCP}. In this broader context, approaches based on generalized probabilistic theories provide a natural setting to investigate complementarity beyond quantum mechanics. This raises an open and conceptually intriguing question: to what extent do these different interpretations and generalized frameworks accommodate, reinterpret, or even bypass complementarity as a fundamental principle? In particular, it remains to be understood whether the quantitative complementarity relations derived in the standard formalism retain their meaning, or acquire a different status, within alternative interpretations of quantum theory and more general probabilistic frameworks.

Interferometric visibility has long been regarded as a signature of the wave-like behavior of quantum systems. However, it is not, in general, a fully reliable quantifier for this purpose. Its use can lead to ambiguities when applied to certain scenarios, as discussed in Sec.~\ref{sec:retro}. Moreover, interferometric visibility is not well suited for formulating generalized complementarity relations in higher-dimensional systems, such as qudits. However, one of its main advantages lies in its experimental accessibility, as it can be directly extracted from interference patterns with relatively low resource cost. In contrast, quantum coherence naturally generalizes the notion of wave-like behavior to qudits and provides a more robust and theoretically grounded quantifier. When formulated within the framework of quantum information theory, quantum coherence allows complementarity relations to be derived directly from the mathematical structure of quantum theory, thereby offering a more formal and systematic characterization of the complementarity principle, for example, completely avoiding retroinference and \textit{ad hoc} constructions of complementarity relations. These generalized complementarity relations, such as the ones presented in Sec.~\ref{sec:QCP}, are not restricted to the traditional wave-particle duality of a quantum system in interferometric setups, but instead apply more broadly to the characterization of local and global properties of quantum systems. The main challenge associated with quantum coherence lies in its experimental implementation: estimating coherence typically requires full or partial quantum state tomography, which becomes increasingly demanding in terms of experimental resources for high-dimensional systems. Developing more efficient and scalable methods to estimate quantum coherence, therefore, remains an important open issue.

As discussed in Sec.~\ref{sec:Rel_comp}, recent results by~\citet{Cepollaro2025} have shown that the combined contribution of entanglement and local coherence remains invariant under transformations between quantum reference frames.
These findings raise the intriguing possibility that such invariance may extend beyond these specific quantities to more general complementarity relations. In particular, one may speculate whether complete complementarity relations, such as Eq.~\eqref{eq:CCRhs}, could also exhibit invariance under quantum reference frame transformations. Whether this intuition can be rigorously shown remains an open question. It is conceivable that the structure of complementarity relations, particularly those expressed in terms of coherence, entropy, and predictability, encodes features that are naturally compatible with the relational character of quantum reference frames.

It is also worth emphasizing that the mixed-state extension of the complementarity relation based on the von Neumann entropy, Eq.~\eqref{eq:CCRBmix}, provides a complete relation in scenarios where the global state $\rho_{AB}$ is not pure, without relying on a purification with a third party. In this case, the balance between local quantities and correlations necessarily involves total correlations, typically quantified by the quantum mutual information, reflecting both classical and quantum contributions. This raises the question of whether similarly complete complementarity relations, as the ones discussed in Sec.~\ref{sec:QCP}, can be formulated using alternative quantifiers, where different measures of quantum and classical correlations, as well as information-based quantities, are considered. An open issue is whether such generalized complete relations exist beyond the von Neumann entropy framework, or whether this structure is, in some sense, unique due to the fundamental role played by von Neumann entropy in quantum information theory.

Continuous-variable systems are ubiquitous in quantum mechanics and quantum information, particularly in quantum optics, where canonical degrees of freedom such as position and momentum, or field quadratures, provide a natural platform for encoding and processing quantum information~\cite{Braunstein2005, Adesso2007}. It would therefore be highly desirable to extend the complementarity relations discussed throughout this review---especially triality relations---to the realm of continuous-variable systems. In this context, the notion of wave-particle duality must be reformulated in terms of phase-space distributions. From a modern perspective, continuous-variable quantum information offers a well-developed framework to address these questions, particularly within the Gaussian formalism~\cite{Weedbrook2012}, where states are fully characterized by first and second moments and admit a clear geometric representation in phase space. This structure suggests that complementarity relations may be recast in terms of covariance matrices, allowing for a natural generalization of duality and triality relations beyond finite-dimensional Hilbert spaces. In this direction, the program of extending complementarity to continuous-variable systems has already been initiated in certain settings. Building on earlier interferometric formulations~\cite{Jaeger1995},~\citet{Georgiev2021} investigated trade-off relations between one- and two-particle visibilities in bipartite entangled Gaussian states, formulated within the framework of continuous-variable quantum optics.

A further and even more challenging direction concerns the extension of complementarity relations to the framework of quantum field theory, where the underlying Hilbert spaces are infinite-dimensional and the standard notions of quantum states and observables require substantial refinement. In this setting, several quantities that play a central role in finite-dimensional formulations---such as entanglement entropy---become ill-defined due to ultraviolet divergences arising from the infinite number of degrees of freedom. Entanglement in quantum field theory is more naturally understood in terms of algebras of local observables rather than in terms of vectors in a Hilbert space (see~\cite{Witten2018} for a review of the subject).
In particular, local algebras are typically von Neumann algebras of type III, for which a reduced density matrix does not exist, rendering the standard definition of entanglement entropy inadequate. Within this algebraic framework, however, well-defined and physically meaningful quantities can still be constructed. Notably, the relative entropy---originally introduced by \citet{Araki1975}---admits a rigorous formulation in quantum field theory through Tomita--Takesaki modular theory~\cite{Summers2005} and satisfies key properties such as positivity, monotonicity, and strong sub-additivity, generalizing the familiar results of finite-dimensional quantum systems. From the perspective of complementarity, these features raise profound conceptual and technical challenges. Any attempt to generalize duality or triality relations to quantum field theory must confront the absence of well-defined measures of quantum coherence, distinguishability, and predictability, as well as the intrinsic divergences of entropy-based quantities. At the same time, the algebraic formulation suggests that new forms of complementarity may emerge, expressed in terms of modular operators, relative entropy, or other information-theoretic quantities adapted to infinite-dimensional settings. Whether a meaningful and general formulation of complementarity---particularly in its quantitative forms---can be established in quantum field theory remains an open question.

%---------------------
\section{Conclusions}
\label{sec:conc}

In this review, we have revisited the principle of complementarity from its original conceptual formulation to its modern quantitative developments, highlighting its central role in shaping our understanding of quantum phenomena. Initially introduced by Bohr as a fundamental limitation on the simultaneous applicability of classical concepts, complementarity emerged as a guiding principle to reconcile the mutually exclusive wave- and particle-like descriptions observed in paradigmatic scenarios such as the two-slit and Mach--Zehnder experiments. 

Over time, this qualitative notion has evolved into a precise and quantitative framework through the development of complementarity relations. In particular, duality relations have provided rigorous trade-offs between wave-like and particle-like properties, where particle behavior is quantified either through predictability---capturing prior knowledge about the system---or through distinguishability, which quantifies the accessible which-path information. These complementary approaches offer distinct yet consistent perspectives on the particle aspect of quantum systems, reflecting different operational scenarios. On the wave side, the traditional role of interferometric visibility has progressively been generalized to more robust and widely applicable measures based on quantum coherence. This shift has allowed for a unified and information-theoretic characterization of wave-like behavior, particularly suited for high-dimensional systems and for formulations that avoid ambiguities primarily associated with interferometric visibility. 

Beyond duality, the framework has been further enriched by the introduction of triality relations, which incorporate quantum correlations---most notably entanglement---alongside predictability and coherence. These relations reveal that complementarity is not merely a trade-off between two competing properties, but rather part of a broader structure in which different quantum resources are quantitatively interconnected. In particular, entanglement has been shown to mediate the relationship between distinguishability and predictability. A key insight emphasized throughout this review is that complementarity relations can be systematically derived from the mathematical structure of quantum mechanics. In this sense, complementarity can be understood as a manifestation of fundamental constraints encoded in the density operator formalism, linking coherence, purity, and correlations in a unified manner. 

We have also discussed a variety of extensions and applications that demonstrate the broad scope of complementarity in contemporary quantum physics. These include its connections with entropic uncertainty relations, contextuality, and incompatible observables, as well as its role within resource-theoretic frameworks, where quantities such as coherence, entanglement, and predictability acquire operational meaning. Furthermore, complementarity has been shown to play a functional role in quantum information processing tasks, including entanglement distribution protocols and quantum thermodynamic settings, where wave-particle duality manifests in the statistical properties of work and heat.

Finally, we have outlined several open questions and directions for future research. Among them, the extension of complementarity relations to relativistic and quantum field-theoretic settings remains a major challenge, due to the lack of well-defined measures of coherence, distinguishability, and entanglement in infinite-dimensional systems. In this context, alternative formulations based on algebraic methods and information-theoretic quantities such as relative entropy may provide promising avenues for generalization.

In summary, complementarity has evolved from a foundational principle into a rich and versatile framework that connects core aspects of quantum theory. Its modern formulations using quantum information science not only deepen our conceptual understanding of quantum mechanics but also provide powerful tools for analyzing and exploiting quantum phenomena in a wide range of physical contexts.

\begin{acknowledgments}
This work was supported by S\~{a}o Paulo Research Foundation (FAPESP), Grant No.~2025/07325-0, by the National Council for Scientific and Technological Development (CNPq) under Grant No. 300083/2025-4, by the Research Support Foundation of the State of Rio Grande do Sul (FAPERGS) under Grant No. 25/2551-0002608-3, by the 
Coordination for the Improvement of Higher Education Personnel (CAPES), 
Grant No. 88887.827989/2023-00, and by the National Institute for the Science and Technology of Applied Quantum Computation (INCT-CQA) under Grant No. 408884/2024-0.
\end{acknowledgments}

%
%---------------------------
%

\end{document}